\documentclass[fleqn, usenatbib, useAMS]{mnras}

\usepackage{ae,aecompl}
\usepackage{amsmath}
\usepackage{amssymb}
\usepackage{bm}
\usepackage{braket}
\usepackage{dcolumn}
\usepackage{diagbox} 
\usepackage{epsfig}
\usepackage{esint}
\usepackage[T1]{fontenc}
\usepackage{slashed}
\usepackage{graphicx}
\usepackage[scr=boondox,  
            cal=esstix]   
           {mathalpha}
\usepackage{widetext}

\numberwithin{equation}{section}

\usepackage{hyperref}
\hypersetup{
linkcolor=blue,         
citecolor=magenta,      
filecolor=magenta,      
urlcolor=magenta}       

\title[Hotspots \& Photon Rings: Static Spacetimes]{Hotspots and Photon Rings in Spherically-Symmetric Spacetimes}

\author{
Prashant Kocherlakota,$^{1,2,3}$
Luciano Rezzolla,$^{3,4,5}$ 
Rittick Roy,$^6$ and
Maciek Wielgus$^7$
\\
$^1$Black Hole Initiative at Harvard University, 20 Garden St., Cambridge, MA 02138, USA\\
$^2$Center for Astrophysics, Harvard \& Smithsonian, 60 Garden St., Cambridge, MA 02138, USA\\
$^3$Institut f{\"u}r Theoretische Physik, Goethe-Universit{\"a}t, Max-von-Laue-Str. 1, 60438 Frankfurt, Germany\\
$^4$School of Mathematics, Trinity College, Dublin 2, Ireland\\
$^5$Frankfurt Institute for Advanced Studies, Ruth-Moufang-Str. 1, 60438 Frankfurt, Germany\\
$^6$Anton Pannekoek Institute for Astronomy, University of Amsterdam, Science Park 904, 1098 XH, Amsterdam, The Netherlands\\
$^7$Max-Planck-Institut f{\"u}r Radioastronomie, Auf dem H{\"u}gel 69, D-53121 Bonn, Germany
}

\date{}

\begin{document}
\label{firstpage}
\pagerange{\pageref{firstpage}--\pageref{lastpage}}
\maketitle


\begin{abstract}
Future black hole (BH) imaging observations are expected to resolve finer features corresponding to higher-order images of hotspots and of the horizon-scale accretion flow. In spherical spacetimes, the image order is determined by the number of half-loops executed by the photons that form it. Consecutive-order images arrive approximately after a delay time of $\approx\pi$ times the BH shadow radius. The fractional diameters, widths, and flux-densities of consecutive-order images are exponentially demagnified by the lensing Lyapunov exponent, a characteristic of the spacetime. The appearance of a simple point-sized hotspot when located at fixed spatial locations or in motion on circular orbits is investigated. The exact time delay between the appearance of its zeroth and first-order images agrees with our analytic estimate, which accounts for the observer inclination, with $\lesssim 20\%$ error for hotspots located about $\lesssim 5M$ from a Schwarzschild BH of mass $M$. Since M87$^\star$ and Sgr A$^\star$ host geometrically-thick accretion flows, we also explore the variation in the diameters and widths of their first-order images with disk scale-height. Using a simple ``conical torus'' model, for realistic morphologies, we estimate the first-order image diameter to deviate from that of the shadow by $\lesssim 30\%$ and its width to be $\lesssim 1.3M$. Finally, the error in recovering the Schwarzschild lensing exponent ($\pi$), when using the diameters or the widths of the first and second-order images is estimated to be $\lesssim 20\%$. It will soon become possible to robustly learn more about the spacetime geometry of astrophysical BHs from such measurements.
\end{abstract}



\section{Introduction}
\label{sec:Sec1_Introduction}

The recent images of M87$^\star$ \citep{EHTC+2019a} and Sgr~A$^\star$ \citep{EHTC+2022a} obtained by the Event Horizon Telescope (EHT) Collaboration show that these supermassive objects can be accurately described as Kerr black holes (BHs) \citep{EHTC+2019f, EHTC+2022f}, the vacuum, spinning BHs of general relativity (GR). 

The Kerr BH spacetime contains a photon shell $\mathrm{S}$, which is the union of a set of spherical surfaces, each of which admits bound photon orbits \citep{Teo2003}. Of these, there are also two planar (circular) orbits at the equator \citep{Bardeen1973} that are fundamentally tied to the symmetries of the spacetime. The (exterior; see, e.g., \citealt{Grenzebach2016}) photon shell of a Kerr BH encloses its horizon and demarcates the fate of photons approaching it into those that (a) fall into its interior and eventually end up inside the BH, (b) remain on one of the spheres in $\mathrm{S}$, or (c) escape to faraway observers, depending on their four-momenta or, equivalently, their impact parameters. 

Photon orbits of type (b) are the bound orbits, and of particular interest since they determine the BH shadow boundary curve or the $n\!=\!\infty$ critical curve $\mathscr{C}_\infty$ on the image plane \citep{Synge1966, Bardeen1973}. This is because $\mathscr{C}_\infty$ is the gravitationally-lensed projection of $\mathrm{S}$ on the observer's image plane. Since photons orbits are described by null geodesics in linear (Maxwell) electrodynamics, the Kerr shadow boundary curve $\mathscr{C}_\infty$ is determined by the Kerr metric as well as the observer's viewing angle. 

When a BH is lit up by a source of emission such as hot inflowing gas, the central intensity depression that appears in the image can typically be related to the BH shadow (region interior to $\mathscr{C}_\infty$): Photon orbits that cross the horizon [type (a)] have shorter paths through the spacetime than those that do not [type (c)] and thus pick up lesser emission from the hot plasma present outside the BH, leading to much smaller intensities on the image plane in the pixels that they arrive in \citep{Jaroszynski+1997, Narayan+2019, Bronzwaer+2021, Bronzwaer+2021a, Bauer+2022, Ozel+2021, Younsi+2021, Kocherlakota+2022, EHTC+2022f}. This does not mean that the intensity maximum in the image coincides exactly with the shadow boundary but that there is a strong association between the two. 

Indeed the diameter of the emission ring in the EHT images depends not just on the spacetime geometry of M87$^\star$ or of Sgr A$^\star$ but also on the arrangement, flow velocity, emissivity, and absorptivity of the accreting material (``non-gravitational physics'') in their vicinity. Nevertheless, the set of sizes of the observed bright emission ring and that of the shadow boundary can be empirically related through synthetic images obtained from simulations, thereby allowing for an inference of the shadow diameter of M87$^\star$ \citep{EHTC+2019f, Psaltis+2020, Kocherlakota+2021} and of Sgr A$^\star$ \citep{EHTC+2022f} from the measured emission ring diameter. An excellent analysis of this point is reported in Sec. 3. of \cite{EHTC+2022f}.

This discussion illustrates how the photon shell and the shadow boundary curve play a fundamental role in determining the properties of BH images. Furthermore, it also elucidates how several confounding effects must be carefully accounted for before extracting information regarding the shadow boundary curve, such as its median diameter. 

One of the exciting prospects for future black hole imaging observations performed at higher angular resolutions and flux-sensitivities, with next-generation microarcsecond resolution instruments achieved through space-based very long baseline interferometry \citep{Gralla+2020b, Gurvits+2022, Kurczynski2022}, is the direct detection of the ``photon ring'' of a BH \citep{Gralla+2019, Johnson+2020}, which forms the focus of our study here. Observables associated with the photon ring are relatively less sensitive to the non-gravitational emission physics and, thus, facilitate robust direct measurements of the BH spacetime geometry.

Photons that appear in a region close to the shadow boundary $\delta\mathscr{C}_\infty$ can have orbits that access the region close to the photon shell $\delta\mathrm{S}$. Since photon orbits close on themselves in $\mathrm{S}$, i.e., they have divergent angular deflections, it is natural to expect that $\delta\mathrm{S}$ is also a region of strong gravitational lensing \citep{Luminet1979, Ohanian1987, Virbhadra+2000, Claudel+2001, Bozza2002, Bozza+2007}. More specifically, in static and spherically-symmetric spacetimes (Sec. \ref{sec:Sec3_Scaling_Relations}), photons that appear in $\delta\mathscr{C}_\infty$ at an image plane radius of $\eta$, and which access $\delta\mathrm{S}$, have orbits whose deflection angles $\slashed{\Delta}\vartheta$ diverge logarithmically in the limit of approach to the shadow boundary radius $\eta_{\mathrm{PS}}$, i.e., $\lim_{\eta \rightarrow \eta_{\mathrm{PS}}} \slashed{\Delta}\vartheta(\eta) \propto \ln{|\eta - \eta_{\mathrm{PS}}|}$ \citep{Luminet1979, Ohanian1987, Bozza+2007, Stefanov+2010, Gralla+2019, Gralla+2020a, Johnson+2020}. The slope of this divergence is given by the lensing Lyapunov exponent $\gamma_{\mathrm{PS}}$, which is determined purely by the spacetime metric. 

The region $\delta\mathscr{C}_\infty$ on the image plane is referred to as the photon ring%
\footnote{Not to be confused with the light ring, which is a bound, planar null geodesic in the bulk of space \citep{Cunha+2018}.} %
and exhibits a rich substructure (barring when the emitting region is perfectly spherical; \citealt{Vincent+2022}). For an extended source of emission in the bulk, such as an accretion disk, the photon ring is comprised of a series of discrete -- and often overlapping -- ``subrings'' that are indexed by the number of half-loops executed by the photons that arrive in them (see Sec. \ref{sec:Sec2dot2_Image_Order} for a precise definition). Furthermore, these subrings are organized self-similarly on the image plane, with the critical exponent $\gamma_{\mathrm{PS}}$ governing this self-similarity. More specifically, the median (over the image plane polar angle) radii of consecutive order images, $\langle \eta_{n+1}\rangle$ and $\langle \eta_n\rangle$, exhibit a scaling relation in their deviation from the shadow radius as $\langle \eta_{n+1} - \eta_{\mathrm{PS}}\rangle \approx \mathrm{e}^{-\gamma_{\mathrm{PS}}}\langle \eta_n - \eta_{\mathrm{PS}}\rangle$. Similarly, their widths, $w_{n+1}$ and $w_n$, are related as $w_{n+1} \approx \mathrm{e}^{-\gamma_{\mathrm{PS}}}w_n$, thus leading to a diminishing subring flux density with increasing subring-order \citep{Gralla+2019, Gralla+2020a, Johnson+2020}. These approximate scaling relations apply very well to the case of equatorial sources viewed by an observer on the BH spin-axis. In Sec. \ref{sec:Sec3_Scaling_Relations} below we obtain approximate scaling relations for arbitrary emitter and observer configurations.

Therefore, the lensing Lyapunov exponent is, in principle, a new observable that can be used to gain additional information about the spacetime geometry. Indeed, the implications of a measurement of this exponent for the black hole ``no hair'' conjecture have recently been explored (see, e.g., \citealt{Wielgus2021, Ayzenberg2022, Broderick+2023, Staelens+2023, Salehi+2023, Ayzenberg+2023, Kocherlakota+2024a}). The lensing Lyapunov exponent is tied to a fundamental ``phase space'' Lyapunov exponent \citep{Cardoso+2009, Stefanov+2010, Yang+2012}, which governs the expansion of critical null congruences at the photon sphere. These Lyapunov exponents are also wonderfully linked to the damping frequencies of BH quasinormal modes \citep{Cardoso+2009, Yang+2012, Chen+2022}.

For the EHT targets, M87$^\star$ and Sgr A$^\star$, their photon subrings are the higher-order images of the horizon-scale accretion flow. Naturally, their structures are determined both by the properties of the hot plasma as well as by their respective spacetime geometries. Thus, in future improved experiments, the close vicinity of the photon shell in the bulk as well as of the critical curve on the boundary of space will soon play a key role in determining BH images, and even BH movies. With these experiments on the horizon, it is imperative to gain a deeper understanding of how the characteristic features of subrings (e.g., their diameters, widths, flux densities, asymmetries) vary with the properties of the emitting region. 

Significant progress has already been made in understanding the appearance of photon subrings cast by geometrically-thin disks in the Kerr metric (see, e.g., \citealt{Gralla+2020b, Palumbo+2022, Paugnat+2022, Tsupko2022, Bisnovatyi-Kogan+2022}) as well as in non-Kerr spacetimes \citep{Wielgus2021, Cardenas-Avendano+2023a}, affording us several valuable insights. 

Astrophysical BHs that are presently relevant such as M87$^\star$ or Sgr A$^\star$, however, do not host geometrically-thin accretion disks \citep{EHTC+2019e, EHTC+2022e}. Instead, the accretion flow extends all the way down to the event horizon, and is geometrically-thick, with typical thicknesses of $|h/r| \lesssim 0.4$ \citep{Narayan+2022}. Here $h$ denotes the scale-height of the flow and $r$ is the radial coordinate. Equivalently, this implies that the ``faces'' of the volume harboring the accretion flow lie at colatitudes $\vartheta = \pi/2 \pm 0.2$. 

Furthermore, the inflowing plasma is magnetized and drags magnetic field lines to the BH and supports them there \citep{Narayan+2003}. This can lead to relativistic outflows, or jets, of highly-magnetised plasma along the spin axis via the \cite{Blandford+1977} mechanism, which describes an electromagnetic Penrose process \citep{Lasota+2014}. These jets have generalized parabolic profiles \citep{Narayan+2022} and their boundaries (``jet sheaths'') can produce a significant amount of emission \citep{Sironi+2021} as well. This is also confirmed by observations (see, e.g., \citealt{Kim+2018, Janssen+2021, Lu+2023}), and constitutes yet another nontrivial source of nonequatorial emission. In particular, while a Schwarzschild BH does not produce powerful jets (due to the absence of an ergoregion), there is still a significant nonequatorial outflow component (``disk wind'').

Motivated by the above considerations, one of our primary goals here is to present a comprehensive survey of the impact of a varying source morphology on the observed photon ring structure (see also \citealt{Vincent+2022}). Another is to estimate the error in inferring the critical lensing exponent $\gamma_{\mathrm{PS}}$ from potential future joint measurements of diameters or widths of a pair of subrings for varied source morphologies. 

We achieve this by working with a simple yet flexible three-parameter model for the shape of the emitting region, which can be used to interpolate smoothly between a geometrically-thin-disk and a sphere. More specifically, we employ conical surfaces, as pictured in the top-right panel of Fig. \ref{fig:FigB2_Schw_Higher_Order_Images}, to generate an axially-symmetric wedge-shaped region -- a ``conical torus'' -- with the three parameters being used to set the locations of its inner $r_{\mathrm{in}}$ and outer $r_{\mathrm{out}}$ surfaces as well as its geometrical-thickness or half-opening angle $\vartheta_{1/2}$ (this defines the latitudes of the bounding surfaces). We note that this setup also allows one to easily model a jet-like feature (see, e.g., \citealt{Papoutsis+2023, Chang+2024}). We will also carefully consider the impact of the observer viewing angle, which may be useful in preparation for observations of the photon ring of Sgr A$^\star$. Finally, to cleanly isolate the impact of source morphology, we fix the spacetime geometry here. For simplicity, in this demonstrative study, we choose it to correspond to that of a Schwarzshild BH, i.e., of a vacuum nonspinning BH in GR. 

The restrictions above ensure that our model complexity is minimal but sufficient for our present purposes, and, crucially, enable intuitive connections between the morphological variations of the emitting region in the bulk and that of the photon subrings on the image plane. For direct practical applications to observations, one must additionally account for realistic astrophysical effects associated with the accreting plasma. Excellent investigations considering the effects of the plasma synchrotron emissivity profile, its velocity profile, and the increased optical depth at submillimeter frequencies experienced by higher-order photons traversing greater path-lengths through the plasma are well underway \citep{Palumbo+2022, Paugnat+2022, Vincent+2022, Chang+2024}. 

While we do not account for the various aforementioned non-gravitational physical effects, we stress that the subring diameters and widths that we report here nonetheless provide a useful \textit{upper bound} for a particular morphology of the emitting region. These are obtained here by finding the inner and outer edges of the higher-order images of the inner and outer boundaries of the bulk emitting region respectively. For example, given two emitting regions $A$ and $B$ of identical morphologies (and in particular of the same radial extent), it is straightforward to see that the inferred diameters and widths of the photon subrings cast by $A$ are larger than those by $B$, if the emissivity falls off more rapidly with radius in $B$ as compared to $A$ (see, e.g., Fig. 7 of \citealt{Kocherlakota+2022} for an analogous conclusion for their direct images). The gravitational and the Doppler redshifts will also introduce additional asymmetry in the image. However, these important physical effects cannot cause their median image sizes or widths to be greater than the upper bounds we report here.

It is worth noting that the plasma emissivity profile should fall off at least as $\sim r^{-4}$ \citep{Narayan+2019, Kocherlakota+2022} from considerations of energy conservation and that it likely falls off much faster (exponentially) in realistic scenarios \citep{Porth+2019, Chael+2021}. The latter establishes a realistic length-scale for the outer boundary of the presently relevant horizon-scale emission zone.

Thus far, we have focussed on the higher-order images of the entire accretion flow. However, compact flux eruption or flaring events associated with Sgr A$^\star$, from within ten gravitational radii, have been observed across a multitude of wavelengths \citep{Baganoff+2001, Eckart+2006, Gravity+2018, Murchikova+2021, Witzel+2021}. Magnetic reconnection in an equatorial current sheet within the accreting plasma can lead to local heating \citep{Ripperda+2021}. Such hot transient compact emission sources -- or ``hotspots'' -- can also move within the equatorial plane (see, e.g., \citealt{Dexter+2020}). Propositions that the time delay between the appearance of the primary ($n=0$) and secondary ($n=1$) images of a hotspot can potentially be measured through observations have been forwarded (see, e.g., \citealt{Tiede+2020, Ball+2021}). This may be possible with future upgrades and ground extensions to the EHT \citep{Doeleman+2019, Johnson+2023}, and has been suggested as a means to infer the spin of Sgr A$^\star$ \citep{Wong2021}. \cite{Sahu+2013} have proposed utilizing the time delay between the appearance of Einstein rings to gather information about the background spacetime geometry, relevant here when a hotspot moves across a caustic.

Below (see eq. \ref{eq:Delay_Time}), we find the \textit{characteristic} delay time between the appearance of higher-order ($n \geq 1$) images to be intimately linked to the shadow size, in general spherically-symmetric spacetimes. Our last goal here is to estimate the error in inferring the shadow size from a measurement of the exact time delay between the appearance of the primary and secondary images of a hotspot in a Schwarzschild BH spacetime. We will also comment on how the bulk hotspot motion is encoded in the evolution of its images on the image plane, which may provide insights into their physical origin.
 
The broad outline of the paper is as follows. We review the mathematical description of higher-order images in general spherically-symmetric spacetimes in Sec. \ref{sec:Sec2_Theory_Review}. The universal photon ring scaling relations are reported in Sec. \ref{sec:Sec3_Scaling_Relations}. The implications of an observation of a secondary ($n=1$) image of a point-sized hotspot orbiting a Schwarzschild BH (relevant for Sgr A$^\star$) are discussed Sec. \ref{sec:Sec4_Hotspots}. Results on the qualitative as well as quantitative variations in the morphological properties of photon subrings cast by geometrically-thin disks for varying viewing angles, and by geometrically-thick disks for a face-on observer (relevant for M87$^\star$) in a Schwarzschild spacetime are described in Sec. \ref{sec:Sec5_Photon_Rings}. Thus, Sec. \ref{sec:Sec4_Hotspots} applies the theoretical framework developed in previous sections towards an investigation of higher-order image formation for point emitters, building on which Sec. \ref{sec:Sec5_Photon_Rings} describes the same but for extended sources (which, for our purposes, are comprised of point sources). We end in Sec. \ref{sec:Sec6_Conclusions} by presenting a summary of our findings. 

We will reserve $\nu$ to denote a frequency throughout. For example, $I_\nu$ will be used to denote specific intensity ($\mathrm{W}~\mathrm{m}^{-2}\mathrm{sr}^{-1}\mathrm{Hz}^{-1}$) and not the component of a one-form $\textbf{I}$. We employ geometrized units throughout, in which $G=c=1$. Some technical but instructive discussions have been relegated to the appendices.


\section{Strong Gravitational Lensing \& Higher-Order Images: Brief Review}
\label{sec:Sec2_Theory_Review}

In this section, we begin with a brief review of salient aspects of strong gravitational lensing in Sec. \ref{sec:Sec2dot1_deltaS}, as needed for a study of the photon ring, and then revisit the ordering of images for general emitter and observer configurations in Sec. \ref{sec:Sec2dot2_Image_Order}.


\subsection{Photon Orbits Close to the Photon Sphere}
\label{sec:Sec2dot1_deltaS}

The line element of an arbitrary static and spherically-symmetric spacetime can be expressed in spherical-polar coordinates $x^\alpha = (t, r, \vartheta, \varphi)$ as,
\begin{equation} \label{eq:Eq2dot1_Static_Spacetime} 
\mathrm{d} s^2 = \mathscr{g}_{\alpha\beta}\mathrm{d}x^\alpha\mathrm{d}x^\beta 
= -f~\mathrm{d}t^2 + \frac{g}{f}~\mathrm{d}r^2 + R^2~\mathrm{d}\Omega_2^2\,,
\end{equation}
where the metric functions $f, g$, and $R$ are functions of $r$ alone, and $\mathrm{d}\Omega_2^2 = \mathrm{d}\vartheta^2 + \sin^2{\vartheta}~\mathrm{d}\varphi^2$ is the standard line element on a unit 2-sphere. We will assume reasonably that $g>0$ everywhere and that $R > 0$ except at the center ($R=0$). Setting $R(r)=r$ yields the metric in areal-polar coordinates. The metric above describes a BH spacetime if $f(r)$ admits real, positive zeroes (with $R>0$), the largest of which locates the event horizon, which we denote by $r_{\mathrm{H}}$. For the Schwarzschild BH metric in particular, we have $f(r) =  1 - 2M/r$, $g(r) = 1$, and $R(r) = r$. Clearly, its event horizon is located at $r_{\mathrm{H}} = 2M$.

Due to spherical symmetry, (1.) all geodesic orbits are simply copies of those that lie, e.g., in the equatorial plane or a meridional plane ($\varphi=\mathrm{const.}$), and (2.) without loss of generality, we can set the observer to lie on the north pole ($\vartheta_{\mathrm{o}} = 0$). The latter has the profoundly simplifying consequence that \textit{all} photons arriving on the screen move only on meridional planes through the bulk of space. Further discussion on this can be found in Appendix \ref{app:AppA_Planarity}. Thus, henceforth, we will restrict our attention to meridional photon orbits. Our analytical methods here derive the crux of their power from these simplifications, and are what attract us to a complete consideration of characterizing higher-order images of hotspots and the entire accretion flow first in nonspinning spacetimes. 

The radial $\mathscr{R}$ and angular $\Theta$ effective potentials for arbitrary meridional null geodesics in the spacetime \eqref{eq:Eq2dot1_Static_Spacetime} are defined from the radial and angular ``energy equations,'' $\mathscr{R}(\eta, r) = (\dot{r}/E)^2$ and $\Theta(\eta, r) = (\dot{\theta}/E)^2$, as (see, e.g., \citealt{Wald1984}; See also Appendix \ref{app:AppA_Planarity})
\begin{align}
\label{eq:r_Effective_Potential}
\mathscr{R}(\eta, r) =&\ g^{-1}\left[1 - \eta^2 f R^{-2}\right]\,, \\
\label{eq:th_Effective_Potential}
\Theta(\eta, r) =&\ \eta^2/R^4.
\end{align}
In the above, $E$ is the energy of the photon, and $\eta = |p_\vartheta|/E$ is its Carter constant, a measure of its angular momentum $p_\vartheta$. The latter is also its impact parameter, i.e., the radius at which the photon appears on the image plane (\citealt{Bardeen1973}; see also Ch. 4 of \citealt{Grenzebach2016}). The overdot represents a derivative w.r.t. the affine parameter $\lambda$ along the geodesic and, thus, $\dot{r}$ and $\dot{\vartheta}$ are the coordinate radial and polar velocities of the photon. 

Due to the strong gravity near ultracompact objects, it is generically possible for photons to move on circular orbits (see, e.g., \citealt{Claudel+2001, Cunha+2020, Ghosh+2021}), where $\dot{r} = \ddot{r} = 0$. These correspond to the bound photon orbits of the spacetime, and such photons do not reach faraway observers. These equations are respectively equivalent to $\mathscr{R} = 0$ and $\partial_r \mathscr{R} = 0$, i.e., \citep{Atkinson1965}
\begin{align} 
\label{eq:Shadow_Size}
\eta_{\mathrm{PS}} =&\ R_{\mathrm{PS}}/\sqrt{f_{\mathrm{PS}}}\,, \\
\label{eq:Photon_Sphere}
0 =&\ (\partial_r f)/f - (\partial_r R^2)/R^2\,.
\end{align}
A solution, $r=r_{\mathrm{PS}}$, to the above differential equation locates a photon sphere in the bulk of space. In the above, and henceforth, the subscript ``$\mathrm{PS}$'' indicates that the function is evaluated at $r=r_{\mathrm{PS}}$. 

The radial deviation $\delta r(\lambda)$ of a photon emitted from a distance $\delta r(0)$ from the photon sphere, with the same (critical) impact parameter as the one on a circular photon orbit $\eta = \eta_{\mathrm{PS}}$, and with either positive ($+_r$) or negative radial velocity ($-_r$), is given as (see, e.g., \citealt{Cardoso+2009})
\begin{equation} \label{eq:Phase_Space_Deviation}
\delta r(\lambda) = \delta r(0)~\exp{\left[\pm_r\hat{\kappa}_{\mathrm{PS}}E\lambda\right]}\,.
\end{equation}
Here $\hat{\kappa}_{\mathrm{PS}}$ is the null geodesic phase space Lyapunov exponent and is given as
\begin{equation} \label{eq:Phase_Space_Lyapunov_Exponent}
\hat{\kappa}_{\mathrm{PS}}^2 := \frac{\partial_r^2 \mathscr{R}(\eta_{\mathrm{PS}}, r_{\mathrm{PS}})}{2} =  -\frac{1}{2g_{\mathrm{PS}}}\left[\frac{\partial_r^2 f_{\mathrm{PS}}}{f_{\mathrm{PS}}} - \frac{\partial_r^2 R^2_{\mathrm{PS}}}{R^2_{\mathrm{PS}}}\right]\,.
\end{equation}
Thus the phase space Lyapunov exponent determines the stability of the radial fixed point as being unstable (stable) if $\hat{\kappa}_{\mathrm{PS}}$ is real (imaginary). For the Schwarzschild BH spacetime, one finds $\hat{\kappa}_{\mathrm{PS}} = 1/(\sqrt{3}M)$. For a loose comparison, we note that the surface gravity $\kappa_{\mathrm{H}}$ of the Schwarzschild horizon is $\kappa_{\mathrm{H}} = 1/(4M)$.

The equation above \eqref{eq:Phase_Space_Deviation} is essentially the $r-$component of the Jacobi or geodesic deviation equation, $\mathrm{d}^2\hat{\zeta}^\delta/\mathrm{d}\lambda^2 = -\mathbb{R}^{\delta}_{\ \alpha\rho\beta}k_{\mathrm{PS}}^\alpha\hat{\zeta}^\rho k_{\mathrm{PS}}^\beta$ (see, e.g., \citealt{Poisson2004}), where $\mathbb{R}$ is the Riemann tensor, $\hat{\zeta}^\alpha$ is an appropriate deviation vector between null geodesics in spacetime, and $k_{\mathrm{PS}}^\alpha$ is tangent to the circular null geodesic in particular (see eq. \ref{eq:NG_kPS}). More explicitly,
\begin{equation} 
\frac{\mathrm{d}^2\hat{\zeta}^r}{\mathrm{d}\lambda^2} = -\left[\mathbb{R}^r_{\ trt}\left(k_{\mathrm{PS}}^t\right)^2 + \mathbb{R}^r_{\ \vartheta r\vartheta}\left(k_{\mathrm{PS}}^\vartheta\right)^2\right]\hat{\zeta}^r = E^2\hat{\kappa}^2_{\mathrm{PS}}\hat{\zeta}^r\,. \nonumber
\end{equation}

While there exist static spacetimes admitting multiple roots to eq. \ref{eq:Photon_Sphere} (see, e.g., \citealt{Wielgus+2020, Gan+2021, Guo+2022}), we restrict our focus here to those that admit a single unstable photon sphere outside the horizon (see \citealt{Cardoso+2014, Keir2016, Cunha+2017b} for related discussion).

The radius of the shadow boundary curve, which is the gravitationally-lensed projection of the (unstable) photon sphere on the image plane, is simply $\eta = \eta_{\mathrm{PS}}$. It is worth noting that the location of the photon sphere as well as the size of the shadow are independent of the metric function $g$ (see, e.g., \citealt{Psaltis+2020}). For the Schwarzschild BH metric in particular, its photon sphere is located at $r_{\mathrm{PS}} = 3M$, and its shadow size is $\eta_{\mathrm{PS}} = \sqrt{27}M$.

More generally, from the effective radial potential \eqref{eq:r_Effective_Potential}, we can see that photons that initially approach the BH ($\dot{r} < 0$) will either reach the horizon or hit a ``radial turning point'' at some $r = r_{\mathrm{tp}}(\eta)$, where their radial velocity vanishes ($\dot{r} = 0$), 
\begin{equation} \label{eq:Radial_Turning_Point_Equation}
\mathscr{R}(\eta, r_{\mathrm{tp}}(\eta)) = 0\,.   
\end{equation}
The latter set of photons will then move outward and reach asymptotic infinity. These are photons with large angular momenta $\eta$ (we set $E=1$ without loss of generality) that simply experience too much ``centrifugal force'' at their closest point of approach. It is straightforward to see from eqs. \ref{eq:r_Effective_Potential} and \ref{eq:Shadow_Size} that these are all photons with angular momenta $\eta > \eta_{\mathrm{PS}}$. It is also clear that every point along a circular photon orbit is a radial turning point, $r_{\mathrm{tp}}(\eta_{\mathrm{PS}}) = r_{\mathrm{PS}}$. For further discussion on radial turning points, one can see Appendix A of \cite{Kocherlakota+2022}.

From the effective angular potential \eqref{eq:th_Effective_Potential}, it is also clear that meridional orbits do not admit any nontrivial polar turning points (where $\dot{\vartheta} = 0 = \Theta$). However, the spherical-polar coordinate system throws up two trivial turning points at $\vartheta = 0 $ and $\vartheta = \pi$, which are understood as follows. A photon approaching the north pole has $\dot{\vartheta} < 0$. After smoothly crossing it, it must have $\dot{\vartheta} > 0$, since the sense of rotation of an orbit about the center ($r=0$) remains invariant. We can ignore these trivial turning points but keep track of the sign of the initial angular velocity of the photon, which determines the sense of rotation or the ``polarity'' of the orbit. 

Thus, the total angular deflection $\slashed{\Delta}\vartheta^\pm$ experienced by a meridional photon with initial positive ($+$) or negative ($-$) polar velocity ($\dot{\vartheta}$) is given as,
\begin{align} \label{eq:Angular_Deflection_Expr}
\slashed{\Delta}\vartheta^\pm = \fint\mathrm{d}\vartheta =&\ \fint\dot{\vartheta}~\mathrm{d}\lambda = \pm E\fint\sqrt{\Theta}~\mathrm{d}\lambda \\
=&\ \fint\left(\dot{\vartheta}/\dot{r}\right)\mathrm{d}r = \pm \fint\sqrt{\Theta/\mathscr{R}}~\mathrm{d}r\,. \nonumber 
\end{align}
In the above, the slash in $\slashed{\Delta}\vartheta^\pm$ is used to denote that it is akin to an angular distance along the photon orbit, as opposed to an angular displacement. The slash in $\fint$ denotes that the integrals are path-dependent, and are evaluated along the worldline of the photon, $x^\mu = x^\mu(\lambda)$ (see also \citealt{Shaikh+2019b, Gralla+2020a}). This convention makes it convenient to introduce the winding number along individual photon orbits, as in eq. \ref{eq:Angle_Deflection_Higher_Order_Images_m}. 

For photons traversing the photon sphere, $\sqrt{\Theta} = \eta_{\mathrm{PS}}/R^2_{\mathrm{PS}}$, and it is clear from eq. \ref{eq:Angular_Deflection_Expr} that their angular deflection diverges. We can also anticipate that it becomes arbitrarily large for some photons ($|\mathscr{R}(\eta, r)| \rightarrow 0$) that closely approach it.

We can ``unslash'' the integrals in eq. \ref{eq:Angular_Deflection_Expr}, and make them path-\textit{independent} by splitting the photon orbit up into pieces over which the map $\lambda \mapsto r$ is bijective. The $r-$bijective pieces are naturally separated by the radial turning points \eqref{eq:Radial_Turning_Point_Equation}. Since we are only interested in photons that make it to a faraway observer present at $r = \infty$, as discussed below, we need only worry about a single radial turning point at most. Therefore, for all observable photons, we can write,
\begin{widetext}
\begin{align} \label{eq:del_vartheta}
\boxed{
\slashed{\Delta}\vartheta^\pm(\eta, r_{\mathrm{e}}) = 
\begin{cases}
\pm\eta\left[+_r \int_{r_{\mathrm{e}}}^\infty 1/(R^2\sqrt{\mathscr{R}})~\mathrm{d} r \right], 
& k^r_{\mathrm{e}} > 0\,; 
\ r_{\mathrm{e}} > \{r_{\mathrm{H}}~\mathrm{if}~\eta < \eta_{\mathrm{PS}},~ r_{\mathrm{tp}}(\eta)~\mathrm{if}~\eta \geq \eta_{\mathrm{PS}}\}\\ 
\pm\eta \left[-_r \int_{r_{\mathrm{e}}}^{r_{\mathrm{tp}}(\eta)} 1/(R^2\sqrt{\mathscr{R}})~\mathrm{d} r +_r  \int_{r_{\mathrm{tp}}(\eta)}^\infty 1/(R^2\sqrt{\mathscr{R}})~\mathrm{d} r \right], 
& k^r_{\mathrm{e}} \leq 0 \,; 
\mathrm{if}~\eta > \eta_{\mathrm{PS}}~\mathrm{and}~\ r_{\mathrm{e}} \geq r_{\mathrm{tp}}(\eta)
\end{cases}
}
\end{align}
\end{widetext}
In the above, $r_{\mathrm{e}}$ and $k^r_{\mathrm{e}}$ denote the emission radius and emission radial velocity of the photon respectively. 

This equation can be understood simply as follows. The top row corresponds to photons with positive radial velocity at emission ($k^r_{\mathrm{e}} > 0$). Since only photons with angular momenta $\eta \geq \eta_{\mathrm{ps}}$ possess radial turning points, these must be emitted from outside their outermost radial turning point to ensure they reach an asymptotic observer. Photons with smaller angular momenta must simply be emitted from outside the horizon. The bottom row corresponds to the case where photons are emitted with initial nonpositive radial velocities. Photons emitted with zero radial velocity ($k^r_{\mathrm{e}} = 0$) are, by definition, emitted from their orbital radial turning points ($r=r_{\mathrm{tp}}(\eta)$). From the next instant onwards, these photons must move away (``retreat'') from the BH to reach an asymptotic observer. Photons with initially negative radial velocities ($k^r_{\mathrm{e}} < 0$) can reach asymptotic observers only if they meet a radial turning point. The two pieces of the integral, thus, simply correspond to the angular deflection accumulated over the approaching ($-_r$) and retreating ($+_r$) legs of the photon orbit. For an intuitive pictorial representation of the above, we direct the reader to see the left panel of Fig. A1 of \cite{Kocherlakota+2022}. Finally, the overall $\pm$ simply makes explicit how photons that differ only in polarity experience identical amounts of angular deflections. For equatorial orbits, one can simply replace $\vartheta$ and $\eta$ by $\varphi$ and $\xi$ respectively in the discussion above.

The polarity of the photon orbit determines its image plane polar angle $\psi$. Two photons emitted from the same meridian (same $\varphi$) but with opposite polarities, appear on opposite sides of the image plane origin. Henceforth, we will use $\psi^\mp(\varphi)$ to denote keep track of this feature (see, e.g., \citealt{Bardeen1973, Luminet1979, Grenzebach2016, Tsupko2022}),
\begin{equation} \label{eq:Image_Plane_Polar_Angle}
\psi^\mp(\varphi) = 
\begin{cases}
 3\pi/2 - \varphi\,, & -\ (\mathrm{i.e.,}\ k^\vartheta_{\mathrm{e}} < 0) \\ 
 \pi/2 - \varphi\,, & +\ (\mathrm{i.e.,}\ k^\vartheta_{\mathrm{e}} > 0)
\end{cases}\,.
\end{equation}
The image plane origin corresponds to the location on the image plane from which the member of the ingoing principal null congruence at the observer's location enters their past light cone.

Putting everything together, photons emitted from an initial location $(r_{\mathrm{e}}, 0 \leq \vartheta_{\mathrm{e}} \leq \pi, 0 \leq \varphi_{\mathrm{e}} < 2\pi)$ that reach our observer, present on the $+z-$axis ($\vartheta_{\mathrm{o}} = 0$), appear on their image plane at $(\eta, \psi)$. The image plane polar angle $\psi$ is determined by eq. \ref{eq:Image_Plane_Polar_Angle}, which requires knowledge also of the initial sign ($\pm$) of the photon polar velocity, $k^\vartheta_{\mathrm{e}} = \dot{\vartheta}_{\mathrm{e}}$. Furthermore, the image plane radius $\eta$ is determined via eq. \ref{eq:del_vartheta} where
\begin{equation} \label{eq:Mod_Equation}
\slashed{\Delta}\vartheta^\pm = \left[\vartheta_{\mathrm{o}} -\vartheta_{\mathrm{e}}\right] \mod{2\pi} = -\vartheta_{\mathrm{e}} \mod{2\pi}\,.
\end{equation}
Eq. \ref{eq:del_vartheta} requires further knowledge of the initial sign of the photon radial velocity $k^r_{\mathrm{e}}$. 

Note, however, that equivalently, and quite wonderfully, the source $(r_{\mathrm{e}}, \vartheta_{\mathrm{e}}, \varphi_{\mathrm{e}})$ and observer $(r_{\mathrm{o}}, \vartheta_{\mathrm{o}}, \varphi_{\mathrm{o}})$ locations along with the image order $n$ (introduced below in eq. \ref{eq:Angle_Deflection_Higher_Order_Images_n}) completely fix the photon orbit (see eq. \ref{eq:Main_Integral_Equation}). The initial coordinate velocities (both $k^\vartheta_{\mathrm{e}}$ and $k^r_{\mathrm{e}}$) as well as the image location $(\eta, \psi)$ of that photon can consequently be obtained uniquely. This is discussed further in Appendix \ref{app:AppB_Lensing_Map}, where, additionally, the nonmonotonicity of the higher-order image radii with source colatitude, for sources present just outside the photon sphere, is also carefully described.


\subsection{Image Order}
\label{sec:Sec2dot2_Image_Order}

The modulo piece in the equation above \eqref{eq:Mod_Equation} is central to the notion of higher-order images. There exist photons emitted from the same initial spatial location and captured by an observer at the same final spatial location that have orbits differing in their total angular deflection
\begin{equation} \label{eq:Angle_Deflection_Higher_Order_Images_m}
\slashed{\Delta}\vartheta^\pm_m = - \vartheta_{\mathrm{e}} \pm 2\pi m\,,\ \ 0 < \vartheta_{\mathrm{e}} < \pi\,,
\end{equation}
where $m$ is a positive integer, called the winding number (see also \citealt{Luminet1979, Ohanian1987, Gralla+2019, Cunha+2020, Wei2020, Xu+2023}), and each distinct orbit is indexed by the pair $(m, \pm)$. 

The total angular deflection for this class of orbits increases in the sequence $\{(0, -), (1, +), (1, -), (2, +), \cdots \}$ taking values in $\{(0, \pi), (\pi, 2\pi), (2\pi, 3\pi), (3\pi, 4\pi), \cdots \}$ respectively. If we index these sets by $n = \{0, 1, 2, 3, \cdots \}$, then $n$ defines the order of the orbit or, equivalently, of the image and we can rewrite eq. \ref{eq:Angle_Deflection_Higher_Order_Images_m} as
\begin{equation} \label{eq:Angle_Deflection_Higher_Order_Images_n}
\boxed{
\slashed{\Delta}\vartheta^\pm_n(\vartheta_{\mathrm{e}}) = \pi/2 - \vartheta_{\mathrm{e}} + (-1)^{n+1}(2n+1)\pi/2\,,\ \ 0 < \vartheta_{\mathrm{e}} < \pi\,.
}
\end{equation}
Thus, we find, naturally, that even$-n$ photons are associated with a negative polarity whereas odd-$n$ photons are associated with a positive polarity.

As discussed above eq. \ref{eq:Angular_Deflection_Expr}, these polarities were introduced to track the sign of $\dot{\vartheta}$ at emission. We can understand the reason for this association as follows. Any ``direct'' or ``primary'' ($n=0$) photon that is emitted from $0 < \vartheta_{\mathrm{e}} < \pi$ and which reaches the observer present at $\vartheta_{\mathrm{o}} = 0$ must have had a negative polar angular velocity, $\dot{\vartheta} < 0$, along its entire orbit (thus, $\dot{\vartheta} < 0$ at emission). On the other hand, for the ``indirect'' or ``secondary'' ($n=1$) photon, which is emitted from $0 < \vartheta_{\mathrm{e}} < \pi$ and which reaches the observer present at the north pole, to experience an angular deflection in $(\pi, 2\pi)$, it must first head toward the south pole, cross it, and then move toward the north pole. Thus, it must have had a positive polar velocity at emission, and so a positive polarity. Similar reasoning applies to all even-order and odd-order photons, and can be readily verified from Fig. \ref{fig:FigB1_Schw_Higher_Order_Images}.

Photons emitted from the same (static) source appear on an observer's screen on a single line. This is because photon orbits are planar in spherically-symmetric spacetimes, and the source location, the observer location, and the center of space define this plane. Furthermore, due to their alternating polarities, consecutive-order images appear on opposite sides of the origin (cf. eq. \ref{eq:Image_Plane_Polar_Angle}). This shows that in spherically-symmetric spacetimes the rotation parameter $\delta_0$, introduced in eq. 72 of \cite{Gralla+2020a}, is trivial, i.e., $\delta_0 = \pi$. By extension, we can define the $n^{\mathrm{th}}-$order image of an extended source of emission as being formed by the set of photons that have index $n$.  

Since different-order photons experience different amounts of angular deflections, they must appear at different radii on the image plane, $\eta = \eta_n$. The latter are obtained simply by solving eq. \ref{eq:del_vartheta} with eq. \ref{eq:Angle_Deflection_Higher_Order_Images_n}, i.e., the integral equation,
\begin{equation} \label{eq:Main_Integral_Equation}
\boxed{
\slashed{\Delta}\vartheta^\pm(\eta_n, r_{\mathrm{e}}) = \slashed{\Delta}\vartheta_n^\pm(\vartheta_{\mathrm{e}})\,.
}
\end{equation}
Furthermore, the total time $\slashed{\Delta}t^\pm$ elapsed along a photon orbit and its total affine length $\slashed{\Delta}\lambda^\pm$, given respectively as
\begin{align} 
\label{eq:Total_Elapsed_Time}
\slashed{\Delta}t^\pm =&\ \fint\mathrm{d}t = \fint\left(\dot{t}/\dot{r}\right)\mathrm{d}r = \fint(1/f)/\sqrt{\mathscr{R}}~\mathrm{d}r\,, \\
\label{eq:Total_Affine_Length}
\slashed{\Delta}\lambda^\pm =&\ \fint\mathrm{d}\lambda = \fint\left(1/\dot{r}\right)\mathrm{d}r = \fint(1/E)/\sqrt{\mathscr{R}}~\mathrm{d}r\,, 
\end{align}
must also differ for different-order photons. We note that these (slashed) path-integrals can be made (unslashed) path-independent integrals in precisely the same way as in eq. \ref{eq:del_vartheta}. We will see this explicitly in Sec. \ref{sec:Sec3_Scaling_Relations} below. We can then find that photons forming increasingly higher-order images take concomitantly longer before getting to the observer and have longer paths, when emitted from the same event. Thus, photon orbits of different orders connect the same initial and final spatial locations but not the same two events in spacetime. 

Finally, for the case when $\vartheta_{\mathrm{e}} = 0$, pairs of photon orbits of orders $2n+1$ and $2n+2$ have identical orbits under reflections across the axis of the observer. In fact, due to the associated symmetries of this configuration, a point source present at such a location (called a caustic) does not form just two images but an entire ring, called an Einstein ring or a critical curve \citep{Virbhadra+2000}. All of the photons that form this ring on the image plane arrive at precisely the same time when emitted from the same event due to identical orbits. Caustics are defined as locations in the past light cone of the observer that have divergent magnifications and critical curves are the maps of these points on the image plane \citep{Virbhadra+2000, Frittelli+2000}. There is another (half-)line of caustics at $\vartheta_{\mathrm{e}} = \pi$ and pairs of photon orbits of orders $2n$ and $2n+1$ are identical. Motivated by Fig. \ref{fig:FigB2_Schw_Higher_Order_Images}, we introduce the convention of indexing the critical curves by $n$, where the total gravitational lensing experienced by the photons that form it is given as $\slashed{\Delta}\vartheta = n\pi$. Then odd-order critical curves are formed by the caustics with $\vartheta_{\mathrm{e}} = \pi$ whereas the even-order critical curves are formed by the other caustics at $\vartheta_{\mathrm{e}} = 0$. With this, it becomes clear that the $n=\infty$ critical curve \citep{Gralla+2020a} is simply the shadow boundary curve.


\section{Universal Photon Ring Scaling Relations}
\label{sec:Sec3_Scaling_Relations}

Following \cite{Gralla+2020a} and \cite{Johnson+2020}, here we will see how the photon ring can be \textit{quantitatively} identified as a region on the image plane where the total deflection angle increases logarithmically as $\slashed{\Delta}\vartheta \propto \ln{|\eta -\eta_{\mathrm{PS}}|} \propto \ln{|\bar{\eta}|}$ in arbitrary static and spherically-symmetric spacetimes which possess photon spheres.

An excellent analysis for the analytic approximation of the angular deflection of photons on orbits that approach the vicinity of the photon sphere in arbitrary static and spherically-symmetric spacetimes was reported in \cite{Bozza+2007}. We will now extend their analysis to include all of the aforementioned quantities (elapsed affine time, elapsed coordinate time, total deflection angle) by developing a simple unifying framework.

Note that the lensing Lyapunov exponent (Sec. \ref{sec:Sec3dot1_Lensing_Lyapunov_Exponent}) and the corresponding angular deflection scaling relation \eqref{eq:SecIIID_Scaling_Relations} have been obtained in \cite{Bozza2002}. The delay time (Sec. \ref{sec:Sec3dot2_Delay_Time}), the Lyapunov time (Sec. \ref{sec:Sec3dot3_Lyapunov_Time}) and the corresponding elapsed time scaling relation \eqref{eq:SecIIID_Scaling_Relations} have been obtained in \cite{Bozza+2003}. The Lyapunov time has also been independently obtained as an instability timescale for photon orbits at the photon sphere in \cite{Cardoso+2009}, where it was also connected to the damping frequency of eikonal quasinormal mode perturbations. The connection to gravitational lensing was made explicit in \cite{Stefanov+2010}. This connection has also been made for the Kerr metric in \cite{Yang+2012}.

We include a review of the above to demonstrate how our framework can be applied to straightforwardly obtain the various spacetime-specific critical or Lyapunov exponents and their associated image scaling relations. This allows us to cleanly demonstrate how these are measurable (reparametrization/gauge-invariant) versions of the phase space Lyapunov exponent \eqref{eq:Phase_Space_Lyapunov_Exponent}.

In Appendix \ref{app:AppD_Flux}, using this framework, we also obtain the photon ring intensity scaling relation (cf. also Sec. 5 of \citealt{Bauer+2022}) and recover the consecutive subring flux-density scaling relation obtained in \cite{Johnson+2020}, for generic emission zones.

In this section, the main new contributions include a description of how the critical parameters control various aspects of image formation. For example, we obtain an approximate relationship between the screen radial positions ($\eta_n$) of consecutive order images \eqref{eq:n_n+1_Lensing_Lyapunov_Exp_Inclination} and also an expression for the time delay between their appearance \eqref{eq:n_n+1_Time_Delay} for \textit{arbitrary} source locations in the bulk. The former is applied to the description of the image scaling of nonequatorial sources of emission (such as jets; Sec. \ref{sec:Sec3dot1dot1_Jets}) and the latter is used to better approximate the exact time delay between the appearance of the \textit{primary} and \textit{secondary} images of a point-sized emitter (hotspot) in Sec. \ref{sec:Sec4_Hotspots}.

We first introduce the general path-\textit{independent} integral (with $\dot{Q}(\eta, r)$ a regular function),
\begin{align} \label{eq:general_Path_Independent_Integral}
& \Delta Q(\eta, r_1, r_2)  := 
\int_{r_1}^{r_2} \dot{Q}(\eta, r)/\sqrt{\mathscr{R}(\eta, r)}~\mathrm{d}r\,,\\
&\quad\quad \mathrm{where}\quad r_2 \geq r_1 \geq \{r_{\mathrm{H}}^+~\mathrm{if}~\eta < \eta_{\mathrm{PS}},~ r_{\mathrm{tp}}(\eta)~\mathrm{if}~\eta \geq \eta_{\mathrm{PS}}\}\,. \nonumber
\end{align}
Since by definition $\mathscr{R}(\eta, r_{\mathrm{tp}}(\eta)) = 0$, all integrals of the type $\Delta Q(\eta, r_{\mathrm{tp}}(\eta), r_2)$ have divergent integrands with poles at precisely the same location. 

Depending on the radial-falloff of $\dot{Q}$, these may or may not be finite but the dominant contribution to these assuredly comes from the region close to $r_{\mathrm{tp}}(\eta)$. We demonstrate in Appendix \ref{app:AppC_Analytic_Approximations} how these dominant pieces can be characterized by elliptic functions in general, for arbitrary static and spherically-symmetric spacetimes. 

The path-independent integral expression above \eqref{eq:general_Path_Independent_Integral} enables the following compact notation for general path-\textit{dependent} integrals along photon orbits,
\begin{widetext}
\begin{align} \label{eq:general_Path_Dependent_Integral}
\slashed{\Delta}Q(\eta, r_{\mathrm{e}}) :=&\ 
\fint_{\lambda_1}^{\lambda_2} \dot{Q}(\eta, r(\lambda))~\mathrm{d}\lambda =
\fint_{r_{\mathrm{e}}}^{\infty} \dot{Q}/\sqrt{\mathscr{R}}~\mathrm{d}r \\
=&\
\begin{cases}
\Delta Q(\eta, r_{\mathrm{e}}, \infty)\,,
& k^r_{\mathrm{e}} > 0\,;
\ r_{\mathrm{e}} > \{r_{\mathrm{H}}~\mathrm{if}~\eta < \eta_{\mathrm{PS}},~ r_{\mathrm{tp}}(\eta)~\mathrm{if}~\eta \geq \eta_{\mathrm{PS}}\} \\ 
\Delta Q(\eta, r_{\mathrm{tp}}(\eta), r_{\mathrm{e}}) + \Delta Q(\eta, r_{\mathrm{tp}}(\eta), \infty)\,, 
& k^r_{\mathrm{e}} \leq 0 \,;
\mathrm{if}~\eta > \eta_{\mathrm{PS}}~\mathrm{and}~\ r_{\mathrm{e}} \geq r_{\mathrm{tp}}(\eta)
\end{cases} \,.
\end{align}
\end{widetext}
Equations \ref{eq:general_Path_Independent_Integral} and \ref{eq:general_Path_Dependent_Integral} yield the total angular deflection $\slashed{\Delta}\vartheta^\pm$ \eqref{eq:del_vartheta} for $\dot{Q} = \dot{\vartheta}/E = \eta/R^2$, the total path length $\slashed{\Delta}\lambda^\pm$ \eqref{eq:Total_Affine_Length} for $\dot{Q} = 1/E$, and the total elapsed time $\slashed{\Delta}t^\pm$ \eqref{eq:Total_Elapsed_Time} for $\dot{Q} = \dot{t}/E = 1/f$.

Due to the anticipated conformal symmetry of the photon ring (see also \citealt{Hadar+2022}), it is best to work with a pair of more natural conformal variables, namely the fractional deviations of the radial coordinates in the bulk and on the image plane from their critical values respectively, 
\begin{equation} \label{eq:Conformal_Variables_Phase_Space}
\bar{r} := r/r_{\mathrm{PS}} - 1\,, \ \ \bar{\eta} := \eta/\eta_{\mathrm{PS}} - 1\,.
\end{equation}
The limit $\bar{r} \rightarrow 0$ takes us to the photon sphere in the bulk whereas the limit $\bar{\eta} \rightarrow 0$ sends us to the shadow boundary on the image plane. 


\begin{figure*}
\begin{center}
\includegraphics[width=2\columnwidth]{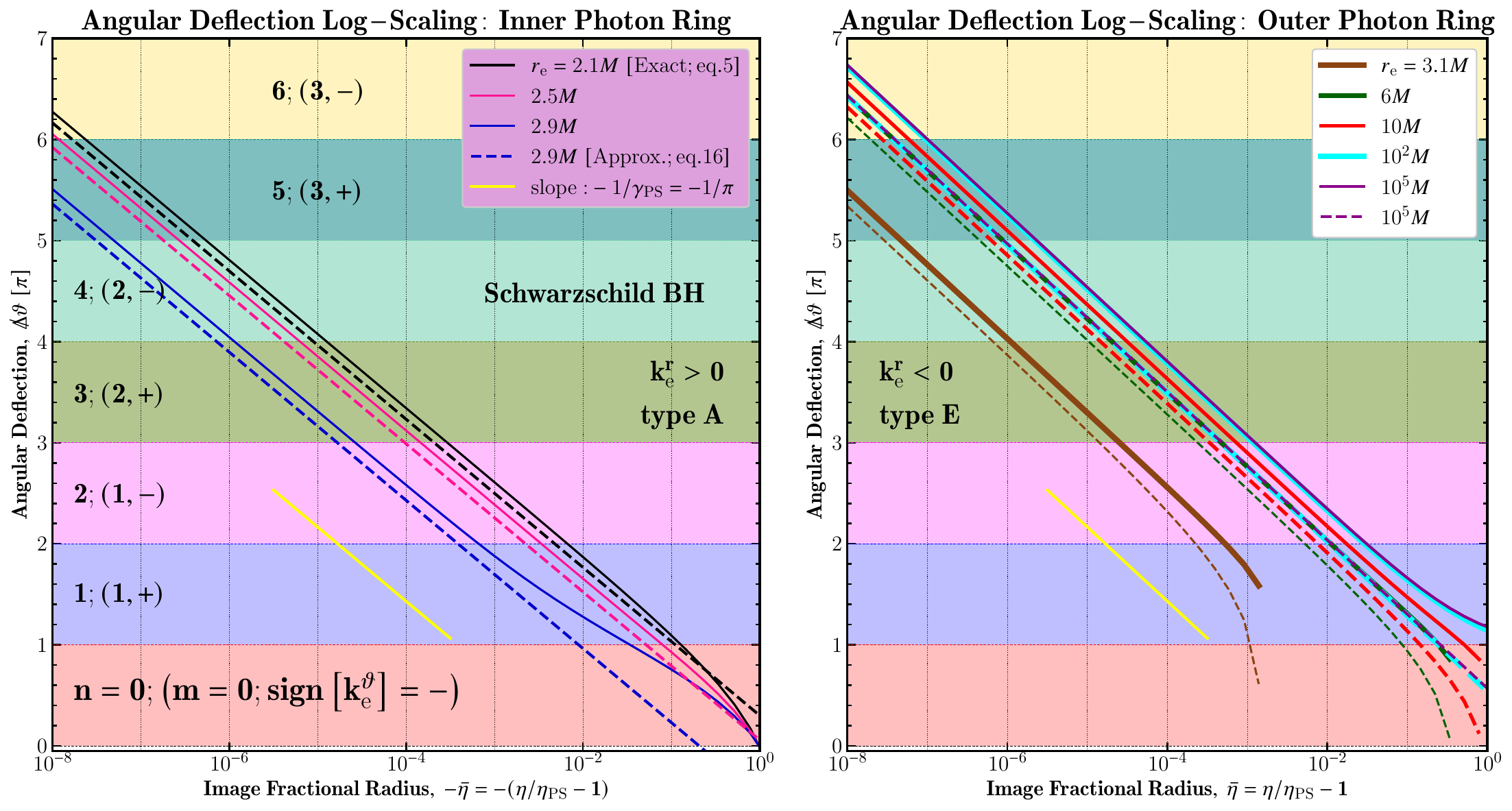}
\caption{\textit{Logarithmic divergence of the deflection angle with impact parameter locates the photon ring on the image plane.} The  angular deflection ($\slashed{\Delta}\vartheta$) experienced by a photon depends on its emission location ($r_{\mathrm{e}}$), the sign of its initial radial velocity ($k^r_{\mathrm{e}}; \pm$), and its apparent impact parameter ($\eta$). The latter is also the radius on the image plane at which it appears. A photon that is emitted at the photon sphere in a Schwarzschild BH spacetime ($r_{\mathrm{e}} = r_{\mathrm{PS}} = 3M$) with the critical impact parameter ($\eta = \eta_{\mathrm{PS}} = \sqrt{27}M$) undergoes infinite deflection due to gravitational lensing. One that appears in the close vicinity of the shadow boundary ($|\bar{\eta}| \ll 1$) experiences large but finite deflections and necessarily accesses the close vicinity of the photon sphere ($|\bar{r}| \ll 1$) along its orbit. The exact angular deflection experienced by a photon \eqref{eq:del_vartheta} and its approximation \eqref{eq:SecIIID_Scaling_Relations} are shown here in solid and dashed lines respectively. The logarithmic divergence occurs for small $|\bar{\eta}|$ values. The left panel shows photons appearing inside the shadow boundary, $\bar{\eta}<0$, i.e., in the ``inner photon ring,'' whereas the right panel shows photons appearing in the ``outer photon ring.'' The lensing Lyapunov exponent, $\gamma_{\mathrm{PS}}$, here takes value $\gamma_{\mathrm{PS}}=\pi$.}
\label{fig:Fig1_Schw_Strong_Lensing_Approximation}
\end{center}
\end{figure*}


Photon orbits that terminate on the image plane in the close vicinity of the critical curve ($|\bar{\eta}| \ll 1$) and which access the close vicinity of the photon sphere somewhere along the orbit ($|\bar{r}(\lambda)| \ll 1$ for some $\lambda$) experience strong gravitational-lensing. More specifically, these are photons that were either emitted from%
\footnote{We introduce the somewhat technical definitions for these types of orbits in Appendix \ref{app:AppC_Analytic_Approximations}. The reader can verify from Fig. \ref{fig:FigB1_Schw_Higher_Order_Images} and the top left panel of Fig. \ref{fig:FigB2_Schw_Higher_Order_Images} that the properties of the orbits listed here are the ones that yield the appropriate divergence ($\slashed{\Delta}Q \propto \ln{|\bar{\eta}|}$).}
\begin{itemize}
\item[[A]] well inside the photon sphere ($r_{\mathrm{H}} < r_{\mathrm{e}} \lnsim r_{\mathrm{PS}}$), and in the radially-outward direction ($k^r_{\mathrm{e}} > 0; \bar{\eta} < 0$),
\item[[C]] close to the photon sphere ($r_{\mathrm{e}} \simeq r_{\mathrm{PS}}$) or a radial turning point ($r_{\mathrm{e}} \simeq r_{\mathrm{tp}}$), with initially nonnegative radial velocity ($k^r_{\mathrm{e}} \geq 0$), or
\item[[E]] well outside the photon sphere ($r_{\mathrm{e}} \gnsim r_{\mathrm{tp}}(\eta)$), and in the radially-inward direction ($k^r_{\mathrm{e}} < 0; \bar{\eta} > 0$). 
\end{itemize}

For such strongly-lensed photon orbits, the leading-order behavior in $\bar{\eta}$ of arbitrary path-integrals $\slashed{\Delta}Q$ for small $|\bar{\eta}|$ is given as (see Appendix \ref{app:AppC_Analytic_Approximations}),
\begin{equation} \label{eq:Universal_Relation}
\slashed{\Delta}Q(\bar{\eta}) 
\approx  \slashed{\Delta}Q_{\mathrm{D}}(\bar{\eta}) =
\left\{
\begin{alignedat}{3}
& -\frac{\dot{Q}(0,0)}{\hat{\kappa}_{\mathrm{PS}}}\left[\ln{|\bar{\eta}|} + \tilde{c}\right]\,,\
&& \mathrm{[types\ A,\ E]} \\
& -\frac{\dot{Q}(0,0)}{2\hat{\kappa}_{\mathrm{PS}}}\left[\ln{|\bar{\eta}|} + \tilde{c}_{\mathrm{o}}\right]\,,\
&& \mathrm{[type\ C]}
\end{alignedat}
\right.
\end{equation}
where $\dot{Q}(0,0) := \dot{Q}(\bar{\eta}=0, \bar{r}=0) = \dot{Q}(\eta=\eta_{\mathrm{ps}}, r=r_{\mathrm{ps}})$ and we have absorbed the dependence on the locations of the emitter ($r_{\mathrm{e}}$) and the observer ($r_{\mathrm{o}}$) locations into some constants $\tilde{c}$ (defined in Appendix \ref{app:AppC_Analytic_Approximations}). These can be ignored for our present purposes since we are primarily interested in the logarithmic scaling and the scaling constant.

The scaling constant for some observable $Q$ is determined by a piece that is specific to the observable $\dot{Q}(0,0)$ and a piece that is independent of the observable. This latter universal scaling constant $\hat{\kappa}_{\mathrm{PS}}$ is the null geodesic phase space Lyapunov exponent that we introduced above (eq. \ref{eq:Phase_Space_Lyapunov_Exponent}), and is defined purely by the spacetime geometry. While the location of the photon sphere \eqref{eq:Photon_Sphere} as well as the size of the shadow \eqref{eq:Shadow_Size} are determined independently of the metric function $g$, $\hat{\kappa}_{\mathrm{PS}}$ depends on all three metric functions ($f, g, R$) but only on the derivatives of $f$ and $R$. \cite{Jacobson2007} discusses the physical implications underlying the metric function $g$ (see also Sec. III C of \citealt{Kocherlakota+2024c}).

Equation \ref{eq:Universal_Relation} presents a powerful closed-form expression for the logarithmic-scaling behavior of arbitrary observables in the photon ring. This expression was made general by being able to pull $\dot{Q}$ entirely outside the integral in eq. \ref{eq:Universal_Relation}. The universal behavior of various important quantities can now be obtained trivially. For the total angular deflection $\slashed{\Delta}\vartheta^\pm$, the total coordinate time $\slashed{\Delta}t^\pm$, and the total affine time $\slashed{\Delta}\lambda^\pm$, we obtain%
\footnote{In writing equation \ref{eq:SecIIID_Scaling_Relations} and henceforth, we restrict to photon orbits that are $\mathrm{type\ A}$ and $\mathrm{E}$ respectively for brevity. The analysis to cover $\mathrm{type\ C}$ orbits follows straightforwardly (e.g., we have to account for the factor of $1/2$). All qualitative statements however apply to all three cases.}
\begin{equation} \label{eq:SecIIID_Scaling_Relations}
\boxed{
\begin{aligned} 
\slashed{\Delta}\vartheta^\pm \approx 
&\ \mp \frac{\eta_{\mathrm{PS}}}{R^2_{\mathrm{PS}}\hat{\kappa}_{\mathrm{PS}}}\left[\ln{|\bar{\eta}|} + \tilde{c}_\vartheta\right]\
=: \mp \frac{\pi}{\gamma_{\mathrm{PS}}}\left[\ln{|\bar{\eta}|} + \tilde{c}_\vartheta\right]\,; \\
\slashed{\Delta}t^\pm \approx 
&\  -\frac{1}{f_{\mathrm{PS}}\hat{\kappa}_{\mathrm{PS}}}\left[\ln{|\bar{\eta}|} + \tilde{c}_t\right]\ 
=: -t_{\ell; \mathrm{PS}}\left[\ln{|\bar{\eta}|} + \tilde{c}_t\right]\,; \\
\slashed{\Delta}\lambda^\pm \approx 
&\  -\frac{1}{E\hat{\kappa}_{\mathrm{PS}}}\left[\ln{|\bar{\eta}|} + \tilde{c}_\lambda\right]\,.
\end{aligned}
}
\end{equation}
Here $\tilde{c}$ are some constants that depend on the radial locations of the observer and emitter. We remind the reader that, we had introduced the sign convention in eq. \ref{eq:Angular_Deflection_Expr} to track the polarity of the photon orbit, or, equivalently, the sign of its polar velocity $\dot{\vartheta}$ at emission.

In the above, we have introduced the lensing Lyapunov exponent, $\gamma_{\mathrm{PS}}$, as well as the Lyapunov time, $t_{\ell; \mathrm{PS}}$, which are the critical exponents that control the logarithmic divergences of the deflection angle and the elapsed time respectively, as 
\begin{equation} \label{eq:Lensing_Lyapunov_Exponent_Lyapunov_Time}
\boxed{
\begin{aligned} 
\gamma_{\mathrm{PS}} &:=\ \frac{\pi R^2_{\mathrm{PS}}}{\eta_{\mathrm{PS}}}\hat{\kappa}_{\mathrm{PS}} = \pi\left[\frac{f_{\mathrm{PS}}R^2_{\mathrm{PS}}}{2g_{\mathrm{PS}}}\left(\frac{\partial_r^2R^2_{\mathrm{PS}}}{R^2_{\mathrm{PS}}} - \frac{\partial_r^2f_{\mathrm{PS}}}{f_{\mathrm{PS}}}\right)\right]^{1/2}\,;\\
t_{\ell; \mathrm{PS}} &:=\ \frac{1}{f_{\mathrm{PS}}\hat{\kappa}_{\mathrm{PS}}} = \left[\frac{f^2_{\mathrm{PS}}}{2g_{\mathrm{PS}}}\left(\frac{\partial_r^2R^2_{\mathrm{PS}}}{R^2_{\mathrm{PS}}} - \frac{\partial_r^2f_{\mathrm{PS}}}{f_{\mathrm{PS}}}\right)\right]^{-1/2}\,.
\end{aligned}
}
\end{equation}
Again, these are determined purely by the spacetime geometry. This shows how the circular null geodesic, $(\eta, r) = (\eta_{\mathrm{PS}}, r_{\mathrm{PS}})$ plays a key role in determining the behavior of a number of quantities associated with photon orbits that arrive in the photon ring $\delta\mathscr{C}_\infty$ and which access the region close to the photon sphere $\delta \mathrm{S}$. We will soon see how these determine observable features such as the sizes and widths of photon subrings, as well as the time delay between the appearance of different-order images.

We now report in Fig. \ref{fig:Fig1_Schw_Strong_Lensing_Approximation} the deflection angles experienced by photons in a Schwarzschild BH spacetime with different impact parameters that access the close vicinity of the photon sphere in the bulk, when computed exactly via eq. \ref{eq:del_vartheta} as well as when computed approximately via eq. \ref{eq:SecIIID_Scaling_Relations}. This figure locates the onset of the logarithmic scaling of $\slashed{\Delta}\vartheta$ with $|\bar{\eta}|$, and thus the photon ring on the image plane as being the region $|\bar{\eta}| \lesssim 10^{-1}$ for $\bar{\eta} > 0$ and $|\bar{\eta}| \lesssim 10^{-2}$ for $\bar{\eta} < 0$. We also note that the constant offset between the exact and approximate curves is accounted for by the regular piece $\Delta Q_{\mathrm{R}}~(\dot{Q} = \eta/R^2)$, as discussed in Appendix \ref{app:AppC_Analytic_Approximations}. 


\subsection{The Lensing Lyapunov Exponent, $\gamma_{\mathrm{PS}}$}
\label{sec:Sec3dot1_Lensing_Lyapunov_Exponent}

The angular deflections $\slashed{\Delta}\vartheta^\pm_n$ experienced by two photons that form the $n^{\mathrm{th}}$ and $(n+2)^{\mathrm{th}}$ order images of \textit{any} point source on the image plane of \textit{any} observer differ by $2\pi$, i.e., (see eq. \ref{eq:Angle_Deflection_Higher_Order_Images_n})
\begin{equation} 
\slashed{\Delta}\vartheta^\pm_{n+2} - \slashed{\Delta}\vartheta^\pm_n = \pm 2\pi\,.    
\end{equation}
We remind the reader that, in our convention, even-order photons appear with a negative sign whereas odd-order photons appear with a positive sign (see the discussion below eq. \ref{eq:Angle_Deflection_Higher_Order_Images_n}).

With the equation above, and the universal angular deflection scaling relation \eqref{eq:SecIIID_Scaling_Relations}, we can immediately write,
\begin{equation} \label{eq:n_n+2_Angular_Deflections}
\slashed{\Delta}\vartheta^\pm_{n+2} - \slashed{\Delta}\vartheta^\pm_n = \pm 2\pi \approx \mp\frac{\pi}{\gamma_{\mathrm{PS}}}\ln{\left[\frac{\bar{\eta}_{n+2}}{\bar{\eta}_n}\right]}\,.
\end{equation}
This is equivalent to the ``image radius scaling relation,''
\begin{equation} \label{eq:n_n+2_Scaling}
\frac{\bar{\eta}_{n+2}}{\bar{\eta}_n} = \frac{\eta_{n+2} - \eta_{\mathrm{PS}}}{\eta_n - \eta_{\mathrm{PS}}} \approx \mathrm{e}^{-2\gamma_{\mathrm{PS}}}\,.
\end{equation}
Notice how these equations are entirely independent of the spatial locations of the emitter and observer.

For an extended stationary source of emission, if we denote the outer and inner boundaries of the order$-n$ image by $\eta_{n; \mathrm{out}}(\psi)$ and $\eta_{n; \mathrm{in}}(\psi)$ respectively, then its width is given as $w_n := \eta_{n; \mathrm{out}} - \eta_{n; \mathrm{in}}$. We can then apply eq. \ref{eq:n_n+2_Scaling} to the outer and inner boundaries of the $(n+2)^{\mathrm{th}}$ and $n^{\mathrm{th}}$ order images,%
\footnote{
Except when the outer boundary of the emission zone is extremely close to the photon sphere such that the lensing map becomes nonmonotonic (see Sec. \ref{app:AppB_Lensing_Map}). In this case, there is no guarantee that the outer boundary of all order images is sourced by emission from the same set of spatial points.} %
to obtain the following ``image width scaling relation,''
\begin{equation}
\frac{w_{n+2}}{w_n} = \frac{\eta_{n+2; \mathrm{out}} - \eta_{n+2; \mathrm{in}}}{\eta_{n; \mathrm{out}} - \eta_{n; \mathrm{in}}} \approx \mathrm{e}^{-2\gamma_{\mathrm{PS}}}\,.
\end{equation}
In this way, we understand why $\gamma_{\mathrm{PS}}$ is called the \textit{lensing} Lyapunov exponent. For the Schwarzschild BH spacetime, $\gamma_{\mathrm{PS}} = \pi$. The image scaling relations, as well as the lensing Lyapunov exponent, have been obtained also for the Kerr metric \citep{Johnson+2020, Gralla+2020a}. This exponent has also been discussed in various non-Kerr spacetimes (\citealt{Wielgus2021, Broderick+2023, Salehi+2023, Kocherlakota+2024a}).

The flux densities of the $(n+2)^{\mathrm{th}}$ and the $n^{\mathrm{th}}$ subrings also obey a scaling behavior as (see Appendix \ref{app:AppD_Flux})
\begin{equation}
\frac{F_{\nu; {n+2}}}{F_{\nu; n}} \approx \frac{w_{n+2}}{w_n} \approx \mathrm{e}^{-2\gamma_{\mathrm{PS}}}\,.
\end{equation}

We can also explain in full generality how the relationship between the locations of appearance of \textit{consecutive}-order images (opposite polarities) is dependent only on the lensing Lyapunov exponent and the relative inclination of the source and emitter, but not on their radial locations. From (eqs. \ref{eq:Angle_Deflection_Higher_Order_Images_n} and \ref{eq:SecIIID_Scaling_Relations})
\begin{equation} 
\slashed{\Delta}\vartheta^\pm_{n+1} + \slashed{\Delta}\vartheta^\mp_n = -2\vartheta_{\mathrm{e}} + \pi[1 \pm 1]  \approx \mp \frac{\pi}{\gamma_{\mathrm{PS}}}\ln{\left[\frac{\bar{\eta}_{n+1}}{\bar{\eta}_{n}}\right]}\,,
\end{equation}
we find the general image scaling relation,
\begin{equation} \label{eq:n_n+1_Lensing_Lyapunov_Exp_Inclination}
\boxed{
\frac{\bar{\eta}_{n+1}}{\bar{\eta}_n} \approx \mathrm{e}^{-\gamma_{\mathrm{PS}}} \cdot \mathrm{e}^{\pm\gamma_{\mathrm{PS}}(2\vartheta_{\mathrm{e}}/\pi-1)}\,.
}
\end{equation}
This is the central equation that can be used to obtain scaling relations for any higher-order image observable, for arbitrary spatial locations of emitter and observer in arbitrary spherically-symmetric and static spacetimes. In the equation above, as well as in eqs. \ref{eq:n_n+1_Flux_Scaling_Photon_Subring} and \ref{eq:n_n+1_Time_Delay} below, we should choose the upper sign when $n$ is odd (and $n\!\!+\!\!1$ is even) and the lower sign when $n$ is even (and $n\!\!+\!\!1$ is odd). 

We can recover eq. \ref{eq:n_n+2_Scaling} from eq. \ref{eq:n_n+1_Lensing_Lyapunov_Exp_Inclination} by seeing that, e.g., $(\slashed{\Delta}\vartheta^+_{n+2} + \slashed{\Delta}\vartheta^-_{n+1}) - (\slashed{\Delta}\vartheta^-_{n+1} + \slashed{\Delta}\vartheta^+_n) = (-2\vartheta_{\mathrm{e}} + 2\pi) - (-2\vartheta_{\mathrm{e}})$.

One can also easily obtain the general width and flux density scaling relations for extended sources at fixed colatitude ($\vartheta_{\mathrm{e}}$), such as a cone or a ring, from eq. \ref{eq:n_n+1_Lensing_Lyapunov_Exp_Inclination} above, as
\begin{align} \label{eq:n_n+1_Flux_Scaling_Photon_Subring}
\frac{F_{\nu; n+1}}{F_{\nu; n}} &\approx\ 
\frac{w_{n+1}}{w_n} \approx 
\mathrm{e}^{-\gamma_{\mathrm{PS}}}\cdot\mathrm{e}^{\pm\gamma_{\mathrm{PS}}(2\vartheta_{\mathrm{e}}/\pi-1)}\,. 
\end{align}

For the special case of an equatorially-located emitter ($\vartheta_{\mathrm{e}} = \pi/2$) viewed by our observer on the $z-$axis  (``face-on'') in particular, eq. \ref{eq:n_n+1_Lensing_Lyapunov_Exp_Inclination} simplifies to (see also \citealt{Johnson+2020, Bisnovatyi-Kogan+2022}),
\begin{equation} \label{eq:n_n+1_Lensing_Lyapunov_Exp_Face_On}
\frac{\bar{\eta}_{n+1}}{\bar{\eta}_n} \approx \mathrm{e}^{-\gamma_{\mathrm{PS}}}\,.
\end{equation}
Concomitantly, for equatorially-located extended sources of emission, we can write,
\begin{align} \label{eq:n_n+1_Flux_Scaling_Photon_Subring_Face_On}
\frac{F_{\nu; n+1}}{F_{\nu; n}} &\approx\ 
\frac{w_{n+1}}{w_n} \approx 
\mathrm{e}^{-\gamma_{\mathrm{PS}}}\,. 
\end{align}


\subsubsection{Image Scaling Relations for Nonequatorial Sources}
\label{sec:Sec3dot1dot1_Jets}

We now highlight the direct astrophysical implications of our general image scaling relation \eqref{eq:n_n+1_Lensing_Lyapunov_Exp_Inclination}. Emission from an astrophysical jet sheath can be modeled approximately by a conical surface (constant$-\vartheta_{\mathrm{e}}$). Clearly, for such a conical source of emission, the scaling relations that relate image radii ($\eta_n$), widths ($w_n$), and flux densities ($F_{\nu; n}$) for consecutive order images depend on its half-opening angle ($\vartheta_{\mathrm{e}}$). Let us now consider four jet models specifically. $\mathtt{Model-I}$: A jet of opening angle $\pi/10$ that is pointed towards our line of sight (forward or approaching jet), i.e., $\vartheta_{\mathrm{e}}/\pi = 0.1$. $\mathtt{Model-II}$: A jet of opening angle $\pi/10$ that is pointed away our line of sight (backward or retreating jet), i.e., $\vartheta_{\mathrm{e}}/\pi = 0.9$. $\mathtt{Model-III}$: A forward jet of opening angle $\pi/4$, i.e., $\vartheta_{\mathrm{e}}/\pi = 0.25$. $\mathtt{Model-IV}$: A backward jet of opening angle $\pi/4$, i.e., $\vartheta_{\mathrm{e}}/\pi = 0.75$. The relations between the $n=1$ and $n=2$ image radii, widths, and flux densities for each of these jet models are given as,
\begin{align} \label{eq:Jet_Scaling}
\frac{\eta_2-\eta_{\mathrm{PS}}}{\eta_1-\eta_{\mathrm{PS}}} \approx 
\frac{w_2}{w_1} \approx 
\frac{F_{\nu; 2}}{F_{\nu; 1}} \approx 
\mathrm{e}^{-1.8\gamma_{\mathrm{PS}}}\ &\ [\mathtt{Jet~Model-I}] \\
\frac{\eta_2-\eta_{\mathrm{PS}}}{\eta_1-\eta_{\mathrm{PS}}} \approx 
\frac{w_2}{w_1} \approx 
\frac{F_{\nu; 2}}{F_{\nu; 1}} \approx 
\mathrm{e}^{-0.2\gamma_{\mathrm{PS}}}\ &\ [\mathtt{Jet~Model-II}] \nonumber \\
\frac{\eta_2-\eta_{\mathrm{PS}}}{\eta_1-\eta_{\mathrm{PS}}} \approx 
\frac{w_2}{w_1} \approx 
\frac{F_{\nu; 2}}{F_{\nu; 1}} \approx 
\mathrm{e}^{-1.5\gamma_{\mathrm{PS}}}\ &\ [\mathtt{Jet~Model-III}] \nonumber \\
\frac{\eta_2-\eta_{\mathrm{PS}}}{\eta_1-\eta_{\mathrm{PS}}} \approx 
\frac{w_2}{w_1} \approx 
\frac{F_{\nu; 2}}{F_{\nu; 1}} \approx 
\mathrm{e}^{-0.5\gamma_{\mathrm{PS}}}\ &\ [\mathtt{Jet~Model-IV}] \nonumber
\end{align}
Since the image of an astrophysical BH could contain an imprint of a strong jet component, our inference of the lensing Lyapunov exponent could be corrupted, if not properly taken into account. This is a simple demonstration of the importance of considering the properties of nonequatorial sources of emission.


\subsection{The Delay Time, $t_{d; \mathrm{PS}}$}
\label{sec:Sec3dot2_Delay_Time}

From the universal scaling relations obtained above \eqref{eq:SecIIID_Scaling_Relations}, we can also obtain the time delay between the arrival of the $n^{\mathrm{th}}$ and $(n+2)^{\mathrm{th}}$ order images on the image plane. Using eq. \ref{eq:n_n+2_Scaling}, we find,
\begin{equation} \label{eq:n_n+2_Time_Delay}
\slashed{\Delta}t_{n+2}^\pm - \slashed{\Delta}t_n^\pm \approx -t_{\ell; \mathrm{PS}}\ln{\left[\frac{\bar{\eta}_{n+2}}{\bar{\eta}_n}\right]} 
= 2t_{\ell; \mathrm{PS}}\gamma_{\mathrm{PS}} = 2\pi\eta_{\mathrm{PS}}\,. \nonumber
\end{equation}
Similarly, we can also obtain the time delay between the arrival of the consecutive order images on the image plane using eqs. \ref{eq:SecIIID_Scaling_Relations} and \ref{eq:n_n+1_Lensing_Lyapunov_Exp_Inclination} as being,
\begin{equation} \label{eq:n_n+1_Time_Delay}
\boxed{
\slashed{\Delta}t_{n+1}^\pm - \slashed{\Delta}t_n^\mp 
\approx \ \pi\eta_{\mathrm{PS}}\left[1 \mp \left(\frac{2\vartheta_{\mathrm{e}}}{\pi} - 1\right)\right]\,.
}
\end{equation}

Using the two equations above, we now introduce the characteristic delay time, $t_{d; \mathrm{PS}}$, which yields an approximate measure of the time elapsed between the appearance of consecutive order images on the image plane, as being
\begin{equation} \label{eq:Delay_Time}
\boxed{
t_{d; \mathrm{PS}} := \pi\eta_{\mathrm{PS}}\,.
}
\end{equation}
This is a remarkable relation: A clean detection of the time delay between higher-order images can yield an independent estimate of the shadow size $\eta_{\mathrm{PS}}$ of any spherically-symmetric ultracompact object (see also Sec. \ref{sec:Sec4_Hotspots} below). Current measurements of the shadow size are inferred through a careful calibration procedure, as mentioned above, and an independent estimate of the same -- perhaps obtained ``more directly'' from the data -- can yield a vital sanity check.

For a Schwarzschild BH, $t_{d; \mathrm{PS}} = \pi\sqrt{27}M$. This corresponds to the Schwarzschild BH $\tau_0$ parameter in eq. 72 of \cite{Gralla+2020a}. For a closely related discussion see also eq. 20 of \cite{Stefanov+2010}.

We can understand this delay time simply as being the half-orbital time of a photon moving on a circular meridional orbit,
\begin{equation}
\frac{t_{orb; \mathrm{PS}}}{2} = \frac{\pi}{\Omega_{\mathrm{PS}}} = \pi\eta_{\mathrm{PS}}\,,
\end{equation}
where $\Omega_{\mathrm{PS}}$ is its angular velocity (see eq. \ref{eq:NG_kPS}),
\begin{equation}
\Omega_{\mathrm{PS}} = \frac{k_{\mathrm{PS}}^\vartheta}{k_{\mathrm{PS}}^t} = \frac{1}{\eta_{\mathrm{PS}}}\,.
\end{equation}


\begin{figure*}
\begin{center}
\includegraphics[width=2\columnwidth]{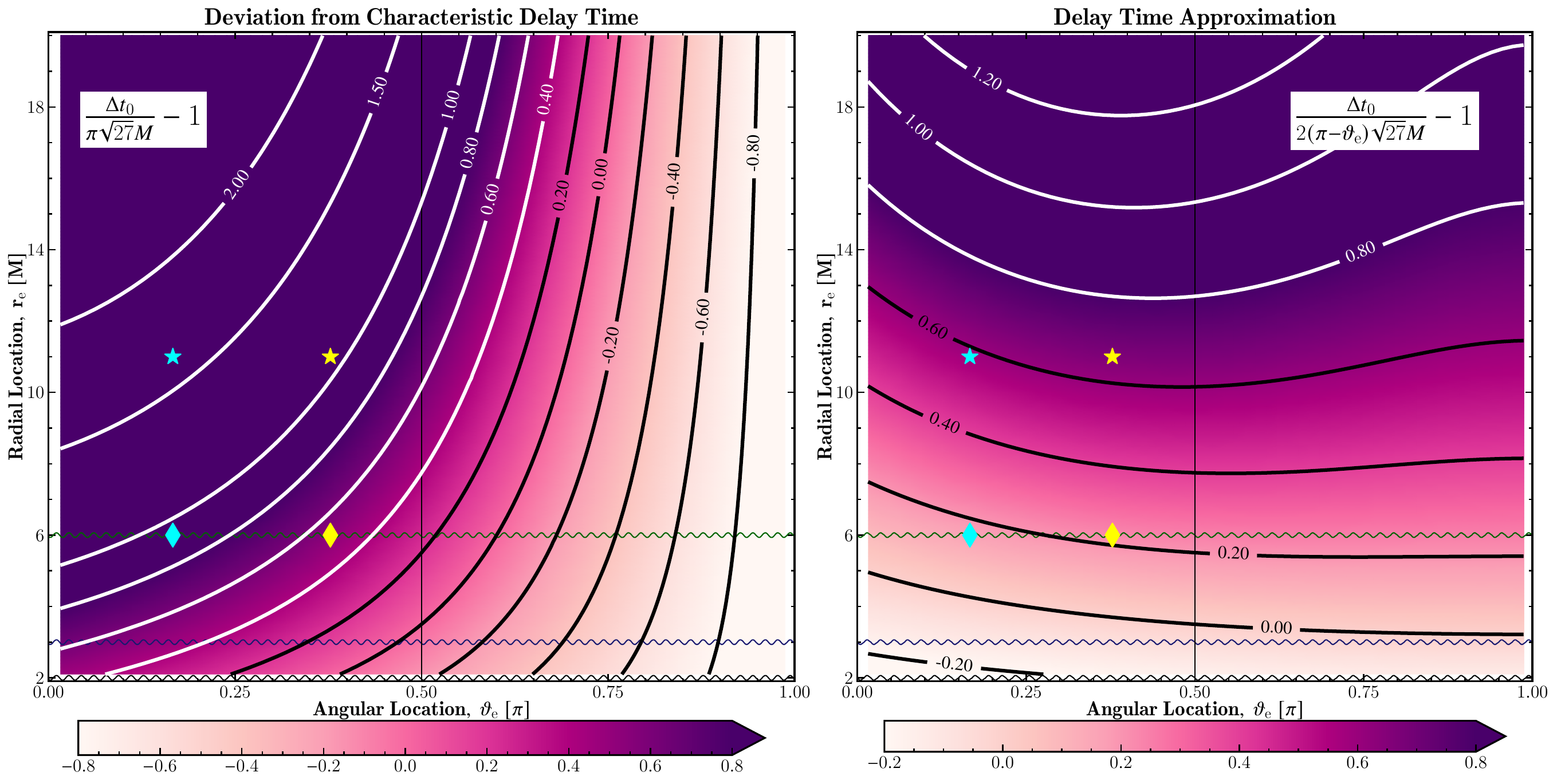}
\caption{\textit{Time delay between the primary and secondary images of a hotspot.} 
The delay time between the appearance of the $n^{\mathrm{th}}$ and $(n+1)^{\mathrm{th}}$ order images on the image plane, of a static source, is approximately given by $t_{d; \mathrm{PS}} = \pi\eta_{\mathrm{PS}}$, where $\eta_{\mathrm{PS}}$ is the BH shadow size. We compute the exact time delay $\Delta t_0$ between the appearance of the primary ($n=0$) and secondary ($n=1$) images of a hotspot (here, a point source) located at a radius $r=r_{\mathrm{e}}$ and angular coordinate $\vartheta= \vartheta_{\mathrm{e}}$ in a Schwarzschild black hole spacetime ($\eta_{\mathrm{PS}} = \sqrt{27}M$). The panel on the left shows the fractional error between the exact time delay and the characteristic one ($t_{d; \mathrm{PS}}$). In the right panel, we use our analytic prediction from eq. \ref{eq:Primary_Secondary_Time_Delay}, which corrects for the viewing angle ($\vartheta_{\mathrm{e}}$), and find lower errors relative to the actual value. We identify a hotspot located at $r_{\mathrm{e}} = 11M$ and viewed at an inclination of $22^\circ$ by the yellow star \citep{Wielgus+2022}. By similarly accounting for the (roughly linear) dependence of the time delay on the hotspot distance from the BH, we expect a measurement of such a time delay to yield an independent estimate of the BH shadow size.}
\label{fig:Fig2_Schw_Delay_Time_Shadow_Size_Err}
\end{center}
\end{figure*}


\subsection{The Lyapunov Time, $t_{\ell; \mathrm{PS}}$}
\label{sec:Sec3dot3_Lyapunov_Time}

We introduced above the characteristic delay time associated with the appearance of higher-order images. Here we emphasize the existence of yet another distinct fundamental instability timescale associated with the photon sphere (see $\lambda$ in Appendix A of \citealt{Cardoso+2009}).

The radial distance between two photons present close to the photon sphere grows exponentially in affine time as given by eq. \ref{eq:Phase_Space_Deviation}. We can rewrite this growth in terms of the coordinate time by starting with $\dot{\bar{r}}/\dot{t} = -\pm_r  \mathscr{g}_{tt}(r_{\mathrm{PS}})\hat{\kappa}_{\mathrm{PS}}\bar{r}$, i.e.,
\begin{align} \label{eq:Dev_Eq_Sol_t}
\bar{r}(t) &=\ \bar{r}(0)\exp{\left[-\pm_r  \mathscr{g}_{tt}(r_{\mathrm{PS}})\hat{\kappa}_{\mathrm{PS}}t\right]}  \nonumber \\
&=\ \bar{r}(0)\exp{[\pm_r t/t_{\ell; \mathrm{PS}}]}\,.
\end{align}
Thus, the Lyapunov time $t_{\ell; \mathrm{PS}}$, given in eq. \ref{eq:Lensing_Lyapunov_Exponent_Lyapunov_Time}, measures the characteristic instability timescale for photons present close to the photon sphere. Adopting the language of dynamical systems, it is the time, as measured by an asymptotic ($r\rightarrow\infty$) static observer ($\textbf{u} \propto \mathbf{\partial_t}$), for the radial coordinate between nearby photon orbits to increase by a factor of $\mathrm{e} \approx 2.72$. For the Schwarzschild BH spacetime, $t_{\ell; \mathrm{PS}} = \sqrt{27}M$. 

Recently, in \cite{Cardoso+2021} (see also \citealt{Ames+1968}), it was shown that this time scale plays an important role in determining the late-time characteristics of the observed luminosity evolution (light curve) of an infalling star. Thus, the Lyapunov time may be measurable, in principle, from future black hole imaging measurements of light-curves of infalling gas clouds (see, e.g., \citealt{Moriyama+2019}). 

Finally, note that \cite{Cardoso+2009} establish the following pleasing connection with the frequencies of eikonal ($l\gg 1$) quasinormal mode perturbations of angular momentum number $l$ and overtone number $n$, in arbitrary spherically-symmetric and static BH spacetimes, 
\begin{equation}
\omega_{\mathrm{QNM}} = l~\Omega_{\mathrm{PS}} - \mathrm{i}\left(n+\frac{1}{2}\right)\left(\frac{1}{t_{\ell; \mathrm{PS}}}\right)\,.
\end{equation}


\section{Hotspots in Schwarzschild Black Hole Spacetimes}
\label{sec:Sec4_Hotspots}

Recent observations of flaring events associated with Sgr A$^\star$ have been modeled in \cite{Wielgus+2022} (see also \citealt{Yfantis2023}). It has been found that a hotspot moving on a Keplerian orbit of radius $r_{\mathrm{K}} \approx 11M$ in a Schwarzschild BH spacetime, and viewed at an inclination of $\mathscr{i} \approx 22^\circ$, yields a good fit for the millimeter wavelength data (see Sec. 3.3 there). While the data available so far need to be interpreted under additional assumptions and with strongly restrictive models, the near-future developments should allow us to directly reconstruct angle- and time-resolved movies of flaring Sgr~A$^\star$ \citep{Emami2023, Johnson+2023}. Such development would enable unambiguous studies of the hotspot dynamics, and inference of the orbital properties such as radius and inclination. Theoretical aspects of hotspot production have been carefully investigated in \cite{Ripperda+2021}.

In this section, we will explore the implications of a possible detection of the secondary ($n=1$) image of such a hotspot. While hotspots are expected to have sizes of up to $1M$ (see, e.g., Sec. 3.4 of \citealt{Ripperda+2020}), for our present purposes, it suffices to consider the hotspot to be a point source. This toy model greatly simplifies the problem and we can infer several useful insights. It can be argued that our results may be directly useful for observational purposes if the point source is interpreted as tracking the location of the hotspot electron number density maximum (cf. \citealt{Tiede+2020}). In the context of currently available data, \citet{Wielgus+2022} established a high level of consistency between observables calculated with a semi-analytic polarized point source model \citep{Gelles2021} and a numerical model with an extended source and full radiative transfer \citep{Vos+2022}. While the observables that we consider in this section are even more robust, as they do not depend on plasma properties and follow directly from the fact that photons travel along null geodesics, some deviation from the point-source model is to be expected for a large and nonstationary hotspot.


\subsection{Time Delay between Primary and Secondary Images}
\label{sec:Sec4dot1_Time_Delay_Hotspots}

We begin with an investigation of whether the time delay between the appearance of the primary ($n=0$) and secondary ($n=1$) images of a hotspot, $\Delta t_0$, can also be used to infer the BH shadow size. As discussed above in Sec. \ref{sec:Sec3dot2_Delay_Time}, the characteristic delay time $t_{d; \mathrm{PS}}$ is related to the shadow size $\eta_{\mathrm{PS}}$ as $t_{d; \mathrm{PS}} = \pi\eta_{\mathrm{PS}}$. For a Schwarzschild BH, we obtain $t_{d; \mathrm{PS}} = \pi\sqrt{27}M$. For Sgr A$^\star$, this corresponds to $t_{d; \mathrm{PS}} \approx \pi\sqrt{27}(GM_{\mathrm{Sgr A}^\star}/c^3) \approx 5~\mathrm{min}$, whereas for M87$^\star$, we find $t_{d; \mathrm{PS}} \approx 6~\mathrm{days}$.

In the left panel of Fig. \ref{fig:Fig2_Schw_Delay_Time_Shadow_Size_Err} we show the fractional deviation of $\Delta t_0$ from the characteristic delay time $t_{d; \mathrm{PS}} = \pi\sqrt{27}M$ for varying hotspot locations $(r_{\mathrm{e}}, \vartheta_{\mathrm{e}})$ in a Schwarzschild BH spacetime. As can be seen from the figure, this error can be quite large depending on the hotspot location.

In eq. \ref{eq:n_n+1_Time_Delay}, we obtained an expression for the time delay between the appearance of a pair of higher-order images (e.g., $n=1$ and $n=2$) that corrected for the angular position of the hotspot. We now forward a proposal that we can also use it to approximate $\Delta t_0$, i.e., 
\begin{equation} \label{eq:Primary_Secondary_Time_Delay}
\Delta t_{0; \mathrm{approx}} \approx 2(\pi-\vartheta_{\mathrm{e}})\eta_{\mathrm{PS}} \,.
\end{equation}
The right panel of Fig. \ref{fig:Fig2_Schw_Delay_Time_Shadow_Size_Err} shows that this approximation does a reasonably good job of obtaining the exact time delay between the appearance of the primary ($n=0$) and secondary ($n=1$) images. As can be seen from the equation above, the time delay $\Delta t_0$ has a linear dependence on the angular position of the hotspot. We find that accounting for the (approximately linear) dependence on the radial position of the hotspot yields an even better approximation (not described here).

Measurements of such time delays at wavelengths \textit{other} than the current EHT observing one ($1.3\mathrm{mm}$) could lead to the possibility of measuring the shadow size at multiple different frequencies. Since the shadow size of a BH, in GR, is independent of the frequency at which these observations are conducted, (nonsimultaneous) multifrequency observations of flaring events can potentially be used to set up null tests of the achromaticity of the shadow size.


\subsection{Angle Offset between Primary and Secondary Images}
\label{sec:Sec4dot2_Angle_Offset_Hotspots}

In Appendix D of \cite{Wielgus+2022}, the authors also reported the angular offset on the image plane between the primary ($n=0$) and secondary ($n=1$) images using full ``slow-light'' numerical simulations for their best-fit hotspot model to be $\approx 140^\circ$. We will now attempt to offer simple arguments to estimate this offset. 

Let us consider a point source moving on a Keplerian orbit (i.e., a circular timelike geodesic) of radius $r$ viewed by an observer face-on ($\mathscr{i}=0$). The $n=1$ photon appears on the screen an additional $\Delta t_0 \approx \pi\eta_{\mathrm{PS}}$ \eqref{eq:Primary_Secondary_Time_Delay} amount of time after the $n=0$ photon has appeared. Thus, at any particular observer time, the source of the $n=0$ image has evolved by $\Delta t_0$ relative to the source of the $n=1$ image. Since the Keplerian angular velocity, $\Omega_{\mathrm{K}}$, is given by (from eq. \ref{eq:NG_k})
\begin{equation} \label{eq:Keplerian_Angular_Velocity}
\Omega_{\mathrm{K}}(r) = [\partial_r f/\partial_r R^2]^{1/2}\,,    
\end{equation}
the angle between the primary and secondary images $\Delta\psi_0(t)$ is given as $\Delta\psi_0(t) := \psi_1^+(t) - \psi_0^-(t) = \psi_1^+(\varphi_{\mathrm{e}}(t)) - \psi_0^-(\varphi_{\mathrm{e}}(t+\Delta t_0)) = \psi_1^+(\varphi_{\mathrm{e}}(t)) - \psi_0^-(\varphi_{\mathrm{e}}(t) + \Omega_{\mathrm{K}}(r)\Delta t_0)$, where $\varphi_{\mathrm{e}}(t)$ is the azimuthal coordinate of the source. This reduces to 
\begin{equation} \label{eq:Higher_Order_Images_Angle_Slow_Light}
\Delta\psi_0(r) =
- \pi + \Omega_{\mathrm{K}}(r)\Delta t_0 = 
- \pi + \left[\frac{\partial_r f(r)}{\partial_rR^2(r)}\right]^{1/2}\pi\eta_{\mathrm{PS}}\,.
\end{equation}
We remind the reader that the meaning of the superscript decorations on $\psi$ is given in eq. \ref{eq:Image_Plane_Polar_Angle}. Notice also that for this special configuration of source and observer, the offset angle becomes a function of the source orbit radius $r$ alone. In a Schwarzschild BH spacetime, the Keplerian angular velocity is given as $\Omega_{\mathrm{K}} = \sqrt{M/r^3}$.

For the best model parameters of \cite{Wielgus+2022} ($r=11M$), we find $|\Delta\psi_0| \approx 154^\circ$, which is close to their value of $140^0$. 
To compare, for a hotspot moving on the innermost stable circular orbit ($r = r_{\mathrm{ISCO}} = 6M$), we find $|\Delta\psi_0| \approx 116^\circ$. 
Thus, eq. \ref{eq:Higher_Order_Images_Angle_Slow_Light} does a reasonable job, considering that we neglected the observer inclination used there and have used an approximation for the time delay.


\begin{figure*}
\begin{center}
\includegraphics[width=2\columnwidth]{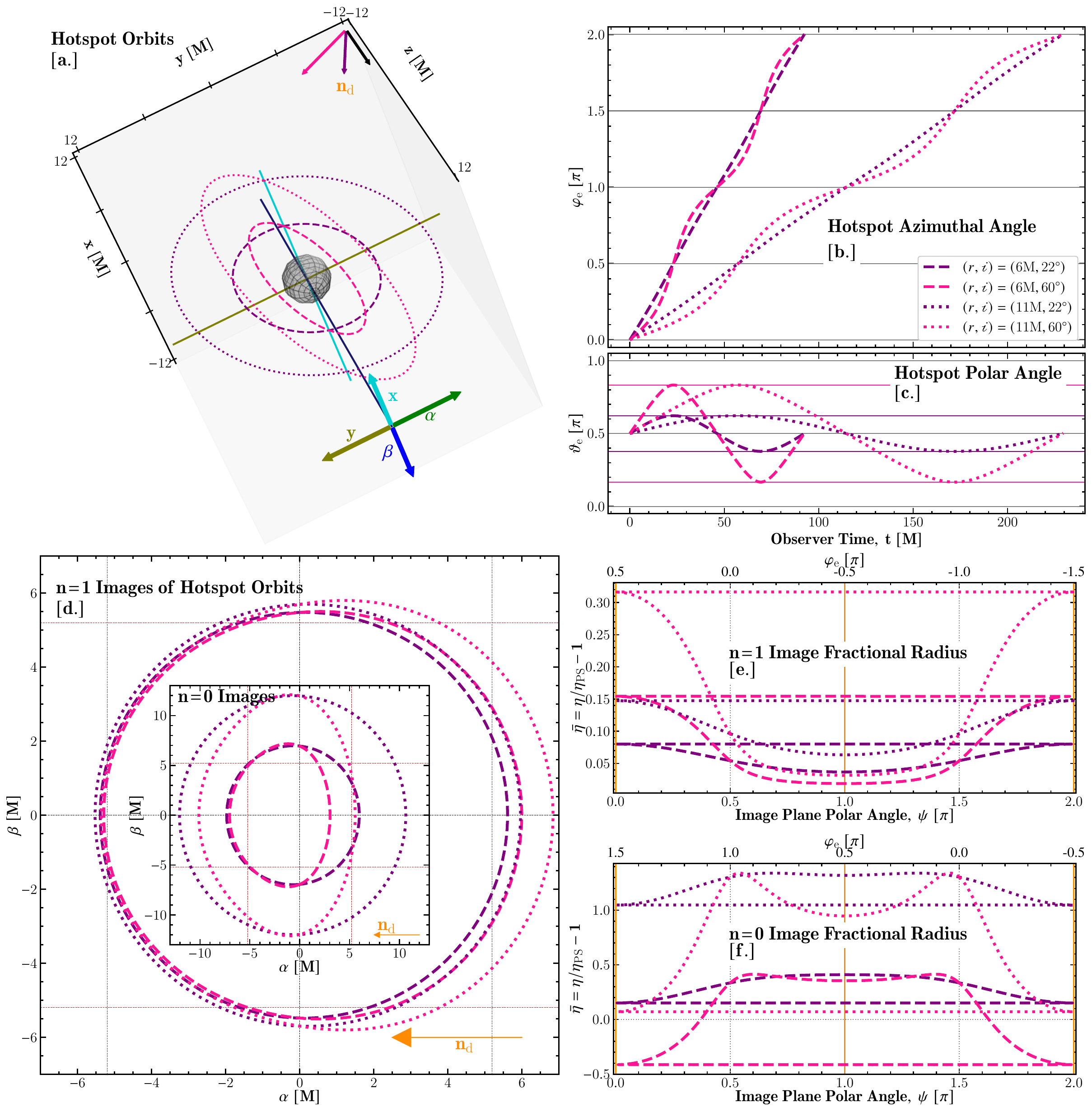}
\caption{\textit{Evolution of hotspots in a Schwarzschild black hole spacetime, and their secondary images.} Panel [a.] shows hotspot orbits of two radii and two inclinations relative to the observer, located on the $+z-$axis. The normals to the orbital planes are displayed in the upper corner. The relative orientation of the coordinate ($x, y$) and the image plane ($\alpha, \beta$) axes is also displayed. The time evolution of the hotspot azimuthal and polar angles are shown in panels [b.] and [c.] respectively, for one time period of each orbit. Panel [d.] shows the orbits of their primary (inset) and secondary images on the observer's screen. The asymmetry is maximal along the projected normals since these correspond to photons undergoing maximal ($\alpha < 0$) and minimal ($\alpha > 0$) angular deflections (see also Fig. \ref{fig:Fig7_Schw_Rings_Angular_Deflection_Asymmetry}). Panel [e.] shows the variation of the fractional radii of the $n=1$ images, $\bar{\eta} = \eta/\eta_{\mathrm{PS}} - 1$, with the image plane polar angle $\psi$. Finally, panel [f.] shows the same as panel [e.] but for the $n=0$ images. The $n=1$ images appear close to the Schwarzschild shadow radius, $\eta_{\mathrm{PS}}=\sqrt{27}M$, due to strong lensing ($\bar{\eta}_1 \lesssim 0.33$). The $y-$axis scales of panels [e.] and [f.] demonstrate the level of image demagnification due to strong lensing.}
\label{fig:Fig3_Schw_Hotspots}
\end{center}
\end{figure*}

\begin{figure*}
\begin{center}
\includegraphics[width=2\columnwidth]{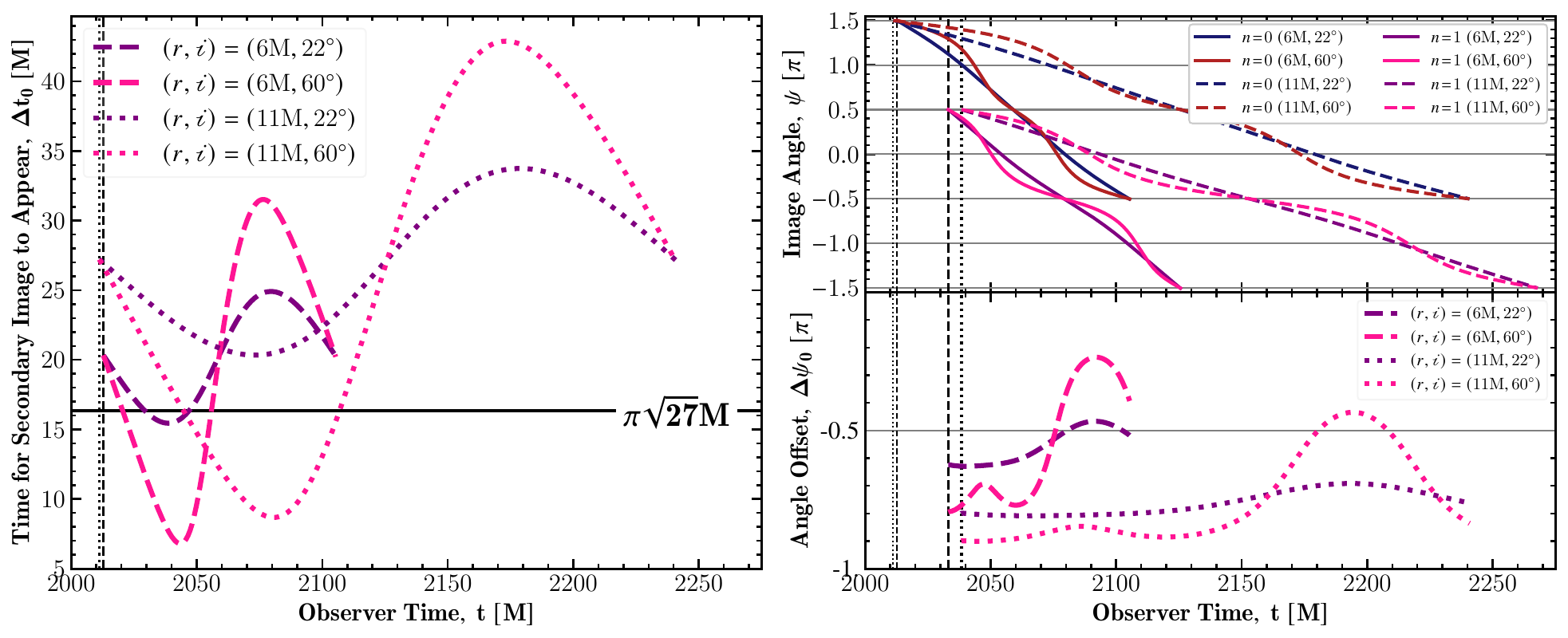}
\caption{\textit{Evolution of the time delay and the angular offset between the primary and secondary images of an orbiting hotspot.} The panel on the left shows the time for the secondary images to appear once the primaries have appeared, of hotspots on Keplerian orbits around a Schwarzschild BH (see previous figure). While the hotspot lights up at a time $t=0$, the primary image appears at a time $t\approx 2000M$ since we have set here the observer to be located at $r = 2000M$. This travel time depends also on the radius of emission, as indicated by the vertical lines. The delay time, $\Delta t_0$, itself is periodic, depending strongly on the polar angle evolution of the source (see panel [c.] of Fig. \ref{fig:Fig3_Schw_Hotspots}) as well as the source radius. The fluctuations in the time delay are amplified for highly inclined orbits. The top right panel shows the evolution of the image plane angle of both the primary and the secondary images. The slopes of these curves are indicative of the angular velocity of the hotspot and the amplitude of the fluctuations are indicative of the inclination of the orbital plane. The bottom right panel shows the angular offset, $\Delta\psi_0(t) = \psi_1(t) - \psi_0(t)$, between the primary and secondary images, over one orbit, when both are visible on the screen. In these panels on the right, the thicker vertical lines indicate the time of appearance of the $n=1$ image.}
\label{fig:Fig4_Schw_Hotspots_Angular_Offset_Time_Delay}
\end{center}
\end{figure*}


\subsection{Primary and Secondary Image Orbits}
\label{sec:Sec4dot3_Secondary_Image_Orbit}

We will now explore properties of the trajectories, on the screen, of primary and secondary images of hotspots on circular orbits around a Schwarzschild BH.

The time evolution of the azimuthal angle $\varphi_{\mathrm{e}}(t)$ of a hotspot on a Keplerian orbit of inclination $\mathscr{i} (\neq \pi/2)$ can be obtained from eq. \ref{eq:Euler_Solution} as
\begin{align} \label{eq:Azimuthal_Orbit_Hotspot}
\varphi_{\mathrm{e}}(t) =&\ \pm_\varphi\arctan{\left[\cos{\mathscr{i}}\cdot\tan{\left(\frac{\eta f}{R^2}t\right)}\right]} + p\pi\,, \nonumber \\
=&\ \pm_\varphi\arctan{\left[\cos{\mathscr{i}}\cdot\tan{\left(\Omega_{\mathrm{K}}t\right)}\right]} + p\pi\,,
\end{align}
where $\Omega_{\mathrm{K}}$ depends on the radius of the orbit \eqref{eq:Keplerian_Angular_Velocity}. Further, since the polar angle of such an orbit is given in terms of the azimuthal angle \eqref{eq:Euler_Solution}, we can write 
\begin{equation} \label{eq:Polar_Orbit_Hotspot}
\vartheta_{\mathrm{e}}(t) = -\mathrm{arctan}\left[\cot{\mathscr{i}} \cdot \csc{\varphi_{\mathrm{e}}(t)}\right] + q\pi\,.
\end{equation}
Our choice of the integration constants above means that $\varphi_{\mathrm{e}}(t=0) = 0$ and $\vartheta_{\mathrm{e}}(t=0) = \pi/2$, and that the normal to the orbit lies in the $yz-$plane. We will use the two integers introduced above, $p$ and $q$, to ensure that the hotspot angular coordinates lie in the principal sheet of the spherical-polar coordinate system, i.e., $0 \leq \varphi_{\mathrm{e}}(t) < 2\pi$ and $0 < \vartheta_{\mathrm{e}}(t) < \pi$. The latter, in particular, is useful because it enables us to directly employ eq. \ref{eq:Angle_Deflection_Higher_Order_Images_n} to determine image formation via eq. \ref{eq:Main_Integral_Equation}. We discuss this procedure in some more detail below, in Sec. \ref{sec:Sec5dot2_Thin_Disks_Inclination}.

Without loss of generality, we will consider below orbits at inclinations $0 \leq \mathscr{i} < \pi/2$%
\footnote{When the observer lies in the plane of the orbit, $\mathscr{i} = \pi/2$, most of its image appears on a line through the image plane origin. In addition, however, when the hotspot crosses the line of sight (at the caustics), the image contains Einstein rings (or critical curves) of all orders. For inclinations in $(\pi/2, \pi]$, the negative sign on the right-hand side of eqs. \ref{eq:Orbital_Inclination} and \ref{eq:Polar_Orbit_Hotspot} should be changed to positive signs. This is equivalent to a passive rotation of the bulk and boundary coordinate systems around the $+z-$axis by $\pi$.} %
and having positive azimuthal angular velocities ($+_\varphi$). Over a complete orbit of this kind, we can see that the hotspot polar velocity ($\dot{\vartheta}_{\mathrm{e}}$) must be positive for the first and last quarters (in time period) of the orbit \eqref{eq:Polar_Orbit_Hotspot}. The positive and negative polar velocity sections are separated by the polar turning points, $\vartheta_\pm = \pi/2 \pm \mathscr{i}$, at which $\dot{\vartheta}_{\mathrm{e}} = 0$. Along the entire orbit, we have $\vartheta_- \leq \vartheta_{\mathrm{e}}(t) \leq \vartheta_+$.

In panel [a.] of Fig. \ref{fig:Fig3_Schw_Hotspots}, we show four hotspot orbits, of different radii ($r=6M, 11M$) and orbital inclinations ($\mathscr{i} = 22^\circ, 60^\circ$), around a Schwarzschild BH (shown as a sphere). Panel [b.] shows the evolution of the azimuthal angle \eqref{eq:Azimuthal_Orbit_Hotspot} and panel [c.] similarly shows the evolution of the polar angle \eqref{eq:Polar_Orbit_Hotspot} of each hotspot, over one respective orbital time period, $T_{\mathrm{K}} = 2\pi/\Omega_{\mathrm{K}}$ (cf. \ref{eq:Keplerian_Angular_Velocity}). The polar turning points for each orbit are shown in horizontal lines of the same color in panel [c.].

Panel [d.] shows the $n=1$ images of these orbits. For all of these orbits, an $n=1$ photon emitted from the point closest to the observer (positive $z$, negative $y$) is emitted in the negative $z$ direction, executes about one full loop around the BH in the bulk, and appears on the image plane on the negative-half of the $\alpha-$axis. One that is emitted from the point furthest away from the observer (negative $z$, positive $y$) is also emitted in the negative $z$ direction but appears on the positive half of the $\alpha-$axis. The first of these two photons undergoes larger angular deflection and thus appears closer to the shadow boundary curve, whose radii are marked in horizontal and vertical red lines. In this way, we understand that the asymmetry of the $n=1$ image orbits is maximal in the direction of the normal projected onto the screen. The inset shows the $n=0$ image orbits.

To see this better, panel [e.] shows the variation of the fractional radii of the $n=1$ image, $\bar{\eta}_1 = \eta_1/\eta_{\mathrm{PS}} - 1$, with the image plane polar angle, $\psi$. Since $\bar{\eta}_1 = 0$ implies that the image lies on the shadow boundary curve, this quantity measures the distance of the image from the latter. Indeed, since $n=1$ photons are strongly lensed, they appear close to the shadow boundary, with $\bar{\eta}_1 \lesssim 0.33$ across all orbits and times. As discussed above, we see that images appearing on the negative $\alpha-$axis (vertical orange line here) are always the closest to the shadow boundary curve. Furthermore, the $n=1$ photons that appear on the $\beta-$axis, were emitted either from $\varphi_{\mathrm{e}} = \pi$ or $0$ \eqref{eq:Image_Plane_Polar_Angle}, and from the equatorial plane \eqref{eq:Polar_Orbit_Hotspot}. Thus, this image plane axis collects photons that undergo gravitational deflections of exactly $\slashed{\Delta}\vartheta_n = n\pi + \pi/2$ \eqref{eq:Angle_Deflection_Higher_Order_Images_n}, i.e., these are photons that execute precisely an integer number of half-loops around the black hole. Finally, panel [f.] shows the equivalent of panel [e.] but for the $n=0$ images. The $y-$axis scales of these two panels show the effect of image demagnification due to strong lensing \eqref{eq:n_n+1_Lensing_Lyapunov_Exp_Inclination}. We should highlight also that the image asymmetry of the $n=0$ and $n=1$ orbits is roughly aligned (see Fig. \ref{fig:Fig7_Schw_Rings_Angular_Deflection_Asymmetry}).

In the previous two sections, we discussed useful approximations of the time delay between the appearance of the primary and secondary images as well as the angular offset between them respectively. We now show, in Fig. \ref{fig:Fig4_Schw_Hotspots_Angular_Offset_Time_Delay}, the exact trends for these observables, for the hotspot orbits discussed above.

The left panel shows the time delay, i.e., the time for the secondary image to appear once the primary has appeared. The hotspot lights up in the bulk at an observer time $t=0$ (cf. panel [b.] of Fig. \ref{fig:Fig3_Schw_Hotspots}). However, due to finite travel time, the primary image appears at a time $t\approx 2000M$ since we have set the observer to be located at $r = 2000M$. This $n=0$ photon travel time depends weakly on the source radius but not on the orbital inclination here since, as discussed above, we have chosen $\varphi_{\mathrm{e}}(t=0) = 0$ and $\vartheta_{\mathrm{e}}(t=0) = \pi/2$. This is indicated in the vertical lines. The delay time itself depends strongly on the source colatitude (cf. panel [c.] of Fig. \ref{fig:Fig3_Schw_Hotspots}), and is clearly periodic. We show also the characteristic delay time, $t_{d;\mathrm{PS}} = \pi\sqrt{27}M$ \eqref{eq:Delay_Time} for reference.

In the top right panel, we show the time evolution of the image angular coordinate for both the $n=0$ as well as the $n=1$ images. The slopes of these curves are indicative of the orbital period and the amplitude of fluctuations are a characteristic of the orbital inclination. The thin vertical lines show the arrival time of the $n=0$ photons whereas the thicker ones show the same for the $n=1$ photons. Notice how the $n=0$ photon emitted from $r=6M$ reaches the screen after the one emitted from $r=11M$ and also how this trend is reversed for the $n=1$ photon. We can understand this as follows. The first $n=0$ photons are both emitted in the radially outward direction but the first $n=1$ photons are both emitted toward the BH (see the top left panel of Fig. \ref{fig:FigB2_Schw_Higher_Order_Images}). The $n=0$ photon emitted from further away travels a smaller distance whereas the $n=1$ photon emitted from further away travels a larger distance. These trends hold reasonably well for emitter locations outside the photon sphere. Finally, the bottom right panel shows the nontrivial evolution of the angle offset between the primary and secondary images. For related ``slow light'' descriptions of the entire accretion flow, see \cite{Bronzwaer+2018}.

In concluding our analysis of hotspots, we note that there is much that we can learn about the source of flux eruption events (see, e.g., \citealt{Ripperda+2021}), and potentially about spacetime geometry, from future images and movies, resolved or otherwise. 

Thus far, we have focussed our attention on point sources. We move now to a discussion of the image formation of the entire accretion flow.


\begin{figure*}
\begin{center}
\includegraphics[width=2\columnwidth]{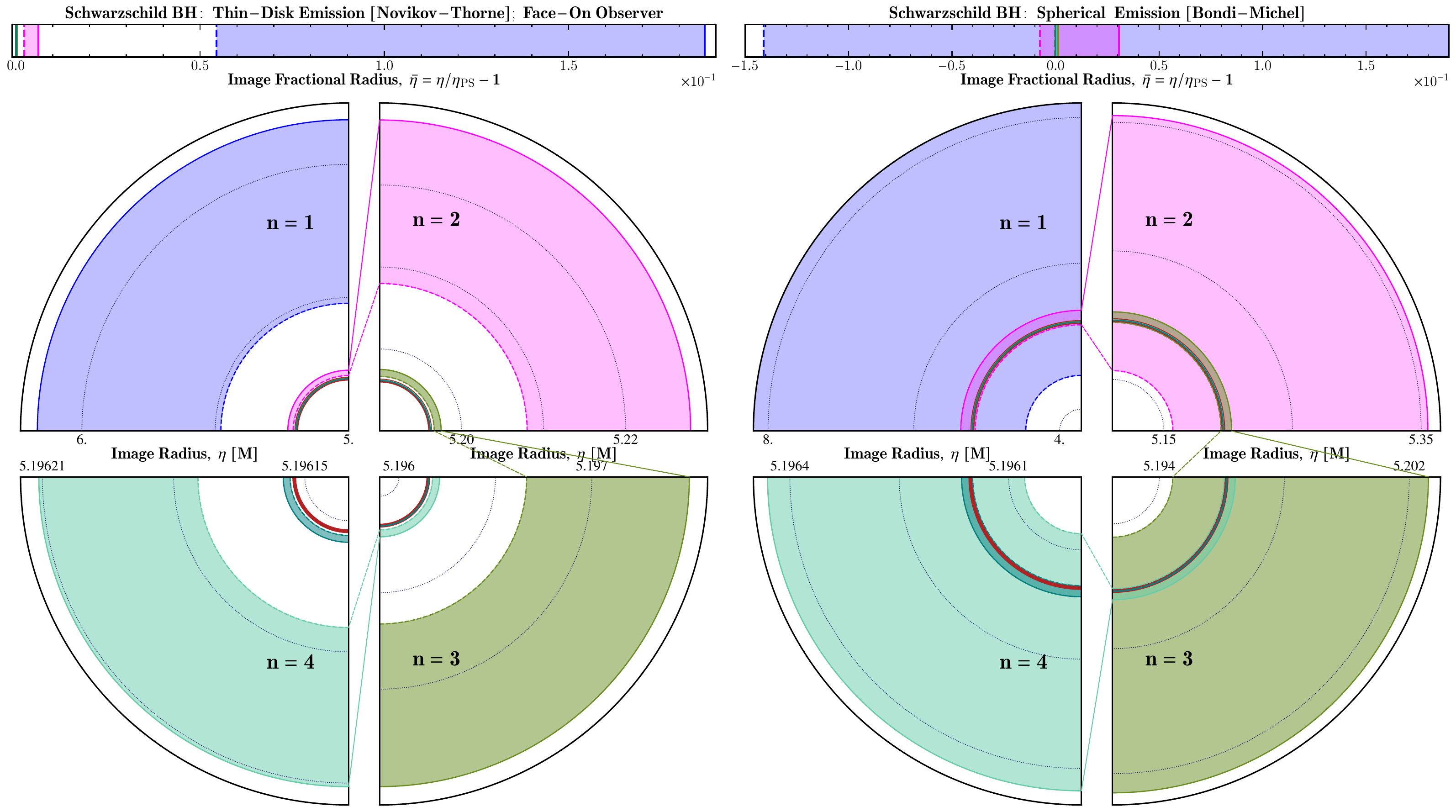}
\caption{\textit{Higher-order images of a thin-disk and a spherical emission zone in a Schwarzschild BH spacetime.} 
The left panel shows the first four ($n=1-4$) photon subrings (higher-order images) of an equatorial-thin disk when viewed face-on. We choose the inner and outer boundaries of the disk to be located at $6M$ and $\approx 2\times 10^4M$ respectively, to mimic an Novikov-Thorne accretion disk, which successfully explains quasar spectra. In the panel on the right, we show the subrings cast by a spherically-symmetric emitting region that extends down to the horizon ($2M$) from the same outer radius, to mimic a Bondi-Michel accretion process. While the universal self-similar scaling of the subrings is clear to see, the structures of the two photon rings are also significantly different. The subrings cast by the thin-disk are well separated and appear outside the shadow boundary, which is shown as a red line in all sectors. For the spherical emission zone, all order subrings overlap and straddle the shadow boundary. In this case, due to emission from the line of caustics $\vartheta_{\mathrm{e}} = 0, \pi$, we also see the formation of Einstein rings or critical curves, shown here as the bounding dashed and solid lines. If sources of emission are present close to the photon sphere ($3M$), or closer to our line of sight, we should expect the photon ring to look qualitatively similar to the panel on the right. Reducing the outer boundary of the emission zone will cause the outer edges of the subrings to move inwards.}
\label{fig:Fig5_Schw_Photon_Ring_Topology}
\end{center}
\end{figure*}


\section{Photon Rings in Schwarzschild Black Hole Spacetimes}
\label{sec:Sec5_Photon_Rings} 

The images of extended sources of emission such as accretion disks around black holes can also be decomposed into the leading-order ($n=0$) and higher-order ($n > 0$) images. We will refer to the latter as (photon) \textit{subrings}.

The properties of these subrings depend on the morphology of the emitting region as well as the inclination $\mathscr{i}$ of the observer. In this section, we will be interested in understanding, both qualitatively and quantitatively, the variation in the diameters $d_n$, widths $w_n$, and asymmetries $\mathscr{A}_n$ of the first few higher-order images (see also \citealt{Cardenas-Avendano+2023}). While the first-higher order ($n=1$) image will likely be accessible in the near future, with space-based very long baseline interferometry, our interest in the $n=2$ image is in exploring the potential for a measurement of the lensing Lyapunov exponent (Sec. \ref{sec:Sec3dot1_Lensing_Lyapunov_Exponent}).

To study the variations in these characteristic subring features, we will employ a simple geometric model for the morphology of the emitting region. This will be a conical torus concentric with the BH, as pictured in the top right panel of Fig. \ref{fig:FigB2_Schw_Higher_Order_Images}. The inner and outer boundaries of the torus are located at $r=r_{\mathrm{in}}$ and $r=r_{\mathrm{out}}$ respectively, and its scale height is characterized by the half-opening angle $0 \leq \vartheta_{1/2} \leq \pi/2$. Finally, the angle between the axis of the wedge and the $z-$axis determines the viewing inclination, $0 \leq \mathscr{i} < \pi/2$. Since this configuration is composed of a series of conical surfaces with half-opening angles $\vartheta$ in $\pi/2-\vartheta_{1/2} \leq \vartheta \leq \pi/2+\vartheta_{1/2}$, this simplistic model is also suitable for analyzing the images of jets as well.


\begin{figure*}
\begin{center}
\includegraphics[width=2\columnwidth]{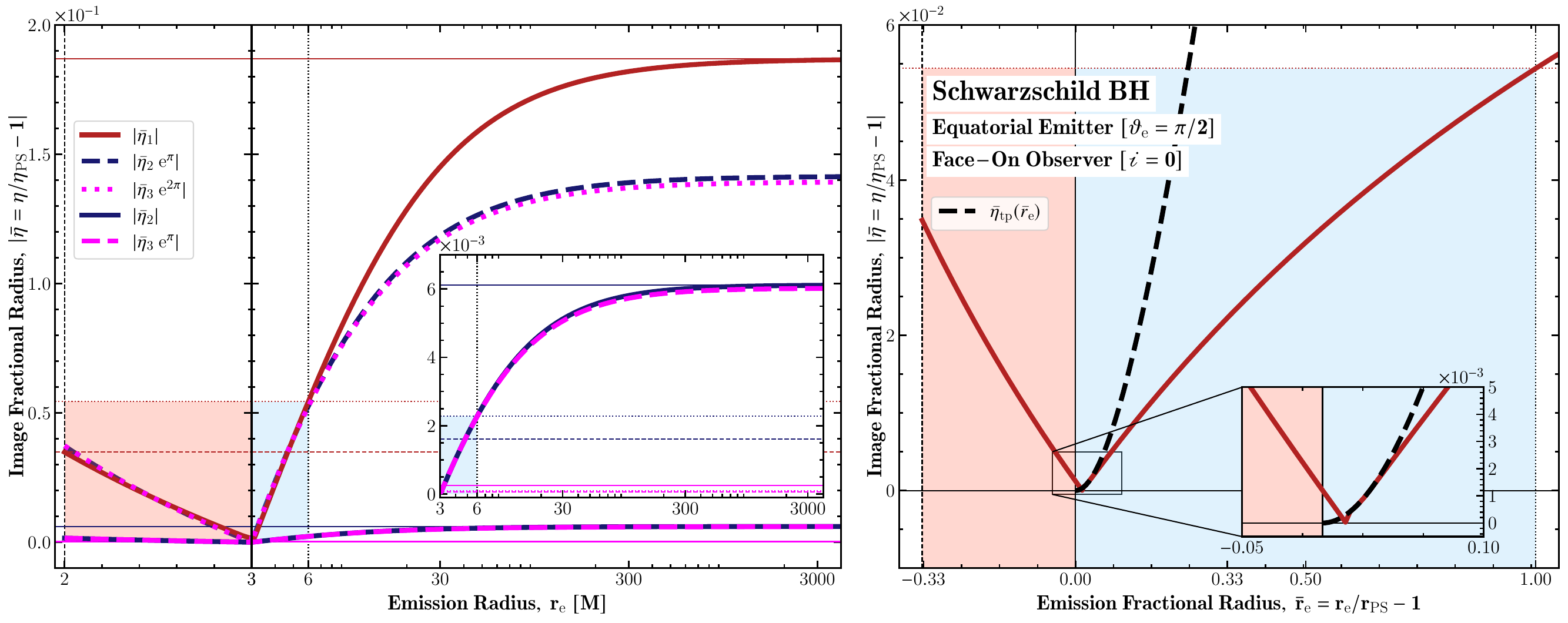}
\caption{\textit{Higher-order image radii of equatorial emitters in a Schwarzschild black hole spacetime.} The left panel shows the fractional radial distance from the black hole shadow boundary ($\eta_{\mathrm{PS}} = \sqrt{27}M$), $\bar{\eta}_1$, at which the first-order image of an equatorial emitter appears. The $n=1$ image always appears close to the shadow boundary $|\bar{\eta}| \lesssim 0.2$, i.e., its image radius exhibits an overall variation of $\approx 20\%$. In particular, for a ``maximal'' thin-disk, extending from the horizon to infinity, the diameter and width of its $n=1$ subring are $\approx 12.4M$ and $\approx 1.3M$ respectively. The scaling relation \eqref{eq:n_n+1_Lensing_Lyapunov_Exp_Face_On} obeyed by the image fractional radii of consecutive-order images, $\bar{\eta}_{n+1} = \bar{\eta}_n\mathrm{e}^{-\pi}$, is also clear to see. The inset zooms the $y-$axis by $\sim\!\mathrm{e}^{\pi} \approx 23$ to highlight this further. In the panel on the right, we zoom in on the shaded regions in the left panel and switch to the fractional radial coordinate, $\bar{r} = r/r_{\mathrm{PS}} - 1$, for the $x-$axis. The dashed line shows the angular momentum $\bar{\eta}_{\mathrm{tp}}$ of a photon that has zero radial velocity at emission (see eq. \ref{eq:Radial_Turning_Point_Equation}). Photons emitted with smaller angular momenta, i.e., $\bar{\eta} < \bar{\eta}_{\mathrm{tp}}$, are emitted toward the black hole (right of the intersection point) and vice versa. Finally, it is also clear that while \textit{most} $n=1$ photons emitted from outside the photon sphere ($\bar{r} > 0$) appear outside the shadow boundary ($\bar{\eta}_1 > 0$), this is not true for all. Those that are emitted from just outside the photon sphere \textit{can} appear inside the shadow boundary. This is generically true for all order photons (see Appendix \ref{app:AppC_Analytic_Approximations}).}
\label{fig:Fig6_Schw_Thin_Disk_Face_On}
\end{center}
\end{figure*}


We begin by considering, qualitatively, the images of a geometrically-thin disk ($\vartheta_{1/2} = 0$), viewed face on (i.e., $\mathscr{i} = 0$), and a spherical emission zone ($\vartheta_{1/2} = \pi/2$), in a Schwarzschild BH spacetime. We model the former following \cite{Novikov+1973} and, thus, choose $(r_{\mathrm{in}}, r_{\mathrm{out}}) = (6M, \infty)$. For the latter, we set $(r_{\mathrm{in}}, r_{\mathrm{out}}) = (2M, \infty)$, to mimic \cite{Michel1972} accretion. We will take a closer look at the thin-disk in Sec. \ref{sec:Sec5dot1_Thin_Disks_Face_On} below.

Fig. \ref{fig:Fig5_Schw_Photon_Ring_Topology} shows the first four subrings ($n=1-4$) for the thin-disk and the spherical zone in the left and right panels respectively. The outer (inner) edges of the subring correspond to the order$-n$ images of the outer (inner) boundaries of the emission zone. Since their images are circularly-symmetric on the sky, we show only a quarter of each photon subring in each sector. As we go clockwise from the top-left sector (which shows the $n=1$ subring) the radial scale is successively zoomed by a factor of $\approx\mathrm{e}^\pi \approx 23$. We show both the image plane radial coordinate $\eta$ as well as the fractional or conformal radial coordinate $\bar{\eta} = \eta/\eta_{\mathrm{PS}} - 1$, which measures the radial distance from the shadow boundary curve, $\eta = \eta_{\mathrm{PS}}$.

This immediately showcases the qualitative self-similar scaling exhibited by the subrings as well as the quantitative scaling discussed above, in eqs. \ref{eq:n_n+1_Lensing_Lyapunov_Exp_Inclination} and \ref{eq:n_n+1_Flux_Scaling_Photon_Subring}. The former applies to the conformal radii of the edges of the subrings and the latter to the widths of the subrings. Since all of the photons collected from the thin-disk execute an exact number of half-loops before appearing on the image plane, the image self-similarity is explained by even simpler scaling relations (\ref{eq:n_n+1_Lensing_Lyapunov_Exp_Face_On}, \ref{eq:n_n+1_Flux_Scaling_Photon_Subring_Face_On}).

We can also see clear qualitative differences in the higher-order images of these two emission zone morphologies. There are gaps between the subrings cast by the thin-disk whereas those cast by the spherical zone overlap. The shadow boundary curve is shown as a red circle in all sectors. We can use this to see that the photon ring lies entirely outside the shadow boundary curve ($\bar{\eta} > 0$) for the thin-disk whereas, for the spherical model, it straddles the shadow boundary curve. For the thin disk, the ``inner photon ring'' ($-1 \ll \bar{\eta} < 0$) is empty because of the observer inclination as well as absence of emission from close to the photon sphere (located at $r = r_{\mathrm{PS}} = 3M$).

We move now towards a quantitative estimation of the variation in subring characteristics by considering, systematically, a sequence of configurations. In Sec. \ref{sec:Sec5dot1_Thin_Disks_Face_On} we will consider the variation in the subring characteristics for geometrically-thin ($\vartheta_{1/2} = 0$) emitting regions viewed face-on ($\mathscr{i} = 0$) due to varying inner $r_{\mathrm{in}}$ and outer $r_{\mathrm{out}}$ boundaries. In Sec. \ref{sec:Sec5dot2_Thin_Disks_Inclination} we will consider the scenario of thin-disks viewed at an inclination. Finally, in Sec. \ref{sec:Sec5dot3_Thick_Disks_Face_On} we allow the geometrical-thickness ($\vartheta_{1/2} \neq 0$) and the boundaries of the disk to vary, while keeping the inclination fixed to zero, and analyze how it affects the structure of the photon ring. 


\subsection{Geometrically-Thin Disk viewed Face-on}
\label{sec:Sec5dot1_Thin_Disks_Face_On}

For this configuration of a thin-disk viewed from zero inclination, the image is circularly symmetric, and the subrings are perfect rings (cf. also \citealt{Bisnovatyi-Kogan+2022}). Furthermore, the image contains no Einstein rings due to the absence of emission from the line of caustics (line of sight). 

The quantitative relation between the order$-n$ image radius $\eta_n(r_{\mathrm{e}})$ and the source radius $r_{\mathrm{e}}$ for an equatorial emitter is obtained straightforwardly by solving the integral equation \eqref{eq:Main_Integral_Equation}
\begin{equation}
|\slashed{\Delta}\vartheta^\pm(\eta_n(r_{\mathrm{e}}), r_{\mathrm{e}})| = (2n+1)\pi/2\,,
\end{equation}
In the above, we have used the understanding that photons of all orders ($n$) execute an exact number of half-loops around the BH \eqref{eq:Angle_Deflection_Higher_Order_Images_n}. Remember also that $\slashed{\Delta}\vartheta^\pm$ is negative for even-order photons and positive for odd-order photons (cf. eq. \ref{eq:Angle_Deflection_Higher_Order_Images_n}). 

The radii of the outer and inner edges of the subrings, $\eta_{n; \mathrm{out}}$ and $\eta_{n; \mathrm{in}}$, for a thin-disk can then be obtained from the above by setting $r_{\mathrm{e}} = r_{\mathrm{out}}$ and $r_{\mathrm{e}} = r_{\mathrm{in}}$ respectively. The former radius ($r_{\mathrm{out}}$) corresponds to the outer boundary of the disk and the latter to its inner boundary. The width of a subring is naturally given as $w_n = \eta_{n; \mathrm{out}} - \eta_{n; \mathrm{in}}$.

The left panel of Fig. \ref{fig:Fig6_Schw_Thin_Disk_Face_On} shows the image \textit{fractional} radius, $\bar{\eta} := \eta/\eta_{\mathrm{PS}} - 1$, of the first-order image ($n=1$) for an equatorial source present at a radius $r_{\mathrm{e}}$ from a Schwarzschild BH. We use the fractional radius here to conveniently measure the distance at which the image appears from the BH shadow boundary, $\eta_{\mathrm{PS}} = \sqrt{27}M$. Indeed, if we denote the $n=1$ subring diameter by $d_1 (=2\eta_1)$ and the shadow diameter by $d_{\mathrm{sh}} (=2\eta_{\mathrm{PS}})$, then the deviation of the former from the latter is given as $d_1/d_{\mathrm{sh}} - 1 = \bar{\eta}_1$. Thus, we can infer from the figure that, for this configuration, the $n=1$ subring diameter appears very close to the shadow, is $-4\times 10^{-2} \lesssim \bar{\eta_1} \lesssim 19 \times 10^{-2}$. The overall variation in the $n=1$ subring diameter is about $20\%$. 

The maximal diameter and the maximal width of the $n=1$ subring are given as $d_1 = 2~\mathtt{max}[\eta_1] \approx 12.4M$ and $w_1 = \eta_{\mathrm{PS}}(\mathtt{max}[\bar{\eta}_1] - \mathtt{min}[\bar{\eta}_1]) \approx 1.3 M$ respectively. The maximal sets of the equatorial emission photon subrings, corresponding to the emission region located anywhere between the event horizon and infinity, are referred to as (equatorial) lensing bands (see, e.g., \citealt{Paugnat+2022, Wielgus2021}).


\begin{figure*}
\begin{center}
\includegraphics[width=2\columnwidth]{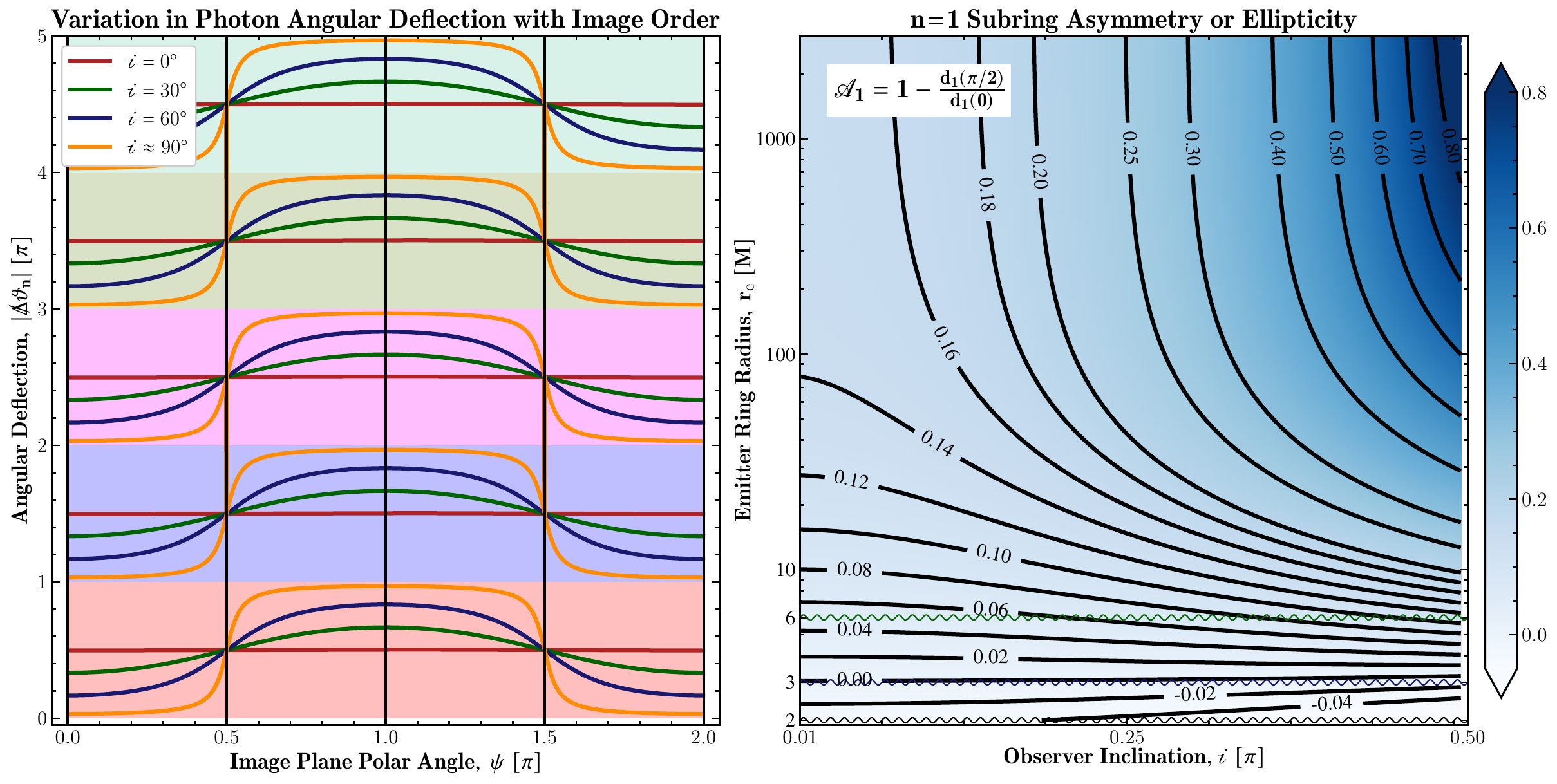}
\caption{\textit{Angular deflection experienced by photons emitted by a ring, and image asymmetry}. In the panel on the left, we show the angular deflection experienced by photons emitted from a ring before reaching an observer at an inclination of $\mathscr{i}$ w.r.t. the normal to the ring. Such rings have been pictured in the top left panel of Fig. \ref{fig:Fig3_Schw_Hotspots}. The radius of the ring is irrelevant for this panel. Order$-n$ photons undergo angular deflections in the range $(n\pi,  (n\!+\!1)\pi)$. Photons appearing at the same image plane polar angle, $\psi$, but of different orders, $n_1$ and $n_2$, differ in their angular deflections by precisely $(n_1 - n_2)\pi$. This is easy to understand for a ring viewed face-on ($\mathscr{i} = 0$). However, this remains true for rings viewed from arbitrary inclinations because the $n$ and $n\!+\!1$ photons appearing at the same $\psi$ were emitted from antipodal points on the ring, i.e., $\varphi_{\mathrm{e}; n+1} - \varphi_{\mathrm{e}; n} = \pi$ \eqref{eq:Image_Plane_Polar_Angle}. Thus, if we denote their source colatitudes as $\vartheta_{\mathrm{e}; n}$ and $\vartheta_{\mathrm{e}; n\!+\!1}$ respectively, then $\vartheta_{\mathrm{e}; n+1}$ - $\vartheta_{\mathrm{e}; n} = \pi$ \eqref{eq:Source_Angular_Coordinate_Image_Plane_Polar_Angle}, i.e., their angular deflections are also simply offset by $\pi$ \eqref{eq:Angle_Deflection_Higher_Order_Images_n}. The panel on the right shows how the asymmetry, $\mathscr{A}_1$, (or ellipticity, axis-ratio) of the $n=1$ image of a ring of radius $r_{\mathrm{e}}$ changes with the observer inclination. For small ring radii, $r_{\mathrm{e}} \lesssim 10M$, and moderate inclinations $0 \leq \mathscr{i} \lesssim \pi/4$, the asymmetry varies essentially with radius of the ring and remains quite small.}
\label{fig:Fig7_Schw_Rings_Angular_Deflection_Asymmetry}
\end{center}
\end{figure*}

\begin{figure*}
\begin{center}
\includegraphics[width=2\columnwidth]{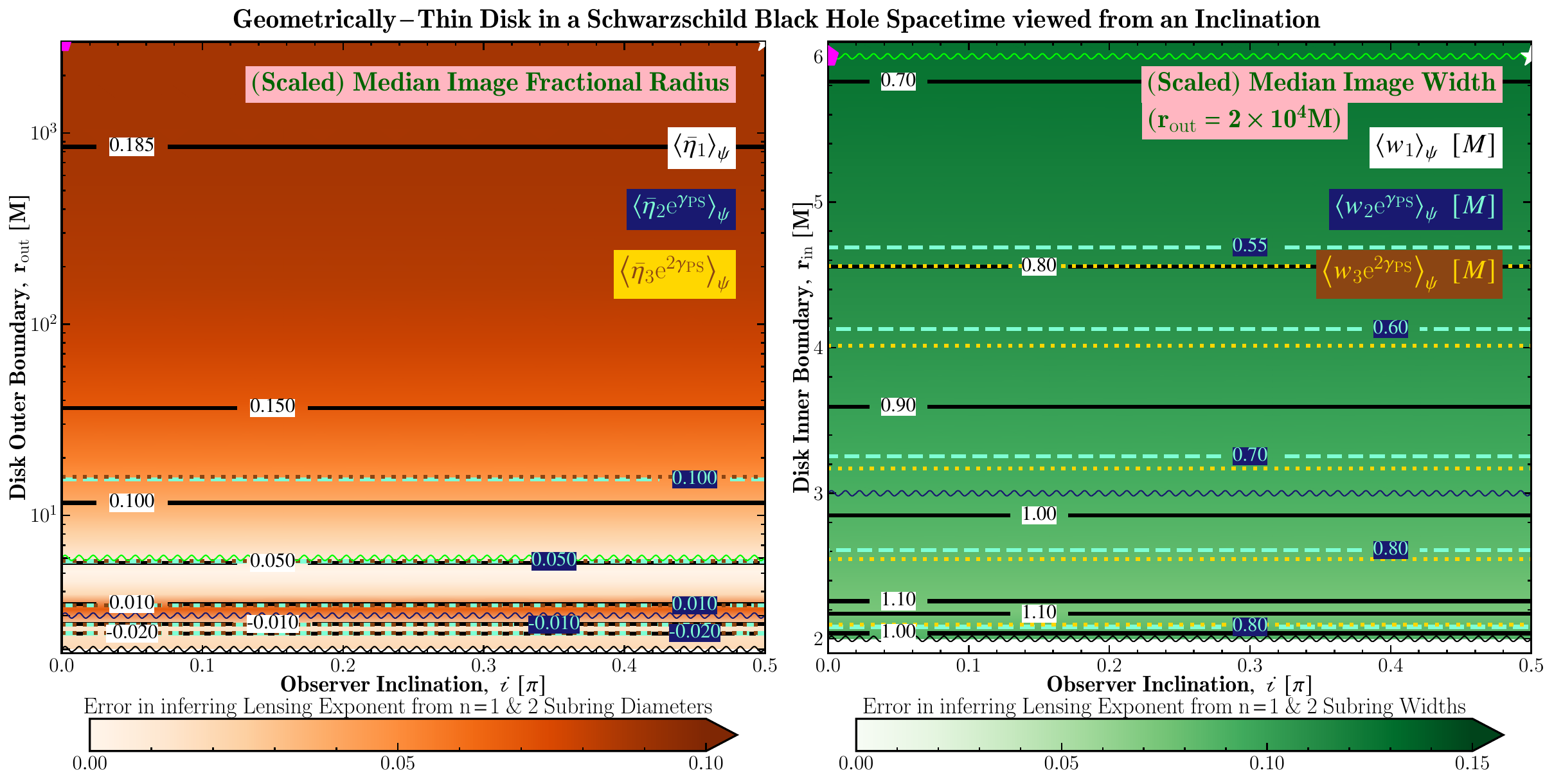}
\caption{\textit{Higher-order images of a geometrically-thin disk in a Schwarzschild BH spacetime, viewed at inclination.} The panel on the left shows the fractional median radii of the first three photon subrings whereas the one on the right shows their median widths. The squiggly lines indicate the locations of the event horizon ($2M$; black), the photon sphere ($3M$; blue), and the innermost stable circular orbit ($6M$; green). This figure shows that the deviations from circularity in the shapes of the higher-order images due to the observer's inclination do not impact their median diameters or widths. This is reminiscent of the independence of the shadow boundary with the observer inclination $\mathscr{i}$ in static and spherically-symmetric spacetimes. We see that the $n=1$ subring diameter always remains close to the shadow boundary remains $|\Braket{\bar{\eta}_1}_\psi| < 0.2$. Thus, a measurement of the subring diameter yields an accurate estimate of the shadow diameter. The right panel estimates the widths of the $n=1$ subrings to be at most $\approx 1M$. Finally, the color bar conveniently shows the error in recovering the lensing Lyapunov exponent from a joint measurement of the first two subrings, when using either their diameters (left) or their widths (right), to be $\lesssim 10\%$ and $\lesssim 15\%$ respectively. Subring asymmetry due to viewing angle is discussed in Figs. \ref{fig:Fig3_Schw_Hotspots} and \ref{fig:Fig7_Schw_Rings_Angular_Deflection_Asymmetry}.}
\label{fig:Fig8_Schw_Thin_Disk_Inclination}
\end{center}
\end{figure*}


The fractional (or conformal) radius is also the perfect coordinate with which to display the self-similar (or conformal) scaling symmetry of the photon ring \eqref{eq:n_n+1_Lensing_Lyapunov_Exp_Face_On}. To that end, we also show in this panel the \textit{scaled} fractional radii, i.e., $\mathrm{e}^{(n-1)\pi}\bar{\eta}_n$, of the next pair of higher-order images ($n=2, 3$). 

The scaling between the $n=1$ and $n=2$ image fractional radii is captured very well by the relations obtained above \eqref{eq:n_n+1_Lensing_Lyapunov_Exp_Face_On}, i.e., $\bar{\eta}_2 \approx \mathrm{e}^{-\pi}\bar{\eta}_1$, for small emission radii $r_{\mathrm{e}} \lesssim 6M$. Nevertheless, we find a maximum error of only $\approx 9\%$ when using the two lowest order subrings diameters to infer the lensing Lyapunov exponent, over all radii. Thus, a measurement of the image radii (or widths) of a pair of higher-order images for small thin-disks will yield an accurate measurement of the lensing Lyapunov exponent. 

We can also see that the scaling relations for images of order $n \geq 2$ continue to work very well for arbitrarily large source radii. To emphasize this point, the inset shows the (unscaled) fractional radius of the $n=2$ image as well as the scaled radii of the next pair of higher-order images. We can infer then that the variations in the subring diameters and widths are suppressed by a factor of $\approx e^\pi \approx 23$ per increasing subring order. 

This figure also shows, as expected, how photons emitted from inside the photon sphere ($r < r_{\mathrm{PS}} - 3M$) in the bulk appear always inside the shadow boundary ($\bar{\eta} < 0$). Similarly, $n\geq1$ photons emitted from outside the photon sphere \textit{typically} appear outside the shadow boundary curve. However, and as discussed below in Appendix \ref{app:AppB_Lensing_Map}, this is not always true: Photons emitted from just outside the photon shell can appear inside the shadow boundary. This is clearly demonstrated for the $n=1$ image by the blue solid line in the right panel here. 

We emphasize that the converse of the observation above may be important in the context of ``spacetime tomography,'' i.e., for attempts to map out the spacetime geometry of astrophysical ultracompact objects by observing, e.g., multiple flaring events occurring near M87$^\star$ or Sgr A$^\star$ (see, e.g., \citealt{Tiede+2020}). A holistic qualitative, albeit somewhat technical, description of the different types of photon orbits that participate in image formation is presented in Appendix \ref{app:AppC_Analytic_Approximations}.


\subsection{Geometrically-Thin Disk viewed from an Inclination}
\label{sec:Sec5dot2_Thin_Disks_Inclination}

In this section, we will consider the impact of a nonzero observer inclination on the properties of the photon subrings cast by a thin-disk. \cite{Tsupko2022} has recently described the shapes of the edges of such subrings on the image plane analytically (see also \citealt{Beckwith+2005}). The image and consequently the photon ring is no longer circularly-symmetric. Since the thin-disk can be decomposed into a series of concentric great-circles of different radii, an analysis of the formation of its higher-order images is identical to the image formation of circular hotspot orbits, which we discussed in Sec. \ref{sec:Sec4dot3_Secondary_Image_Orbit}.

The order$-n$ image, $\eta_n(r_{\mathrm{e}}, \psi)$, of a ring of emitters of radius $r_{\mathrm{e}}$ viewed from an inclination, $\mathscr{i}$, can be determined by solving the integral equation \eqref{eq:Main_Integral_Equation}
\begin{align} \label{eq:xi_n_inclined_thin_disk}
\slashed{\Delta}\vartheta^\pm\left(\eta_n(r_{\mathrm{e}}, \psi), r_{\mathrm{e}}\right) = \slashed{\Delta}\vartheta_n^\pm(\vartheta_{\mathrm{e}}(\mathscr{i}, \psi))\,, 
\end{align}
where, in the above we have (see eqs. \ref{eq:Polar_Orbit_Hotspot}, \ref{eq:Image_Plane_Polar_Angle}),
\begin{equation} \label{eq:Source_Angular_Coordinate_Image_Plane_Polar_Angle}
\vartheta_{\mathrm{e}}(\mathscr{i}, \psi) =
\begin{cases}
+\arctan{[\cot{\mathscr{i}} \cdot \sec{\psi}]}\,, & \mathrm{for\ even\ n}\\
-\arctan{[\cot{\mathscr{i}} \cdot \sec{\psi}]}\,, & \mathrm{for\ odd\ n}
\end{cases}
\end{equation}
for rings of inclinations $0 \leq \mathscr{i} < \pi/2$. For rings with inclinations in $(\pi/2, \pi]$, the signs on the right are flipped. For small inclinations $\mathscr{i} \approx 0$, we can find that the emission is sourced from colatitudes $\vartheta_{\mathrm{e}}(\psi) \approx \pi/2 - \mathscr{i}\cos{\psi}$ (cf. Sec. III D of \citealt{Gralla+2019}).

The left panel of Fig. \ref{fig:Fig7_Schw_Rings_Angular_Deflection_Asymmetry} shows the variation in the angular deviation $\slashed{\Delta}\vartheta_n(\vartheta_{\mathrm{e}}(\mathscr{i}, \psi))$ discussed above, with the image plane polar angle, $\psi$, for rings of various inclinations. For photons emitted from a particular ring, the additional angular deflection experienced by two that appear at the same angle but which are of different orders differs by $\pi$, in spherically-symmetric spacetimes. This may be slightly unintuitive but is true because (a) antipodal points on the ring of emitters appear at antipodal points on the image plane and (b) consecutive order images of the same point emitter appear at antipodal points on the image plane.

This means that the orientation of the asymmetry of arbitrary higher-order images is identical. If the normal to the plane of the ring $n_{\mathrm{d}}$ appears on the image plane as pointing along the negative $\alpha-$axis (see Fig. \ref{fig:Fig3_Schw_Hotspots}), this is also the direction on the image plane along which a higher-order image is also most stretched. In the right panel of Fig. \ref{fig:Fig7_Schw_Rings_Angular_Deflection_Asymmetry}, we show the asymmetry,
\begin{equation} \label{eq:Asymmetry}
\mathscr{A}_n = 1-d_n(\psi=\pi/2)/d_n(\psi=0)\,,
\end{equation}
of the secondary ($n=1$) image. This definition measures the deviation between the maximum and minimum image diameters, in the directions parallel and perpendicular to the projected normal $n_{\mathrm{d}}$ respectively. As expected, viewing an axisymmetric emission source from higher inclinations leads to higher image asymmetry. Interestingly, the asymmetry remains small $\mathscr{A}_1 \lesssim 0.10$ independently of the viewing angle for emission coming from close to the BH $r \lesssim 10M$ (cf. also \citealt{Medeiros+2022} for discussion on the asymmetry of the complete image).

We expect the shape asymmetry of the $n=1$ image to closely track the shadow boundary curve (see Fig. \ref{fig:Fig8_Schw_Thin_Disk_Inclination} below), with the latter defined purely by the spacetime geometry and the observer inclination. It should then be possible to infer the spin of a black hole from such a measurement (see, e.g., \citealt{Johannsen+2010}, \citealt{Johnson+2020}). This may be a particularly good avenue to measure the spin of M87$^\star$ since we know its inclination rather well (see, e.g., \citealt{Walker+2018}). 

The left panel of Figure \ref{fig:Fig9_Schw_Thick_Disk_Face_On} shows the change in the median fractional radii (similar to Fig. \ref{fig:Fig6_Schw_Thin_Disk_Face_On}) of the first three ($n=1\!-\!3$) photon subrings respectively, with changing inclination of the observer $\mathscr{i}$ and varying size $r_{\mathrm{out}}$ of a geometrically-thin emission disk. These image radii correspond to the order$-n$ gravitationally-lensed sizes of the outer boundary of the disk. Equivalently, these also describe the median image radii of a single ring of radius $r=r_{\mathrm{out}}$.

The panel on the right shows the variation in the median widths of these subrings for a varying disk inner boundary. For concreteness, we pick the disk outer boundary to be located at $r_{\mathrm{out}} \approx 2\times 10^4M$. Changing the outer boundary to smaller values naturally reduces the width of the ring. We pick this large outer boundary radius to show the magnitude of the maximal width possible in principle.

These panels demonstrate how the median diameters and widths of the subrings depend primarily on the size of the thin disk $r_{\mathrm{out}}$ and are independent of the viewing inclination $\mathscr{i}$. Thus, the maximal diameters and widths reported in the previous section continue to indicate the range of possible \textit{median} diameters and widths, for thin-disks viewed from arbitrary inclinations.

In addition to encoding the extent of the physical region that sources the observed emission, the median subring diameters and widths capture the scaling exponent, which is specific to spacetime geometry. This is seen from the error ($\lesssim 9\%$; shown in the color bars) in inferring the lensing Lyapunov exponent from a joining measurement of the diameters or widths of the $n=1$ and $n=2$ subrings.


\subsection{Geometrically-Thick Disk viewed Face-on}
\label{sec:Sec5dot3_Thick_Disks_Face_On}

There is mounting evidence that the ultracompact objects M87$^\star$ and Sgr A$^\star$ host geometrically-thick accretion flows \citep{EHTC+2019e, EHTC+2022e}, and thus have geometrically-thick emitting regions. The effective scale-heights, $h/r$, for hot accretion flows around Kerr BHs were carefully studied using general relativistic magnetohydrodynamics (GRMHD) simulations, and an upper bound of $h/r \lesssim 0.4$ was obtained (see Fig. 7 of \citealt{Narayan+2022}; See also \citealt{Porth+2019, Chatterjee+2022}). This can be converted into the faces of the emission zone being located at $(h/2)/r = \tan{[\pm(\pi/2 - \vartheta_{\mathrm{e}})]} \approx \pm[\pi/2 - \vartheta_{\mathrm{e}}]$, or, equivalently, $\vartheta_{\mathrm{e}} \approx \pi/2 \pm 0.2$.

With this in mind, we will now explore the impact of a varying disk scale-height on potential inferences of the Schwarzschild photon subring characteristics. Since we understand the impact of the observer inclination on the subring characteristics from the previous section, we will set here, for clarity, the inclination to vanish (face-on observer). The image morphology for this configuration is also circularly symmetric, and higher-order images are perfect annuli concentric with the shadow boundary curve. While a similar analysis was presented in Sec. C of \cite{Gralla+2019}, our goal here is to understand how sensitive these characteristics are to each morphological parameter. Furthermore, as discussed above (Sec. \ref{sec:Sec1_Introduction}), the quantitative values we report here for any given morphology serve as useful upper bounds for the same, independent of other non-gravitational emission physics.

For our present purposes, we will employ a conical torus, which is formed out of the intersection of cones and spheres, to model a thick-disk. It is parametrized by three parameters, two that control its inner $r_{\mathrm{in}}$ and outer $r_{\mathrm{out}}$ spherical surfaces, and a third $\vartheta_{1/2}$ that modifies its conical faces, i.e., its scale-height. It is to be imagined that photons are emitted from the region $r_{\mathrm{in}} \leq r_{\mathrm{e}} \leq r_{\mathrm{out}}$ and between colatitudes $\pi/2 - \vartheta_{1/2} \leq \vartheta_{\mathrm{e}} \leq \pi/2 + \vartheta_{1/2}$ (or equivalently from latitudes between $\pm\vartheta_{1/2}$). The realistic upper bound (from GRMHD simulations) on the scale height is equivalent to $\vartheta_{1/2} \lesssim 0.2$.

In the top-right panel of Fig. \ref{fig:FigB2_Schw_Higher_Order_Images}, we show such a conical torus, with parameters $\{r_{\mathrm{in}}, r_{\mathrm{out}}, \vartheta_{1/2}\} = \{2M, 18M, \pi/10\}$. Its primary ($n=0$) and secondary ($n=1$) images, as seen by a face-on observer, are shown in the lower right panel there. The regions with relatively darker shading in the top left panel there capture the properties of the photons that, as discussed below, form the photon ring ($n \geq 1$) for this source morphology.


\begin{figure*}
\begin{center}
\includegraphics[width=2\columnwidth]{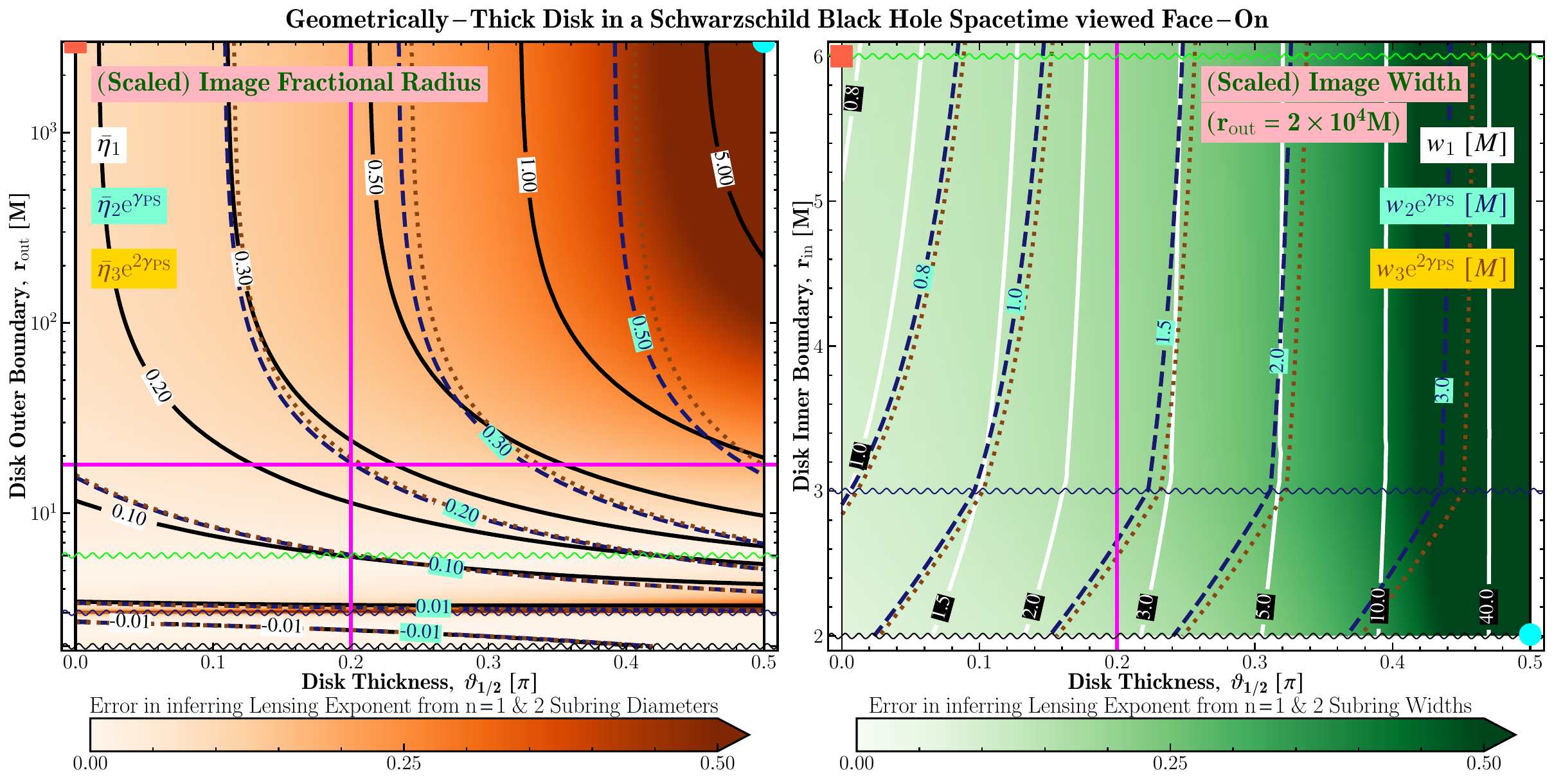}
\caption{\textit{Higher-order images of a geometrically-thick disk in a Schwarzschild black hole spacetime viewed face-on.} 
Same as Fig. \ref{fig:Fig9_Schw_Thick_Disk_Face_On} but for geometrically-thick disks, which we model as a conical torus (see right panel of Fig. \ref{fig:FigB2_Schw_Higher_Order_Images}). The scale-height of the disk is shown on the $x-$axis (i.e., the surfaces of the disk are at latitudes $\pm\vartheta_{1/2}$). The left panel shows the $n=1$ subring diameter to be sensitive to variations in the disk scale-height, even for realistic values of the morphological parameters $\vartheta_{1/2} \lesssim 0.2$ and $r_{\mathrm{out}} \lesssim 20M$ (shown in bright magenta lines; see, e.g., \citealt{Narayan+2022}). Over these ranges, the $n=1$ subring diameter nevertheless closely tracks the shadow diameter, $|\bar{\eta}_1| \lesssim 0.25$. In the panel on the right, while we have chosen a large outer boundary radius to show the maximal possible width variations, we can infer the impact of a varying outer boundary from the left panel (a negligible change for $r_{\mathrm{out}} \gtrsim 100 M$). As with the left panel, the disk geometrical-thickness plays an important role in determining subring widths and, for realistic morphologies, the maximal $n=1$ subring width is approximately $\approx 1.3M$. Since our interferometers measure angular sizes in practice, we note that a ring of width $w_n \approx 1M$ has an angular thickness of one angular gravitational radius $\theta_{\mathrm{g}} = GM/c^2D = M/D$ on the sky. For M87$^\star$ and for Sgr A$^\star$, these have been inferred by the EHT to be $\theta_{\mathrm{g}} = 3.8^{+0.4}_{-0.4}\mu\mathrm{as}$ \citep{EHTC+2019a} and $\theta_{\mathrm{g}} = 4.8^{+1.1}_{-0.7}\mu\mathrm{as}$ \citep{EHTC+2022a} respectively. Finally, the color bars in each panel show that the lensing Lyapunov exponent can be inferred, in principle, with an error of $\lesssim 20\%$ by comparing the $n=1$ and $2$ subring diameters (left) or widths (right).}
\label{fig:Fig9_Schw_Thick_Disk_Face_On}
\end{center}
\end{figure*}


We can understand the higher-order image formation of our conical torus by considering first the image formation of a conical surface. As can be seen from the top left panel of Fig. \ref{fig:FigB2_Schw_Higher_Order_Images}, photons of any particular order that are emitted from increasingly larger radii, $r_{\mathrm{e}}$, but from the same conical surface (fixed $\vartheta_{\mathrm{e}}$; i.e., as we move along a fixed horizontal line in this plot) always appear at increasingly larger radii, $\eta$. On such surfaces, the lensing map, $r_{\mathrm{e}} \mapsto \eta$, is neatly monotonic (see also related discussion in Appendix \ref{app:AppB_Lensing_Map}). Therefore, for emission from any single cone, the lensed size of the emission outer (inner) boundary radius determines the outer (inner) edge of its image. Understanding the images of cones in black hole spacetimes finds an interesting application in the analysis of jet shapes and jet opening angles \citep{Papoutsis+2023, Chang+2024}.

Since the image of a conical torus can be thought of as the sum of the images of single cones, to determine the edges of the higher-order images of the former, we must also understand how the images of different single cones compete. As can be seen from the solid lines in the top left panel of Fig. \ref{fig:FigB2_Schw_Higher_Order_Images}, as we move across a spherical surface (fixed $r_{\mathrm{e}}$) located inside the photon sphere ($r_{\mathrm{e}} < r_{\mathrm{{PS}}}$), the lensing map displays monotonic behaviour. The inner and outer edges of the order$-n (>0)$ image of smaller spheres correspond to the Einstein rings of its north and south poles for odd-$n$, and of its south and north poles for even-$n$. Its primary ($n=0$) image is the entire region inside the first Einstein ring or the first critical curve. 

As we move across a spherical surface (fixed $r_{\mathrm{e}}$) located outside the photon sphere ($r_{\mathrm{e}} > r_{\mathrm{{PS}}}$), the lensing map displays nonmonotonic behaviour. The outer edge of its $n=0$ image is determined by the radius at which the photon that has its radial turning point at the sphere appears, $\eta = \eta_{\mathrm{tp}}(r_{\mathrm{e}})$, which can be obtained as a solution to the turning point equation \eqref{eq:Radial_Turning_Point_Equation}, $\mathscr{R}(\eta_{\mathrm{tp}}(r_{\mathrm{e}}), r_{\mathrm{e}}) = 0$. This is also true for order$-n$ images of spheres of radii $r_{\mathrm{PS}} < r_{\mathrm{e}} < r_{n; \mathrm{C^T}}$ (cf. Footnote \ref{fn:HIO_Nonmonotonicity} for details). While we will account for this below, we note that $\eta_{\mathrm{tp}}(r_{n; \mathrm{C^T}})$ approaches $\eta_{\mathrm{PS}}$ exponentially quickly with each increasing image order. Thus, the higher-order lensing map is essentially a monotonic function of the colatitude over a sphere. Further discussion on this point is presented in Appendix \ref{app:AppC_Analytic_Approximations}. 

We introduce one last ingredient before writing down simple expressions that locate the outer and inner edges of the photon subrings cast by conical torii in spherically-symmetric spacetimes. For an observer on the north pole, the ``front-face'' of the direct ($n=0$) image is the one closer to the north pole, $\vartheta_{\mathrm{e}} = \pi/2 - \vartheta_{1/2}$, whereas that for the indirect ($n=1$) image it is the one closer to the south pole, $\vartheta_{\mathrm{e}} = \pi/2 + \vartheta_{1/2}$. Thus, below, by the front and back face we will mean the conical surfaces $\vartheta_{\mathrm{e}} = \pi/2 - \vartheta_{1/2}$ and $\vartheta_{\mathrm{e}} = \pi/2 + \vartheta_{1/2}$ for even-order images and vice versa for odd-order images.

All of the above can be understood more concretely by considering, e.g., the zeroth and first-order images of a conical torus of outer radius $r_{\mathrm{e}} = 6M$, inner radius $3 M < r_{\mathrm{e}} < 3.5 M$, and large opening angle $\vartheta_{1/2} = 0.8$, and by looking at the top left panel of Fig. \ref{fig:FigB2_Schw_Higher_Order_Images}.

With the equations for the order$-n$ image radii of the front ($\eta_{n-\mathrm{FF}}$) and back ($\eta_{n-\mathrm{BF}}$) faces of the conical torus \eqref{eq:del_vartheta},
\begin{align}
|\slashed{\Delta}\vartheta^\pm\left(\eta_{n-\mathrm{FF}}(r_{\mathrm{e}}), r_{\mathrm{e}}\right)| =&\ n\pi + \left(\pi/2 - \vartheta_{1/2}\right)\,, \\
|\slashed{\Delta}\vartheta^\pm\left(\eta_{n-\mathrm{BF}}(r_{\mathrm{e}}), r_{\mathrm{e}}\right)| =&\ n\pi + \left(\pi/2 + \vartheta_{1/2}\right)\,, 
\end{align}
we find the edges of the order$-n$ images as,
\begin{align} \label{eq:thick_disk_eta_n_in_out}
\eta_{n; \mathrm{in}} :=&\
\mathrm{min.}\left\{\eta_{n-\mathrm{FF}; \mathrm{in}}, \eta_{n-\mathrm{BF}; \mathrm{in}}, \eta_{n-\mathrm{tp}; \mathrm{in}}\right\}\,, \\
\eta_{n; \mathrm{out}} :=&\
\mathrm{max.}\left\{\eta_{n-\mathrm{FF}; \mathrm{out}}, \eta_{n-\mathrm{BF}; \mathrm{out}}, \eta_{n-\mathrm{tp}; \mathrm{out}}\right\}\,. \nonumber
\end{align} 
where we have defined $\eta_{n-\mathrm{FF}; \mathrm{in}} := \eta_{n-\mathrm{FF}}(r_{\mathrm{in}})$ and $\eta_{n-\mathrm{BF}; \mathrm{in}} := \eta_{n-\mathrm{BF}}(r_{\mathrm{in}})$. Also, as promised, we have introduced $\eta_{n-\mathrm{tp}; \mathrm{in}} := \eta_{n-\mathrm{tp}}(r_{\mathrm{in}})$, where the latter is a solution to the turning point equation \eqref{eq:Radial_Turning_Point_Equation}, $\mathscr{R}\left(\eta_{n-\mathrm{tp}}(r_{\mathrm{e}}), r_{\mathrm{e}}\right) = 0$, and is relevant only when $n\pi + \left(\pi/2 - \vartheta_{1/2}\right)  \leq |\slashed{\Delta}\vartheta^\pm\left(\eta_{n-\mathrm{tp}}, r_{\mathrm{e}}\right)|  \leq n\pi + \left(\pi/2 + \vartheta_{1/2}\right)$. The definitions of outer edge radii are analogously defined using $r_{\mathrm{e}} = r_{\mathrm{out}}$. We reiterate that $\slashed{\Delta}\vartheta^\pm$ is negative for even order photons and positive for odd order ones \eqref{eq:Angle_Deflection_Higher_Order_Images_n}. Finally, we will note here that similar equations can be used to determine the edges of the photon subrings when a thick-disk is viewed from an inclination (see eq. \ref{eq:xi_n_inclined_thin_disk}). 

Fig. \ref{fig:Fig9_Schw_Thick_Disk_Face_On} captures the variation in the diameters of the first three subrings cast by thick disks in a Schwarzschild BH spacetime, for varying outer boundary radius $r_{\mathrm{out}}$ and disk-height $\vartheta_{1/2}$. The increase in the subring diameter with the increasing geometrical thickness of the disk is to be expected and can be understood broadly as follows. While photons that appear in the first-order image of a geometrically-thin emission disk ($\vartheta_{1/2} = 0$) all experience a net angular deflection of $\slashed{\Delta}\vartheta = 3\pi/2$, that of a thick-disk contains additionally photons that undergo smaller deflections. Thus, photons emitted from the same outer boundary radial location $r=r_{\mathrm{out}}$ in the thick-disk case can undergo smaller deflections before appearing in the $n=1$ image as compared to the $n=1$ image of the thin-disk. Since $n \geq 1$ photons that are emitted from the same radial location, outside the photon sphere, but which undergo smaller angular deflections typically appear at larger impact parameters (see Fig. \ref{fig:FigB2_Schw_Higher_Order_Images}), the diameter of photon subring increases with scale-height. Therefore, this figure makes clear that overall the disk scale-height plays a significant role in determining the sizes of the photon rings (see also \citealt{Gralla+2019}). For realistic geometrical-thicknesses (to the lower left of the bright green lines), however, we find the fractional diameter of the $n=1$ subring to lie in $-0.01 \lesssim \bar{\eta}_1 \lesssim 0.25$. 

With no prior knowledge of the emission zone morphology, a strong association between the $n=1$ image diameter and that of the shadow boundary is not, in general, possible. This is, of course, trivially true since the $n=1$ image can, in general, be non-compact. However, with indirect prior knowledge inferred, e.g., by comparing synthetic ($n=0-$dominated) images produced from realistic numerical simulations against those obtained by the EHT (see, e.g., Fig. 8 of \citealt{EHTC+2022e}), arguing for a strong association between the $n=1$ image diameter and that of the shadow boundary becomes plausible. \cite{Chang+2024} demonstrate strikingly how information regarding the emission zone morphology as well as the spacetime geometry (there the spin of a Kerr BH) may be simultaneously extracted in this way (see, e.g., Fig 15 there). We note that similar associations between the $n=0$ image diameter and the shadow boundary curve have also been established using both semi-analytic accretion models for a range of emission zone morphologies in a range of non-Kerr spacetimes (see, e.g., \citealt{Ozel+2021, Kocherlakota+2022, Younsi+2021}) as well as with numerical simulations in Kerr spacetimes (see Fig. 7 of \citealt{EHTC+2022f}). It is worth mentioning that while numerical simulations, which also account for all relevant non-gravitational emission physics, in non-Kerr BH spacetimes have been performed \citep{Mizuno+2018, Chatterjee+2023b, Chatterjee+2023a}, the necessary analysis to conclusively establish these associations (both $n=0, 1$) remains to be carried out. Thus, with these caveats in mind, we are able to argue reasonably that a direct and accurate inference of the size of the shadow $\eta_{\mathrm{PS}}$ from a measurement of the $n=1$ subring diameter seems to be within reach, with longer (space-based) baseline radio interferometry. 

Another potential (albeit harder; \citealt{Johnson+2020}) observable characteristic of a photon subring is its width $w_n$. The right panel of Fig. \ref{fig:Fig9_Schw_Thick_Disk_Face_On} displays the variation in the width of the $n=1$ subring cast by a thick disk. We set the disk outer boundary to be at some large distance ($r_{\mathrm{out}} = 2\times 10^4 M$) to obtain a sense of the maximal subring width possible. For realistic geometrical-thicknesses (to the left of the magenta line), the $n=1$ subring width can be as large as one gravitational radius, $w_1 \approx 1.3M$. In \cite{Kocherlakota+2024a}, a companion to this paper, the widths of $n=1$ subrings cast by an emission zone of identical morphological parameters in a large number of spherically-symmetric BH spacetimes were computed, in an identical analysis to the one presented here. We argued there, qualitatively, that, with prior knowledge of the morphology of the emitting region, a measurement of the subring width could be used to set constraints on the spacetime geometry. However, it remains imperative that numerical simulations be used to quantitatively establish the magnitude of dependence of the shape of the emission zone on the underlying spacetime parameters.

In the left panel, we also show the scaled fractional radii of the next pair of higher-order subrings diameters. In particular, the variation in the fractional radius of the $n=2$ subring is $-0.001 \lesssim \bar{\eta}_2 \lesssim 0.03$, making it an incredibly fine feature that may prove elusive to observations in the near future. 

Finally, the color bar in both panels indicates the error in obtaining the Schwarzschild value of the lensing Lyapunov exponent, $\gamma_{\mathrm{PS}} = \pi$, from a joint measurement of the $n=1$ and $n=2$ photon subring diameters (left) and widths (right). For realistic emission region morphologies, we find a maximum error of $\lesssim 20\%$.

In summary, we find from our analysis of photon subring variations that median subring diameters and widths in static and spherically-symmetric spacetimes are roughly independent of the observer inclination. Furthermore, both these characteristics depend acutely on the geometrical thickness of the emitting region, with thicker emitting regions generically leading to larger and wider photon rings. However, for morphologies informed by simulations, we expect to find moderate variations. This analysis provides a solid basis for future work to focus on considerations of the impact of other physical effects that we have ignored here.


\section{Summary \& Conclusions}
\label{sec:Sec6_Conclusions}
 
\subsection{Higher-Order Image Scaling Relations}
\label{sec:Sec6dot1_Theory_Summary}

The region on the sky that collects all higher-order images ($n > 0$) is referred to as the photon ring. The order of an image is determined by the maximum number of half-loops executed around the BH by the photons that form it. Various properties of photons arriving in the photon ring exhibit universal scaling relations (see, e.g., \citealt{Bozza2002, Bozza+2003, Gralla+2019, Johnson+2020, Gralla+2020a}). The photon ring contains the shadow boundary curve (or, equivalently, the $n=\infty$ critical curve), and photons that appear in the photon ring increasingly closer to it have logarithmically-divergent angular deflections, $\slashed{\Delta}\vartheta^\pm$, as well as travel times, $\slashed{\Delta}t^\pm$. Since this is a self-similar or conformal symmetry, these divergences are best represented using the fractional radius or the conformal radial coordinate on the image plane, $\bar{\eta} := \eta/\eta_{\mathrm{PS}} - 1$. These are given respectively as
\begin{align}
\slashed{\Delta}\vartheta^\pm(\bar{\eta}) \approx&\ \mp(\pi/\gamma_{\mathrm{PS}})\ln{|\bar{\eta}|}\,, \\
\slashed{\Delta} t^\pm(\bar{\eta}) \approx&\ -t_{\ell; \mathrm{PS}}\ln{|\bar{\eta}|}\,.
\end{align}
The superscripts denote the sign of the photon's initial polar velocity $\dot{\vartheta}$. The constants $\gamma_{\mathrm{PS}}$ and $t_{\ell; \mathrm{PS}}$ are characteristics of the spacetime, related to the radial instability of photon orbits close to the photon sphere, and are called the lensing Lyapunov exponent and the Lyapunov time respectively (their analytic expressions can be found in eq. \ref{eq:Lensing_Lyapunov_Exponent_Lyapunov_Time}).

Using the exact analytic expression for the angular deflections experienced by photons of different orders (eq. \ref{eq:Angle_Deflection_Higher_Order_Images_n}) 
\begin{equation} 
\slashed{\Delta}\vartheta^\pm_n = \pi/2 - \vartheta_{\mathrm{e}} + (-1)^{n+1}(2n+1)\pi/2\,,\ \ 0 < \vartheta_{\mathrm{e}} < \pi\,,
\end{equation}
when emitted from a point source present at a colatitude of $\vartheta_{\mathrm{e}}$, with the above, we can find the scaling relations for their image plane fractional radii, $\bar{\eta}_n$, and time delays, $\Delta t_n^\mp$, to be 
\begin{align} 
\label{eq:Image_Plane_etabar_Scaling}
\frac{\bar{\eta}_{n+1}}{\bar{\eta}_n} \approx \mathrm{e}^{-\gamma_{\mathrm{PS}}}\cdot\mathrm{e}^{\pm\gamma_{\mathrm{PS}}(2\vartheta_{\mathrm{e}}/\pi - 1)}\,, \\
\label{eq:Time_Delay_n_n+1}
\Delta t_n^\mp \approx \pi\eta_{\mathrm{PS}}\left[1 \mp \left(\frac{2\vartheta_{\mathrm{e}}}{\pi} - 1\right)\right]\,.
\end{align} 
Here the upper sign is chosen if $n$ is even and the lower sign is chosen if $n$ is odd (eq. \ref{eq:Angle_Deflection_Higher_Order_Images_n}).

\subsection{Higher-Order Images of Point Sources}
\label{sec:Sec6dot2_Hotspots_Summary}

Flaring events are frequently observed in Sgr A$^\star$ across the electromagnetic spectrum \citep{Marrone+2008, Do+2019, Haggard+2019, Wielgus+2022LC}. The emergence of compact sources of flux transiently orbiting the central black hole \citep{Gravity+2018, Wielgus+2022} is likely related to flares locally heating the accreting plasma \citep{Dexter+2020, Ripperda+2021}. In future high-resolution movies obtained via interferometry, it might be possible to capture the appearance of the primary ($n=0$) image of the hotspot as well as observe its evolution over time \citep{Johnson+2023}. In such scenarios, we will likely also observe the appearance and evolution of the secondary ($n=1$) image \citep{Tiede+2020}. 

Such detections of higher-order images have the potential to allow us to measure the effects of strong gravitational lensing on horizon-scales. Due to the additional half-loop executed by the photon forming the secondary image of a hotspot appears at a later time than the primary, even when both sets of photons are emitted at the same time. The characteristic delay time is linked to the size of the black hole shadow as $t_{d; \mathrm{PS}} \approx \pi\eta_{\mathrm{PS}} (= \pi\sqrt{27}M$ for a Schwarzschild BH of mass $M$). For Sgr A$^\star$ and M87$^\star$, this is roughly $5~\mathrm{min}$ and $6~\mathrm{days}$ respectively. The exact time delay depends sensitively on the viewing inclination (see eq. \ref{eq:Time_Delay_n_n+1} above) and also, relatively weakly, on the distance of the hotspot from the black hole. We provide a simple analytic expression that captures the first of these effects \eqref{eq:Primary_Secondary_Time_Delay}, and find errors of $\lesssim 20\%$ for hotspots produced close to the light cylinder radius \citep{Ripperda+2021}. Characterizing the various effects that relate the primary and secondary images, including those due to geometric and lensing effects (as here), can help develop observational strategies to detect the latter.

We also consider the orbits of the primary and secondary images on the sky for a point source on a circular Keplerian orbit (a toy model for an orbiting hotspot) in detail (Sec. \ref{sec:Sec4dot3_Secondary_Image_Orbit}) and find that these neatly encode information relating to the angular velocity of the hotspot as well as the orbital inclination relative to the observer. The evolution of the time delay between the images also encodes this information, with the evolution of the angle offset between the two images turning out to be slightly more involved to interpret.

\subsection{Higher-Order Images of Extended Sources}
\label{sec:Sec6dot2_Photon_Rings_Summary}

Supermassive compact objects, such as M87$^\star$ or Sgr A$^\star$, host hot accretion flows, which act as extended sources of emission. Such extended sources cast higher-order images, each of which is simply the union of the higher-order images of individual fluid elements, all of the same order. These are referred to as photon subrings, having diameters $d_n$, widths $w_n$, and flux densities $F_n$, which also satisfy scaling relations qualitatively similar to the ones above.

Recent Event Horizon Telescope (EHT) images of the supermassive compact objects M87$^\star$ and Sgr A$^\star$ are dominated by the primary or the zeroth-order ($n=0$) images of their accretion flows. Future radio very long baseline interferometry, likely radio dishes in low Earth orbits, is expected to reach sufficient angular resolutions to reveal their first-order ($n=1$) images \citep{Johnson+2020, Gralla+2020b, Gurvits+2022, Kurczynski2022}.

For realistic morphologies of the emission zone, we find that the fractional deviation of the $n=1$ subring diameter $\bar{\eta}_1$ in a Schwarzchild BH spacetime roughly takes values $|\bar{\eta}_1| \lesssim 0.3$.%
\footnote{We note that the quantity $\bar{\eta}_1$ is essentially equivalent to the $\alpha_1-$calibration factor introduced in \cite{EHTC+2022f} (see Appendix \ref{app:AppE_PR_Calibration}).} %
Thus, a future measurement of its diameter will likely yield an accurate and direct inference of the shadow size. We comment on some sources of possible degeneracies between gravitational and non-gravitational degrees of freedom in Sec. \ref{sec:Sec5dot3_Thick_Disks_Face_On}. Furthermore, we find that the width of the first subring to be $\lesssim 1.3M$. Therefore, the effective angular resolution required to resolve its width, in the best case scenario, is comparable to the angular gravitational radius $\theta_{\mathrm{g}} = GM/(c^2D)$, where $D$ is the distance to the compact object. For M87$^\star$ and Sgr A$^\star$, these have been inferred from the 2017 EHT observations to be $3.8^{+0.4}_{-0.4}\mu\mathrm{as}$ \citep{EHTC+2019a} and $4.8^{+1.4}_{-0.7}\mu\mathrm{as}$ \citep{EHTC+2022a} respectively. We also find that a joint measurement, in principle, of either the diameters or the widths of a pair of subrings (e.g., $n=1$ and $n=2$) can be used to obtain the lensing Lyapunov exponent with an error of $\lesssim 20\%$. 

Another promising method for measurements of the delay time as well as of the lensing Lyapunov exponent has been proposed in \cite{Hadar+2021}. These involve constructing the autocorrelations either of the light curve or of the intensity fluctuations across the image plane in future high-resolution black hole movies.

Finally, the Lyapunov time could be accessible, in principle, through sensitive measurements of the late time evolution of the luminosity when observing, e.g., gas clouds \citep{Moriyama+2019} or stars \citep{Ames+1968, Cardoso+2021} falling into a supermassive black hole.

There are two obvious limitations to our work. First, we have only considered nonspinning spacetimes here. However, complementary work (see, e.g., \citealt{Johnson+2020, Ayzenberg2022, Vincent+2022, Paugnat+2022}) indicates that many of the qualitative features obtained here should carry forward to the case of stationary and axisymmetric spacetimes. Second, we have not explicitly considered the full variations in the non-gravitational degrees of freedom that are possible. Nonetheless, barring the impact of optical depth (but see also \citealt{Junior+2021, Bisnovatyi-Kogan+2023}), we do not see this as a significant limitation (see Sec. \ref{sec:Sec1_Introduction} and Appendix \ref{app:AppD_Flux}). 

In conclusion, the EHT has already demonstrated that measurements of the spacetime geometry with black hole imaging observations are now becoming possible \citep{EHTC+2019f, Psaltis+2020, Kocherlakota+2021, EHTC+2022f}. With future improvements, detecting the time delays between hotspot primary and secondary images as well as measuring properties of photon subrings will open up interesting new windows into understanding black hole spacetimes. These will also lead to even more stringent and unprecedented tests of the spacetime geometry (see, e.g., \citealt{Kocherlakota+2024a}), as well as of the underlying theory of gravity and fields.


\bigskip


\section*{Acknowledgements}

It is a pleasure to thank Dominic Chang for several insightful discussions and suggestions. We are also grateful to Koushik Chatterjee, Ramesh Narayan, and Michael Johnson for helpful feedback. Lastly, we thank the referee for several constructive comments. PK acknowledges support in part from grants from the Gordon and Betty Moore Foundation (GBMF-8273) and the John Templeton Foundation (62286) to the Black Hole Initiative at Harvard University, and from NSF award OISE-1743747. PK and LR acknowledge support from the ERC Advanced Grant ‘JETSET: Launching, propagation and emission of relativistic jets from binary mergers and across mass scales’ (884631). LR acknowledges the Walter Greiner Gesellschaft zur F\"orderung der physikalischen Grundlagenforschung e.V. through the Carl W. Fueck Laureatus Chair.

\section*{Data Availability}
The data used in the work presented in this article are available upon request to the corresponding author.


\bibliographystyle{mnras}
\bibliography{Refs-Hotspots-Photon-Rings-Schwarzschild-MNRAS}


\appendix
\onecolumn


\section{Planarity of Orbits and Meridional Photon Orbits}
\label{app:AppA_Planarity}

The Lagrangian $\mathscr{L}$ that describes the motion along an arbitrary geodesic $x^\alpha(\lambda)$ is defined as $2\mathscr{L} := g_{\alpha\beta}\dot{x}^\alpha\dot{x}^\beta$, where $\dot{x}^\alpha = \mathrm{d}x^\alpha/\mathrm{d}\lambda =: k^\alpha$ is the tangent and $\lambda$ an affine parameter along it. Of the associated momenta, $p_\alpha = \partial_{\dot{x}^\alpha}\mathscr{L} = k_\alpha$, there are two that are conserved (due to the Euler-Lagrange equations, $\dot{p}_\alpha = \partial_\alpha\mathscr{L}$), namely the energy $E = - p_t$ and the azimuthal angular momentum $L = p_\varphi$ of the orbit. In addition to the three conserved quantities $\{2\mathscr{L}, E, L\}$, a fourth constant of the motion exists, namely the (non-negative) \cite{Carter1968} constant $C = p_\vartheta^2 + p_\varphi^2\csc^2{\vartheta}$. For null geodesics in particular, we have $2\mathscr{L} (= p_\mu p^\mu) = 0$, whereas for timelike geodesics, $2\mathscr{L}=-1$. The Carter constant defined in this way is simply the square of the total angular momentum, and is associated with the existence of an irreducible symmetric Killing tensor (see, e.g., \citealt{Kapec+2020}), which plays a role similar to that of the familiar Laplace-Runge-Lenz vector in Newtonian gravity \citep{Brill+1999, Gibbons+2007,  Cariglia+2014, Cariglia2014}. With four dynamical constants, we can separate the geodesic equation and write the tangent to an arbitrary geodesic as being (see also, e.g., \citealt{Kocherlakota+2022}),
\begin{align} \label{eq:NG_k}
k^\alpha =&\ E\left(f^{-1}, \pm_r g^{-1/2}\sqrt{1 + 2\mathscr{L}E^{-2}f - \eta^2 f R^{-2}}, 
\pm_\vartheta R^{-2}\sqrt{\eta^2 - \xi^2\csc^2{\vartheta}}, 
\pm_\varphi R^{-2}\xi\csc^2{\vartheta}
\right)\,,
\end{align}
where the indices $\pm_r, \pm_\vartheta$, and $\pm_\varphi$ correspond to the signs of the radial, polar, and azimuthal velocities respectively, and we have introduced the useful (energy-rescaled, non-negative) quantities $\xi = |L|/E$ and $\eta = \sqrt{C}/E$. Clearly, we must have $\eta \geq \xi$ along any orbit so that $k^\vartheta$ remains real.

Due to spherical symmetry, \textit{all} geodesics are spatially-planar. To see this, let us first introduce
\begin{equation} \label{eq:Orbital_Inclination}
\mathscr{i} = \pi/2 - \arcsin{(\xi/\eta)}\,,\ \ 0 < \mathscr{i} < \pi/2\,.
\end{equation}
The relevant Euler equation then is 
\begin{equation}
\frac{\mathrm{d}\varphi}{\mathrm{d}\vartheta} =\frac{k^\varphi}{k^\vartheta} = \pm \frac{\cot{\mathscr{i}}\csc^2{\vartheta}}{\sqrt{1 - \cot^2{\mathscr{i}}\cot^2{\vartheta}}}\,,
\end{equation}
the general solution to which, with $\varphi_0$ an integration constant, is
\begin{equation} \label{eq:Euler_Solution}
\varphi(\vartheta) = \mp \arcsin\left(\cot{\mathscr{i}}\cdot\cot{\vartheta}\right) + \varphi_0\,,
\end{equation}
We can put this into a more suggestive form,
\begin{align}
0 =&\ \mp (\sin{\varphi_0}\cdot\sin{\mathscr{i}})\sin{\vartheta}\cos{\varphi} \pm (\cos{\varphi_0}\cdot\sin{\mathscr{i}})\sin{\vartheta}\sin{\varphi} + (\cos{\mathscr{i}})\cos{\vartheta}\,, \nonumber
\end{align}
to immediately recognize that this is the equation of a plane passing through the center ($r=0$), and that $\mathscr{i}$ is the inclination of the normal to the orbital plane (i.e., the angle between the normal and the $z-$axis, $\vartheta = 0$). Clearly, orbits that differ only in the sign of the polar ($\pm_\vartheta$) or the azimuthal ($\pm_\varphi$) velocity lie in the same plane. While our definition of the orbital inclination \eqref{eq:Orbital_Inclination} restricts it to $0 < \mathscr{i} < \pi/2$, it can be seen that orbits with inclinations in $(\pi/2, \pi)$ are obtained by shifting $\varphi_0$ by $\pi$. 

Introducing the general effective polar potential $\tilde{\Theta}$ as (compare with eq. \ref{eq:th_Effective_Potential}),
\begin{equation} \label{eq:Polar_Potential_Gen}
\tilde{\Theta}(\eta, \xi, \theta) :=  (k^\vartheta/E)^2 = \left(\eta^2 - \xi^2\csc^2{\vartheta}\right)/R^4\,,
\end{equation} 
we see that any orbit must stay in $\vartheta_- \leq \vartheta(\lambda) \leq \vartheta_+$, where $\vartheta = \vartheta_\pm$ are the polar turning points ($\tilde{\Theta} = 0$), and are given as
\begin{equation}
\vartheta_\pm = \pi/2 \pm \mathscr{i}\,.
\end{equation}
Thus, it is clear that orbits with $\mathscr{i} = 0$ (i.e., $\eta = \xi \neq 0$) are equatorial orbits, and are, therefore, also planar. Finally, orbits with $\mathscr{i} = \pi/2$ (or $\eta \neq \xi = 0$) correspond to meridional orbits (i.e., constant-$\varphi$ orbits), corresponding to motion along longitudes. For meridional circular photon orbits in particular, we can write (from eq. \ref{eq:Shadow_Size}), 
\begin{align} \label{eq:NG_kPS}
k_{\mathrm{PS}}^\alpha =&\ E\left(f_{\mathrm{PS}}^{-1}, 0, 
\pm_\vartheta \eta_{\mathrm{PS}}R_{\mathrm{PS}}^{-2}, 
0\right)
=\ Ef_{\mathrm{PS}}^{-1}\left(1, 0, 
\pm_\vartheta 1/\eta_{\mathrm{PS}}, 
0\right)\,.
\end{align}

Note also that orbits with $\eta = 0$ necessarily require $\xi = 0$ as well, i.e., $\dot{\vartheta} = \ddot{\vartheta} = \dot{\varphi} = \ddot{\varphi} = 0$, which correspond to radial geodesics. Null geodesics, in particular, of this kind are members of the ingoing ($-_r$) and outgoing ($+_r$) principal null congruences of these \cite{Petrov2000} Type D static spacetimes. 


\section{Nonmonotonicity of The Bulk $\leftrightarrow$ Boundary Lensing Maps: Arbitrary Order Images}
\label{app:AppB_Lensing_Map}


\begin{figure*}
\begin{center}
\includegraphics[width=\columnwidth]{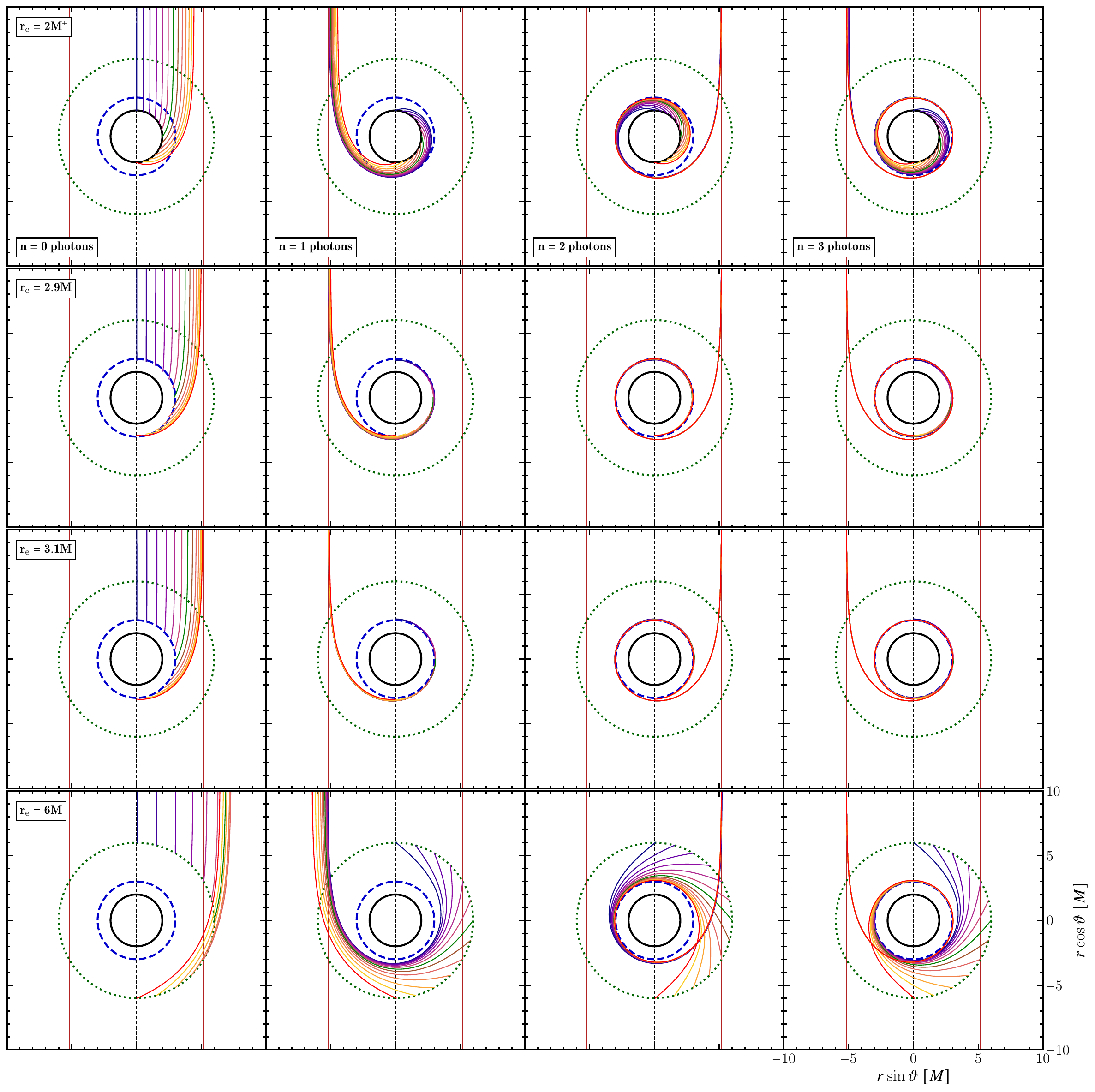}
\caption{\textit{Photon orbits connecting point sources to an observer in a Schwarzschild black hole spacetime}. In the top-left panel, we show orbits of photons emitted from just outside the horizon, $r_{\mathrm{e}} = 2M^+$, which reach an asymptotic observer, present on the $+z-$axis. The top edge of each panel represents a radial slice of the observer's screen. The different colors indicate that they are emitted from different colatitudes $\vartheta_{\mathrm{e}} = 0^\circ, 15^\circ, 30^\circ, \cdots, 180^\circ$. Across the top row, in each panel, lines of the same color represent photon orbits that begin from the same initial spatial positions. Thus, a countably infinite number of photon orbits connect any specific pair of emitter and observer. The order ($n$) of photons with the same line color increases from left to right across columns, as do their path lengths, travel times, and total angular deflections. Across rows, the radial location of the emitters is varied, $r_{\mathrm{e}} = 2M^+, 2.9M, 3.1M, 6M$. The locations of the event horizon, of the photon shell, and of the innermost stable circular orbit (ISCO) are at $r_{\mathrm{H}}=2M$ (black circle), $r_{\mathrm{PS}}=3M$ (dashed blue circle), and $r_{\mathrm{ISCO}}=6M$ (dotted green circle). Larger light bending occurs for photons that appear closer to the shadow boundary on the image plane $\eta_{\mathrm{PS}} = \sqrt{27}M$, shown as vertical red lines in all panels. It is clear that these photons necessarily access the close vicinity of the photon shell. Finally, we see that higher-order images (across each row) occupy increasingly compact regions on the image plane.}
\label{fig:FigB1_Schw_Higher_Order_Images}
\end{center}
\end{figure*}

\begin{figure*}
\begin{center}
\includegraphics[width=\columnwidth]{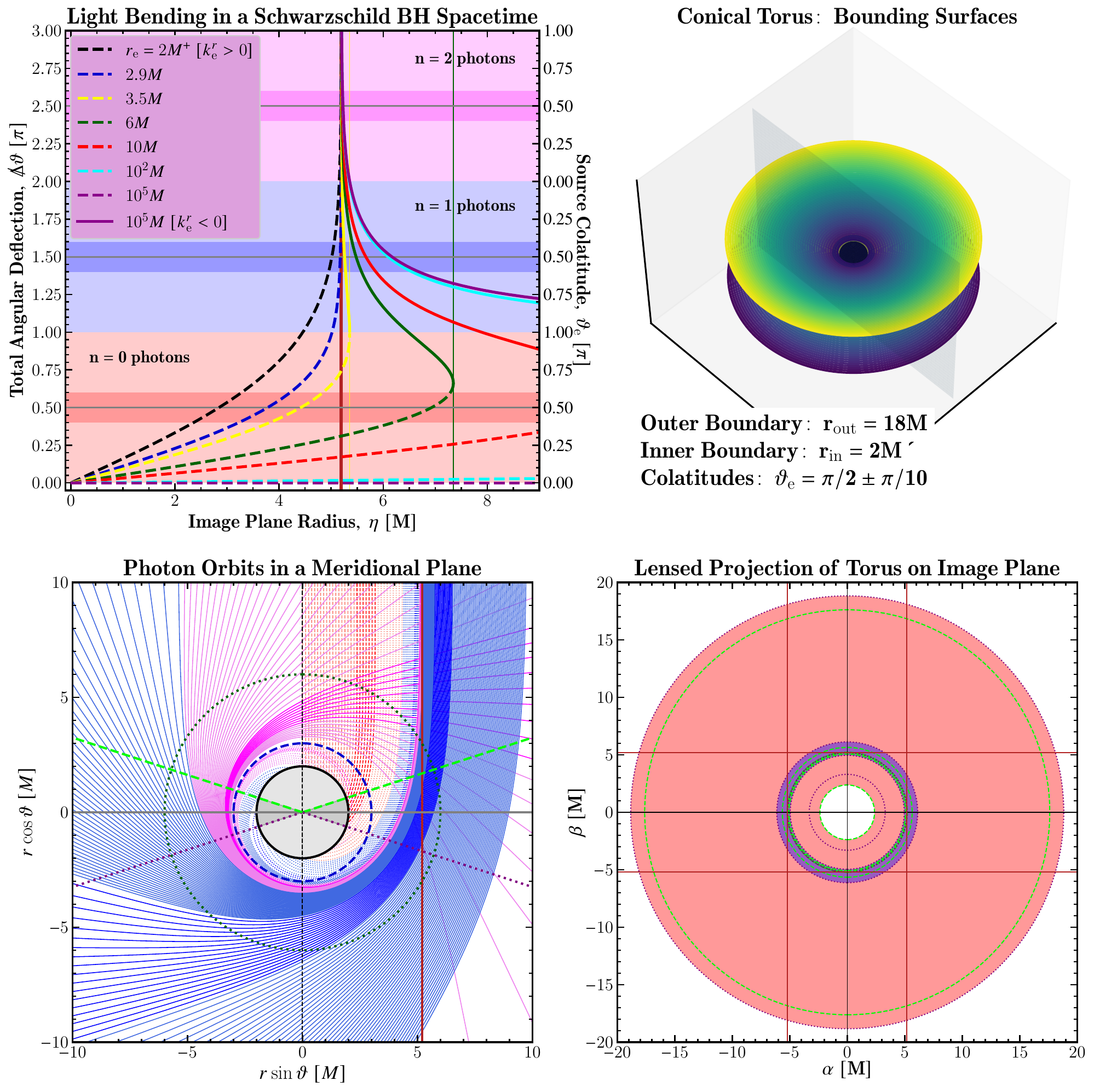}
\caption{\textit{Photon orbits and higher-order images of emitters in a Schwarzschild black hole (BH) spacetime}. The top-left panel presents the angular deflection for photons emitted from different radii and with different angular momenta ($\eta$). The latter corresponds to their apparent impact parameters or screen radii. Photons emitted towards and away from the BH are shown in dashed and solid lines respectively. Photons emitted from the conical torus (see top-right panel) occupy the darker-shaded regions. The dashed black line in the $n=0$ region tracks the size of the ``inner shadow'' \citep{Chael+2021}. In all panels, the vertical red lines show the size of the BH shadow. The top-right panel shows our simple conical torus model. The bottom-left panel displays a set of photon orbits reaching an observer on the $+z-$axis. The event horizon and the photon sphere are shown as black and blue circles respectively. The green and purple lines represent a cross-section of the torus. The bottom-right figure shows the observer's screen image of the torus, with red and blue shading indicating regions collecting $n=0$ and $n=1$ photons, corresponding to the direct image and the first photon subring.}
\label{fig:FigB2_Schw_Higher_Order_Images}
\end{center}
\end{figure*}


In this section, we highlight the nonmonotonicity of lensing maps $\eta_n(r_{\mathrm{e}}, \vartheta_{\mathrm{e}})$, which relate emitter locations $(r_{\mathrm{e}}, \vartheta_{\mathrm{e}})$ in the bulk to order$-n$ image radii $\eta_n$ on the boundary, w.r.t. the source colatitude, $\vartheta_{\mathrm{e}}$. As noted above, these maps are obtained by solving the integral equation \eqref{eq:Main_Integral_Equation}. We can safely omit both $\varphi_{\mathrm{e}}$ and $\psi_n$ from this discussion due to the planarity of (null) geodesic orbits in spherically-symmetric spacetimes (Appendix \ref{app:AppA_Planarity}), which leads to a trivial map $\varphi_{\mathrm{e}} \mapsto \psi_n$ \eqref{eq:Image_Plane_Polar_Angle}.

Fig. \ref{fig:FigB1_Schw_Higher_Order_Images} clearly illustrates how the emitter and observer spatial locations as well as the order of the image uniquely determine a specific photon orbit (see also, e.g., Fig. 16 of \citealt{Broderick+2022b}). Shown are the first four order ($n=0, 1, 2, 3$) images of emitters located at different radii ($r_{\mathrm{e}}$) and colatitudes ($\vartheta_{\mathrm{e}}$) in a Schwarzschild BH spacetime. The photon sphere, $r=r_{\mathrm{PS}}=3M$, is shown as a dashed blue circle and the circular shadow boundary, $\eta = \eta_{\mathrm{PS}}=\sqrt{27}M$, is indicated by the vertical red lines. The photon ring on the image plane is the region that collects all higher-order ($n>0$) photons \citep{Johnson+2020}.

The primary lensing map $\eta_0(r_{\mathrm{e}}, \vartheta_{\mathrm{e}})$ is nonmonotonic (nonbijective) w.r.t. variations in $\vartheta_{\mathrm{e}}$ but only for photons that are emitted from outside the photon sphere $r > r_{\mathrm{PS}}$. This is because such radial locations can source photon orbits that permit radial turning points: Notice how the primary ($n=0$) images of emitters at the innermost stable circular orbit (ISCO), $r_{\mathrm{e}}=6M$, appear either inside or outside the shadow boundary curve, depending on the polar location of the emitter. 

\begin{figure*}
\begin{center}
\includegraphics[width=\columnwidth]{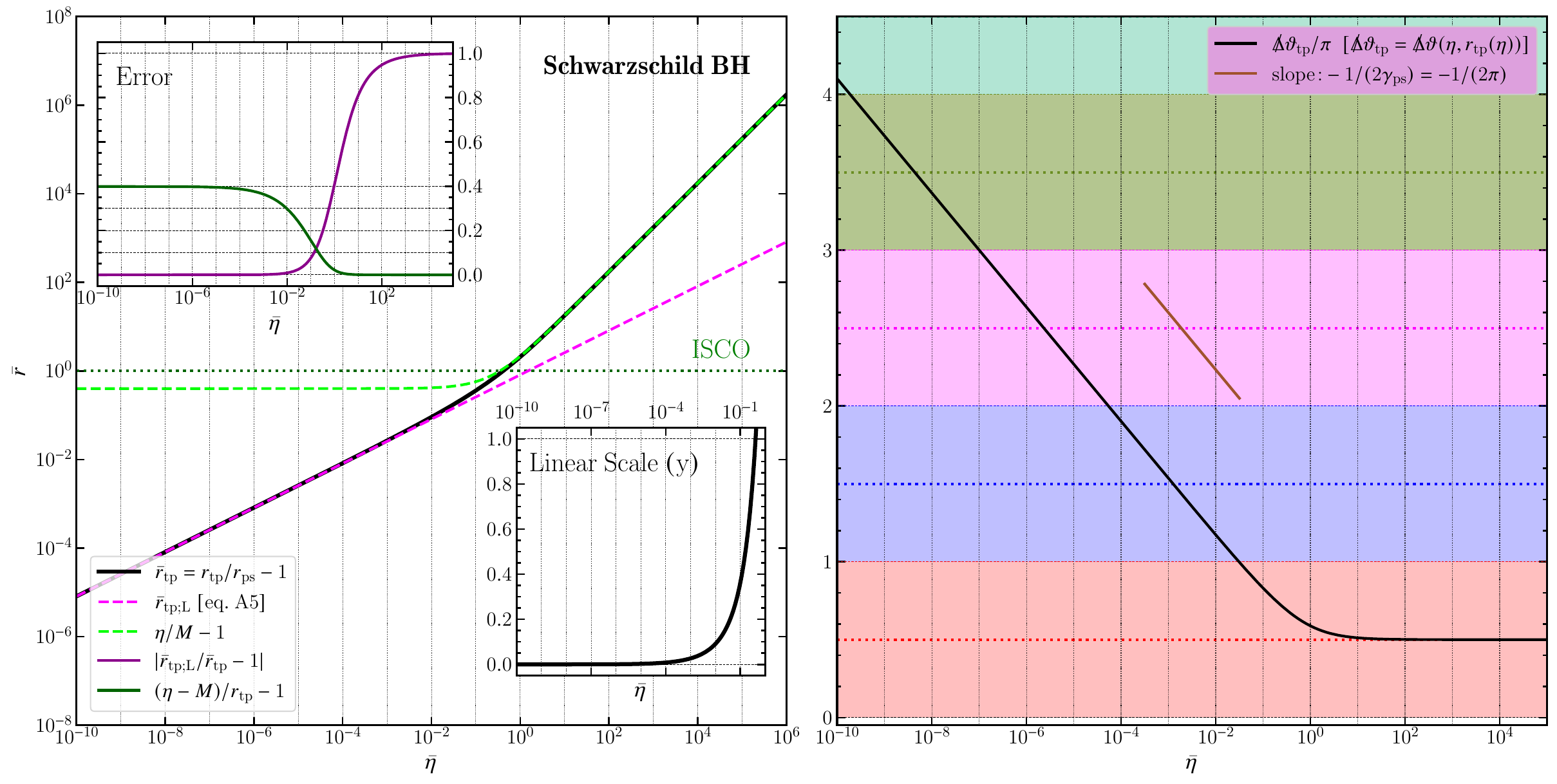}
\caption{\textit{Properties of photon radial turning points.} An arbitrary photon orbit that terminates outside the shadow boundary, $\bar{\eta} > 0$, permits a single radial turning point when the photon radial velocity vanishes, denoted by $\bar{r}=\bar{r}_{\mathrm{tp}}(\bar{\eta})$. Here, $\bar{\eta} := \eta/\eta_{\mathrm{PS}} - 1$ and $\bar{r} := r/r_{\mathrm{PS}} - 1$ measure distances from the shadow boundary curve on the image plane and the photon sphere in the bulk. In the left panel, the black curve shows how the turning point radius changes with the impact parameter in a Schwarzschild BH spacetime. The magenta curve shows how the linearized turning point radius ($\bar{r}_{\mathrm{tp; L}}$; eq. \ref{eq:Linearized_TPE_Sol}) is an excellent approximation to the exact one for small $\bar{\eta}$. We also show, in the bright green line, how the approximation for the weakly-lensed photons ($\bar{\eta} \ll 1$) is qualitatively different, and is given by $\bar{r}_{\mathrm{tp}} \approx \eta - M$. The insets quantify the fractional errors of both approximations. The panel on the right shows the total angular deflection experienced by a photon emitted from its turning point ($\mathrm{type\ C^T}$ orbit; eq. \ref{eq:Divergent_Slashed} and Table \ref{tab:Photon_Orbit_Classification}). Such photons, with $0 < \bar{\eta} \ll 1$ have turning points sufficiently close to the photon sphere, and can contribute to higher-order image formation.}
\label{fig:Fig13_Fig1_Schw_Strong_Lensing_Approximation_Appendix}
\end{center}
\end{figure*}

Furthermore, we emphasize that this is a generic feature of \textit{all} order lensing maps. Therefore, two photons that were emitted from the same radial location $r_{\mathrm{e}}$ in the bulk and which appear at the same radii $\eta$ on the image plane can have different trajectories, experiencing different amounts of angular deflections as well as travel times. However, the band of emission radii outside the photon sphere for which this map is nonmonotonic shrinks exponentially with increasing image order.%
\footnote{\label{fn:HIO_Nonmonotonicity}
For an image of order-$n$ this band is bounded from above ($r_{\mathrm{PS}} < r_{\mathrm{e}} \leq r_{n; \mathrm{C^T}}$) by the radius $r_{n; \mathrm{C^T}}$ at which a photon emitted with zero radial velocity ($k^r_{\mathrm{e}} = 0$) experiences an angular deflection of exactly $n\pi$, i.e., $\slashed{\Delta}\vartheta(\eta_{n; \mathrm{C^T}}, r_{n; \mathrm{C^T}}) = n\pi$, where $\eta_{n; \mathrm{C^T}}$ is a solution of the radial turning point equation, $k^r(\eta, r_{n; \mathrm{C^T}}(\eta)) = 0$ (see also the right panel of Fig. \ref{fig:Fig13_Fig1_Schw_Strong_Lensing_Approximation_Appendix}).} %
Thus, the \textit{nonmonotonicity} of the higher-order lensing maps \textit{may} not lead to significant confusion when inferring properties of the accretion flow from observed patterns in future higher-resolution dynamical movies of M87$^\star$ and Sgr A$^\star$ \citep{Tiede+2020, Levis+2022, Conroy+2023}.

Nevertheless, since the peak of the emissivity profile in the bulk likely lies close to the photon shell \citep{EHTC+2019f, EHTC+2022d, EHTC+2022f}, and we can only access a complicated superposition of the primary and secondary images simultaneously, it is useful to keep in mind these subtle features of gravitational lensing. The nonmonotonicity of the primary and secondary lensing maps also underscores the importance of considering the impact on image formation of nonequatorial emission and non-face-on observer inclinations (Sec. \ref{sec:Sec4_Hotspots}, \ref{sec:Sec5_Photon_Rings}). 

The top left panel of Fig. \ref{fig:FigB2_Schw_Higher_Order_Images} (see also \citealt{Gyulchev+2021}) offers an alternative perspective for understanding how a combination of the source location $(r_{\mathrm{e}}, \vartheta_{\mathrm{e}})$ and the order ($n$) of the image pick out a unique photon orbit, without needing any other initial conditions. It neatly shows the variation in the image radius ($x-$axis) of a source located at a particular radius (a particular line) and a particular colatitude (right $y-$axis). The angular deflection experienced by these photons, of different orders, can be read off from the left $y-$axis. Horizontal lines in this plot correspond to emission coming from conical surfaces, except for $\slashed{\Delta}\vartheta = (2n+1)\pi/2$ which correspond to emission coming from the equatorial plane. The relation between the source colatitude, $\vartheta_{\mathrm{e}}$, and the angular deflection experienced by the photon, $\slashed{\Delta}\vartheta$, is straightforwardly given by eq. \ref{eq:Angle_Deflection_Higher_Order_Images_n}. 

Photons that were emitted in the radially-outward ($k^r_{\mathrm{e}} > 0$) and -inward ($k^r_{\mathrm{e}} < 0$) directions are represented here in dotted and dashed lines respectively. The meeting point of these two line types naturally corresponds to the case when the photon is emitted with zero radial velocity $k^r_{\mathrm{e}} = 0$. That is, the emission radius in the bulk matches the radial turning point of that photon orbit, $r_{\mathrm{e}} = r_{\mathrm{tp}}(\eta)$. This photon appears at a radius $\eta = \eta_{\mathrm{tp}}(r_{\mathrm{e}}) := R(r_{\mathrm{e}})/\sqrt{f(r_{\mathrm{e}})}$ on the image plane. 

While, as discussed above, the lensing map is nonmonotonic when fixing the source radius and varying the source colatitude, it remains monotonic when fixing the source colatitude and varying the source radius. This greatly simplifies finding the edges of an arbitrary order photon subring for geometrically-thick emission sources, as described in eq. \ref{eq:thick_disk_eta_n_in_out} of Sec. \ref{sec:Sec5dot3_Thick_Disks_Face_On}.

To understand the impact of varying emitting region morphologies on inferences of photon ring properties, we model the emitting region as a conical torus in Sec. \ref{sec:Sec5_Photon_Rings}, as shown in the top-right panel of Fig. \ref{fig:FigB2_Schw_Higher_Order_Images}.

The bottom-left panel of Fig. \ref{fig:FigB2_Schw_Higher_Order_Images} shows the spatial orbits of null geodesics in a meridional plane in a Schwarzschild BH spacetime. If we take this to be the $yz-$plane ($\varphi = \pi/2, 3\pi/2$), then these photons appear on the image plane Cartesian ``$\alpha-$''axis (\citealt{Bardeen1973}; See also eq. \ref{eq:Image_Plane_Polar_Angle}). Photons that appear in the photon ring ($\eta \approx \eta_{\mathrm{PS}}$) can be strongly lensed by the BH. These necessarily access the close vicinity of the photon shell $(r_{\mathrm{e}} \approx r_{\mathrm{PS}})$ somewhere along their orbit. 

In the bottom-right panel of Fig. \ref{fig:FigB2_Schw_Higher_Order_Images}, we show the image of the solid conical torus as seen by an observer on the north pole, viewing the emitting region face-on. We have also shown the region on the image plane occupied by the $n=0$ or direct image as well as the $n=1$ or first-order image, which collect photons that undergo deflections between $0$ and $\pi$ and between $\pi$ and $2\pi$ respectively. It is clear to see how a higher-order image is a demagnified (or thinner) version of a lower-order image. 


\section{Analytic Approximations of Elliptic Integrals} 
\label{app:AppC_Analytic_Approximations}

The general path-\textit{dependent} integral introduced above \eqref{eq:general_Path_Dependent_Integral} captures various quantities of interest along any null geodesic that ends on the screen of the observer, such as the angular deflection, the affine length, and the elapsed time along it, as well as the image plane intensity (see \citealt{Kocherlakota+2022} as well as Appendix \ref{app:AppD_Flux} below). To completely circumvent solving the null geodesic equation to determine the path for each photon, we rewrote this using a general path-\textit{independent} integral \eqref{eq:general_Path_Independent_Integral}, 
\begin{align} \label{eq:general_Path_Independent_Integral_App}
& \Delta Q(\bar{\eta}, \bar{r}_1, \bar{r}_2 \geq \bar{r}_1)  := 
\begin{cases}
r_{\mathrm{PS}}\int_{\bar{r}_1}^{\bar{r}_2} \dot{Q}(\bar{\eta}, \bar{r})/\sqrt{\mathscr{R}(\bar{\eta}, \bar{r})}~\mathrm{d}\bar{r}\,,
&\quad \bar{r}_1 \geq \bar{r}_{\mathrm{H}}^+\ \mathrm{if}\ \bar{\eta} < 0 \\
r_{\mathrm{PS}}\int_{\bar{r}_1}^{\bar{r}_2} \dot{Q}(\bar{\eta}, \bar{r})/\sqrt{\mathscr{R}(\bar{\eta}, \bar{r})}~\mathrm{d}\bar{r}\,,
&\quad \bar{r}_1 \geq \bar{r}_{\mathrm{tp}}(\bar{\eta})\ \mathrm{if}\ \bar{\eta} \geq 0\,. 
\end{cases}
\end{align}
In the above, we have introduced the fractional (or conformal) bulk and boundary radii respectively as
\begin{equation}
\bar{r} = r/r_{\mathrm{PS}} - 1\,;\ \bar{\eta} = \eta/\eta_{\mathrm{PS}} - 1\,,
\end{equation}
which respectively measure the distance from the BH photon sphere and from the shadow boundary curve on the image plane. 

Our aim in this section is to obtain a simple approximation to the general path-independent integral above (i.e., without making a choice that leads to it describing any particular observable) for the class of photon orbits that are strongly-lensed by an arbitrary spherically-symmetric black hole (BH). This facilitates obtaining, in one stroke, the specific properties that are exhibited by each specific observable for this class of orbits as well as the properties that are universal across all these observables. We do this by finding the dominant contribution to such an integral and using this dominant piece to approximate the exact integral. 

The dominant piece to the general path-independent integrals above \eqref{eq:general_Path_Independent_Integral_App} is contributed by locations where $|\mathscr{R}(\bar{\eta}, \bar{r})| \ll 1$. Since for a circular null geodesic, $(\bar{\eta}, \bar{r}) = (0, 0)$, we begin with the series expansion of the effective potential, $\mathscr{R}(\bar{\eta}, \bar{r})$, around this critical value
\begin{align} \label{eq:Lyapunov_Exponent_Derivation_Series_Expansion_R}
\mathscr{R}(\bar{\eta}, \bar{r}) =&\ \mathscr{R}(0, 0) 
+ \partial_{\bar{\eta}}\mathscr{R}(0,0)\bar{\eta} + \partial_{\bar{r}}\mathscr{R}(0, 0) \bar{r} + \partial^2_{\bar{\eta}}\mathscr{R}(0,0)\frac{\bar{\eta}^2}{2} + \partial_{\bar{\eta}}\partial_{\bar{r}}\mathscr{R}(0,0)\bar{\eta}\bar{r} + \partial^2_{\bar{r}}\mathscr{R}(0,0)\frac{\bar{r}^2}{2} + \mathscr{O}(3)\nonumber \\
=&\ -\frac{2}{{g_{\mathrm{PS}}}}\bar{\eta} - \frac{1}{{g_{\mathrm{PS}}}}\bar{\eta}^2 + 2r_{\mathrm{PS}}\frac{\partial_{\bar{r}}g_{\mathrm{PS}}}{g_{\mathrm{PS}}^2}\bar{\eta}\bar{r} + \hat{\kappa}_{\mathrm{PS}}^2r_{\mathrm{PS}}^2\bar{r}^2 + \mathscr{O}(3) 
\approx\ -\frac{2}{g_{\mathrm{PS}}}\bar{\eta} + \hat{\kappa}_{\mathrm{PS}}^2r_{\mathrm{PS}}^2\bar{r}^2\,.
\end{align}
In the above, the subscript $``\mathrm{PS}"$ indicates that the function is evaluated at $\bar{r}=0$, and $\hat{\kappa}_{\mathrm{PS}}$ was introduced in eq. \ref{eq:Phase_Space_Lyapunov_Exponent} above. In writing the final approximation, we have retained only the leading-order contributions in the small variables $\bar{\eta}$ and $\bar{r}$. Thus, we expect that the photons that access the close vicinity of the photon sphere $|\bar{r}| \ll 1$ and which appear close to the shadow boundary $|\bar{\eta}| \ll 1$ will experience large angular deflections ($\dot{Q} = \dot{\vartheta}/E$), affine lengths ($\dot{Q} = 1/E$), and elapsed times ($\dot{Q} = \dot{t}/E$).

For a particular photon orbit (fixed-$\bar{\eta}$), the dominant contribution \textit{along it} comes particularly from the radial turning point ($\mathscr{R} = 0$) when one is permitted ($\bar{\eta} > 0$), and from the photon sphere ($\bar{r} = 0$) otherwise ($\bar{\eta} < 0$; See eq. \ref{eq:Lyapunov_Exponent_Derivation_Series_Expansion_R}). For the former set of orbits ($0 < \bar{\eta} \ll 1$), we can obtain a closed-form expression for the (\textit{linearized}) turning point radius as being given by,%
\footnote{Note that this expression is qualitatively different from the case of weakly-lensed photon orbits ($\bar{\eta} \gg 1$) where $\bar{r}_{\mathrm{tp}}(\bar{\eta}) \propto \bar{\eta}$.}
\begin{equation} \label{eq:Linearized_TPE_Sol}
\bar{r}_{\mathrm{tp; L}}(\bar{\eta}) \approx \left[\sqrt{2}/(\sqrt{g_{\mathrm{PS}}}\hat{\kappa}_{\mathrm{PS}}r_{\mathrm{PS}})\right]\sqrt{\bar{\eta}}\,. 
\end{equation}

To compute these dominant contributions, for clarity, we bring $\bar{r}=\bar{r}_{\mathrm{tp; L}}(\bar{\eta})$ for $0 < \bar{\eta} \ll 1$ and $\bar{r} = 0$ for $-1 \ll \bar{\eta} < 0$ to $x = 0$ as follows (cf. also \citealt{Bozza+2007}). We introduce the conformal variable $x$ first as
\begin{align}  \label{eq:Appendix_x_Def}
r = 
\begin{cases}
r_{\mathrm{PS}}/(1-x)\,,\ & \bar{\eta} < 0 \\
r_{\mathrm{tp}}(\eta)/(1-x)\,,\ & \bar{\eta} \geq 0 
\end{cases}\,,
\end{align}
which leads to
\begin{align}
\bar{r} = 
\begin{cases}
x/(1-x)\,,\ & \bar{\eta} < 0 \\
(\bar{r}_{\mathrm{tp}}(\bar{\eta}) + x)/(1-x)\,,\ & \bar{\eta} \geq 0 
\end{cases}\
\approx (\mathrm{for}\ |\bar{\eta}| \ll 1)
\begin{cases}
x/(1-x)\,,\ & \bar{\eta} < 0 \\
(\bar{r}_{\mathrm{tp; L}}(\bar{\eta}) + x)/(1-x)\,,\ & \bar{\eta} \geq 0 
\end{cases}\
\approx (\mathrm{for}\ x \approx 0)
\begin{cases}
x\,,\ & \bar{\eta} < 0 \\
\bar{r}_{\mathrm{tp; L}}(\bar{\eta}) + x\,,\ & \bar{\eta} \geq 0 
\end{cases}\,. \nonumber
\end{align}
The effective radial potential \eqref{eq:Lyapunov_Exponent_Derivation_Series_Expansion_R} then simplifies, for $x \approx 0$, to
\begin{align} \label{eq:Elliptic_Form}
\mathscr{R}(\bar{\eta}, x) \approx
\begin{cases} 
\hat{\kappa}_{\mathrm{PS}}^2r_{\mathrm{PS}}^2\left[\left(-2/(g_{\mathrm{PS}}\kappa_{\mathrm{PS}}^2r_{\mathrm{PS}}^2)\right)\bar{\eta} + x^2\right]\,, \ 
& \bar{\eta} < 0 \\
\hat{\kappa}_{\mathrm{PS}}^2r_{\mathrm{PS}}^2\left[2\bar{r}_{\mathrm{tp; L}}(\bar{\eta})x + x^2\right]\,,\
& \bar{\eta} \geq 0 
\end{cases}
=
\begin{cases} 
\hat{\kappa}_{\mathrm{PS}}^2r_{\mathrm{PS}}^2\left[\bar{r}^2_{\mathrm{tp; L}}(|\bar{\eta}|) + x^2\right]\,, \ 
& \bar{\eta} < 0 \\
\hat{\kappa}_{\mathrm{PS}}^2r_{\mathrm{PS}}^2\left[2\bar{r}_{\mathrm{tp; L}}(\bar{\eta})x + x^2\right]\,,\
& \bar{\eta} \geq 0 
\end{cases}\,.
\end{align}
No radial turning points exist for orbits with $\bar{\eta} < 0$. In the above, we are using $\bar{r}_{\mathrm{tp; L}}(|\bar{\eta}|)$ as a formal function, as defined in eq. \ref{eq:Linearized_TPE_Sol}, primarily to show the difference in the approximations for photon orbits with $\bar{\eta} > 0$ and with $\bar{\eta} < 0$. 


The equation above \eqref{eq:Elliptic_Form} provides the insight that the dominant contributions of the general path-independent integrals of interest \eqref{eq:general_Path_Independent_Integral_App} are elliptic integrals \citep{Press+1990}. More specifically, we can split $\Delta Q(\eta \rightarrow \eta_{\mathrm{PS}}^-, r_1 \rightarrow r_{\mathrm{PS}}, r_2)$ and $\Delta Q(\eta \rightarrow \eta_{\mathrm{PS}}^+, r_1 \rightarrow r_{\mathrm{tp}}(\eta), r_2)$, into pieces containing the dominant term $\Delta Q_{\mathrm{D}}$ and a regular or residual term $\Delta Q_{\mathrm{R}}$, as $\Delta Q := \Delta Q_{\mathrm{D}} + \Delta Q_{\mathrm{R}}$. We define the dominant piece as
\begin{align} \label{eq:Divergent_x2}
\Delta Q_{\mathrm{D}}(\bar{\eta}, 0, x_2) :=
\begin{cases}
\frac{\dot{Q}(0,0)}{\hat{\kappa}_{\mathrm{PS}}}\int_0^{x_2} \frac{\mathrm{d}x}{\sqrt{\bar{r}^2_{\mathrm{tp; L}}(|\bar{\eta}|) + x^2}} = 
\frac{\dot{Q}(0,0)}{\hat{\kappa}_{\mathrm{PS}}}\ln{\left[\frac{x_2 + \sqrt{\bar{r}_{\mathrm{tp; L}}^2(|\bar{\eta}|) + x_2^2}}{\bar{r}_{\mathrm{tp; L}}(|\bar{\eta}|)}\right]}\,, 
& \bar{\eta} < 0 \\
\frac{\dot{Q}(0,0)}{\hat{\kappa}_{\mathrm{PS}}}\int_0^{x_2}\frac{\mathrm{d}x}{\sqrt{2\bar{r}_{\mathrm{tp; L}}(\bar{\eta})x + x^2}} = 
\frac{\dot{Q}(0,0)}{\hat{\kappa}_{\mathrm{PS}}}\ln{\left[\frac{\bar{r}_{\mathrm{tp; L}}(\bar{\eta}) + x_2 + \sqrt{2\bar{r}_{\mathrm{tp; L}}(\bar{\eta})x_2 + x_2^2}}{\bar{r}_{\mathrm{tp; L}}(\bar{\eta})}\right]}\,, & \bar{\eta} \geq 0 
\end{cases}\,.
\end{align}
and the regular piece then defined as,
\begin{align} \label{eq:Regular_x2}
\Delta Q_{\mathrm{R}}(\bar{\eta}, 0, x_2) :=&\ \Delta Q(\bar{\eta}, 0, x_2) - \Delta Q_{\mathrm{D}}(\bar{\eta}, 0, x_2) 
=
\begin{cases}
\int_0^{ x_2} \frac{\dot{Q}(\bar{\eta}, x)r_{\mathrm{PS}}}{\sqrt{\mathscr{R}(\bar{\eta}, x)}}\frac{1}{(1-x)^2}~\text{d}x -\Delta Q_{\mathrm{D}}(\bar{\eta}, 0, x_2)\,, & \bar{\eta} < 0 \\
\int_0^{ x_2} \frac{\dot{Q}(\bar{\eta}, x)r_{\mathrm{PS}}}{\sqrt{\mathscr{R}(\bar{\eta}, x)}}\frac{(1+\bar{r}_{\mathrm{tp}}(\bar{\eta}))}{(1-x)^2}~\text{d}x -\Delta Q_{\mathrm{D}}(\bar{\eta}, 0, x_2)\,, & \bar{\eta} \geq 0 
\end{cases}\,. 
\end{align}
Since $x$ can take negative values on photon orbits with $\bar{\eta} < 0$, i.e., $x_{\mathrm{H}}:= 1 - r_{\mathrm{PS}}/r_{\mathrm{H}} < x < 0$, we also require
\begin{align} \label{eq:Divergent_x1}
\Delta Q_{\mathrm{D}}(\bar{\eta}, x_1, 0) =
\frac{\dot{Q}(0,0)}{\hat{\kappa}_{\mathrm{PS}}}\int_{x_1}^0 \frac{\mathrm{d}x}{\sqrt{\bar{r}^2_{\mathrm{tp; L}}(|\bar{\eta}|) + x^2}} = 
\frac{\dot{Q}(0,0)}{\hat{\kappa}_{\mathrm{PS}}}\ln{\left[\frac{-x_1 + \sqrt{\bar{r}_{\mathrm{tp; L}}^2(|\bar{\eta}|) + x_1^2}}{\bar{r}_{\mathrm{tp; L}}(|\bar{\eta}|)}\right]}\,, 
\end{align}
to consistently define $\Delta Q_{\mathrm{R}}(\bar{\eta}, x_1, 0) = \Delta Q(\bar{\eta}, x_1, 0) - \Delta Q_{\mathrm{D}}(\bar{\eta}, x_1, 0)$. 

We emphasize that we have made no approximations thus far. 


\begin{table}
\centering
\caption{\textit{Classification of photon orbits that appear on the image plane.} 
The angular deflection experienced by an arbitrary photon depends only on its impact parameter $\eta$ (or equivalently, $\bar{\eta} = \eta/\eta_{\mathrm{PS}} - 1$), its radius of emission $r_{\mathrm{e}}$ (or equivalently $x_{\mathrm{e}}$; eq. \ref{eq:Appendix_x_Def}), and the sign of its radial velocity at emission $k^r_{\mathrm{e}}$ (cf. eq. \ref{eq:del_vartheta}). Therefore, the space of initial conditions for null geodesics is three-dimensional. For photons that undergo strong gravitational lensing ($\slashed{\Delta}\vartheta > \pi; \bar{\eta} \rightarrow 0$), we have partitioned this space into the different types of possible orbits in eq. \ref{eq:Divergent_Slashed}. In this summary table, the pattern in our nomenclature is more easily evident. For example, notice how all the $\mathrm{type\ C}$ emanate either from close to the photon sphere or are emitted from just outside the turning point.}
\label{tab:Photon_Orbit_Classification}
\begin{tabular}{||c||c|c|c||c|c|c||c|c|c||c|c|c||c|c|c||}
\hline\hline
$x_{\mathrm{e}}$ 
& \multicolumn{3}{c||}{$\lnsim 0$} 
& \multicolumn{3}{c||}{$\lesssim 0$} 
& \multicolumn{3}{c||}{$0$} 
& \multicolumn{3}{c||}{$\gtrsim 0$} 
& \multicolumn{3}{c||}{$\gnsim 0$} \\
\hline\hline
\diagbox{$k^r_{\mathrm{e}}$}{$\bar{\eta}$} 
& $-$ & $0$ & $+$ 
& $-$ & $0$ & $+$ 
& $-$ & $0$ & $+$ 
& $-$ & $0$ & $+$ 
& $-$ & $0$ & $+$ \\
\hline\hline
$-$ 
& $\times$ & $\times$ & $\times$ 
& $\times$ & $\times$ & $\times$ 
& $\times$ & $\times$ & $\times$ 
& $\times$ & $\times$ & $\mathrm{C^+}$ 
& $\times$ & $\times$ & $\mathrm{E}$ \\
\hline
$0$ 
& $\times$ & $\times$ & $\times$ 
& $\times$ & $\times$ & $\times$ 
& $\times$ & $\times$ & $\mathrm{C^T}$ 
& $\times$ & $\times$ & $\times$ 
& $\times$ & $\times$ & $\times$ \\
\hline
$+$ 
& $\mathrm{A}$ & $\times$ & $\times$ 
& $\mathrm{C^-}$ & $\times$ & $\times$
& $\mathrm{C^-}$ & $\times$ & $\times$
& $\mathrm{C^-}$ & $\mathrm{C^0}$ & $\mathrm{C^+}$
& $\mathrm{B}$ & $\mathrm{D}$ & $\mathrm{D}$ \\
\hline
\end{tabular}    
\end{table}


We can now obtain the leading-order piece, in $\bar{\eta}$, of the dominant part of the path-independent integrals as
\begin{align} \label{eq:Divergent}
\Delta Q_{\mathrm{D}}(\bar{\eta}, x_1, 0) = &\ 
\left(\dot{Q}(0,0)/\hat{\kappa}_{\mathrm{PS}}\right)
\left[-\ln{{\sqrt{-\bar\eta}}} + \ln{(-x_1)} + \ln{[\sqrt{2g_{\mathrm{PS}}}r_{\mathrm{PS}}\hat{\kappa}_{\mathrm{PS}}]} + \mathscr{O}(\bar{\eta})\right]\,,
\ \ \ \bar{\eta} < 0\,,\ x_1 < 0 \\
\Delta Q_{\mathrm{D}}(\bar{\eta}, 0, x_2) \approx&\
\begin{cases}
\left(\dot{Q}(0,0)/\hat{\kappa}_{\mathrm{PS}}\right)
\left[-\ln{{\sqrt{-\bar\eta}}} + \ln{x_2} +  \ln{[\sqrt{2g_{\mathrm{PS}}}r_{\mathrm{PS}}\hat{\kappa}_{\mathrm{PS}}]} + \mathscr{O}(\bar{\eta})\right]\,, 
& \bar{\eta} < 0\,,\ x_2 > 0 \\
\left(\dot{Q}(0,0)/\hat{\kappa}_{\mathrm{PS}}\right)
\left[-\ln{{\sqrt{+\bar\eta}}} + \ln{x_2} + \ln{[\sqrt{2g_{\mathrm{PS}}}r_{\mathrm{PS}}\hat{\kappa}_{\mathrm{PS}}]} + \mathscr{O}(\sqrt{\bar{\eta}})\right]\,, 
& \bar{\eta} \geq 0\,,\ x_2 > 0
\end{cases} \nonumber \,.
\end{align}
With this, we can straightforwardly write the dominant parts of the path-\textit{dependent} integrals, $\slashed{\Delta}Q(\bar{\eta}\rightarrow 0, x_{\mathrm{e}}, x_{\mathrm{o}} > 0)$, defined in eq. \ref{eq:general_Path_Dependent_Integral} as,
\begin{equation} \label{eq:Divergent_Slashed}
\slashed{\Delta} Q_{\mathrm{D}}(\bar{\eta}, x_{\mathrm{e}}, x_{\mathrm{o}}) =
\left\{
\begin{alignedat}{3} 
& +\Delta Q_{\mathrm{D}}(\bar{\eta}, x_{\mathrm{e}}, 0) + \Delta Q_{\mathrm{D}}(\bar{\eta}, 0, x_{\mathrm{o}})\,,\quad 
&& \bar{\eta} < 0\,,\ k^r_{\mathrm{e}} > 0\,,\ x_{\mathrm{H}} < x_{\mathrm{e}} \lnsim 0\quad
&& \mathrm{[type\ A]} \\
& -\Delta Q_{\mathrm{D}}(\bar{\eta}, 0, x_{\mathrm{e}}) + \Delta Q_{\mathrm{D}}(\bar{\eta}, 0, x_{\mathrm{o}})\,,\quad
&& \bar{\eta} < 0\,,\ k^r_{\mathrm{e}} > 0\,,\ x_{\mathrm{e}} \gnsim 0\quad
&& \mathrm{[type\ B]} \\ 
& + \Delta Q_{\mathrm{D}}(\bar{\eta}, 0, x_{\mathrm{o}})\,,\ 
&& \bar{\eta} < 0\,,\ k^r_{\mathrm{e}} > 0\,,\ |x_{\mathrm{e}}| \simeq 0\quad 
&& \mathrm{[type\ C^-]} \\
& + \Delta Q_{\mathrm{D}}(\bar{\eta}, 0, x_{\mathrm{o}})\,,\ 
&& \bar{\eta} = 0\,,\ k^r_{\mathrm{e}} > 0\,,\ x_{\mathrm{e}} \gtrsim 0\quad 
&& \mathrm{[type\ C^0]} \\
& + \Delta Q_{\mathrm{D}}(\bar{\eta}, 0, x_{\mathrm{o}})\,,\ 
&& \bar{\eta} > 0\,,\ k^r_{\mathrm{e}} \neq 0\,,\ x_{\mathrm{e}} \gtrsim 0\quad 
&& \mathrm{[type\ C^+]} \\
& + \Delta Q_{\mathrm{D}}(\bar{\eta}, 0, x_{\mathrm{o}})\,,\ 
&& \bar{\eta} > 0\,,\ k^r_{\mathrm{e}} = 0\,,\ x_{\mathrm{e}} = 0\quad 
&& \mathrm{[type\ C^T]} \\
& -\Delta Q_{\mathrm{D}}(\bar{\eta}, 0, x_{\mathrm{e}}) + \Delta Q_{\mathrm{D}}(\bar{\eta}, 0, x_{\mathrm{o}})\,,\quad
&& \bar{\eta} \geq 0\,,\ k^r_{\mathrm{e}} > 0\,,\ x_{\mathrm{e}} \gnsim 0\quad
&& \mathrm{[type\ D]} \\
& +\Delta Q_{\mathrm{D}}(\bar{\eta}, 0, x_{\mathrm{e}}) + \Delta Q_{\mathrm{D}}(\bar{\eta}, 0, x_{\mathrm{o}})\,,\quad
&& \bar{\eta} > 0\,,\ k^r_{\mathrm{e}} < 0\,,\ x_{\mathrm{e}} \gnsim 0\quad
&& \mathrm{[type\ E]}
\end{alignedat}
\right.\,.
\end{equation}
or more explicitly (compare against \citealt{Bozza+2007}),
\begin{align} \label{eq:Divergent_Slashed_v2}
\slashed{\Delta} Q_{\mathrm{D}}(\bar{\eta}, x_{\mathrm{e}}, x_{\mathrm{o}}) =&\
\begin{cases}
-\frac{\dot{Q}(0,0)}{\hat{\kappa}_{\mathrm{PS}}}
\left[\ln{|\bar{\eta}|} - \ln{(|x_{\mathrm{e}}|x_{\mathrm{o}})} - \ln{\left(2g_{\mathrm{PS}}r_{\mathrm{PS}}^2\kappa_{\mathrm{PS}}^2\right)}\right]\,, 
& \mathrm{[types\ A,\ E]} \\
-\frac{\dot{Q}(0,0)}{\hat{\kappa}_{\mathrm{PS}}}
\left[\ln{|\bar{\eta}|} - \ln{(x_{\mathrm{o}}^2)} -\ln{\left(2g_{\mathrm{PS}}r_{\mathrm{PS}}^2\kappa_{\mathrm{PS}}^2\right)}\right]\,, & \mathrm{[type\ C]} \\
\mathscr{O}(\sqrt{\bar{\eta}})\,,
& \mathrm{[type\ B]} \\
\mathscr{O}(\bar{\eta})\,,
& \mathrm{[type\ D]} 
\end{cases}\,.
\end{align}
In the above, we have used the subscripts $``\mathrm{e}"$ and $``\mathrm{o}"$ to suggest that these variables can be associated with the radial locations of the emitter and the observer respectively.

Figure \ref{fig:Fig1_Schw_Strong_Lensing_Approximation} shows the accuracy of the approximation obtained above \eqref{eq:Divergent_Slashed_v2} for the case of the angular deflection, for which $\dot{Q} = \dot{\vartheta}/E = \eta/R^2$, in recovering the exact value as per eq. \ref{eq:del_vartheta}. The constant offset is simply explained by the constant terms obtained above. We see the dependence on the source and observer radial locations to be quite weak for high-order photons.


\subsection{Role of Different Classes of Photon Orbits in Image Formation}

To concretely understand the different classes of photon orbits introduced above, let us consider their role in image formation. From Fig. \ref{fig:FigB2_Schw_Higher_Order_Images}, we see that the $n=1$ photons that appear sufficiently well outside the shadow boundary ($\bar{\eta}_{1; \mathrm{e}} \gnsim 0$) were necessarily emitted from well outside the photon sphere ($\bar{r}_{\mathrm{e}} \gnsim 0$) towards the BH ($k^r_{\mathrm{e}} < 0$), i.e., the $\mathrm{type\ E}$ orbits. On the other hand, the $\mathrm{type\ A}$ photons appear inside the shadow boundary and were necessarily emitted from well inside the photon sphere in the radially outward direction ($k^r_{\mathrm{e}} > 0$). 

Furthermore, while all of the photons that appear inside the shadow boundary were all emitted with initial positive radial velocity ($k^r_{\mathrm{e}} > 0$; dashed lines in Fig. \ref{fig:FigB2_Schw_Higher_Order_Images}), the photons that appear outside the shadow boundary need not have been emitted in the radially-inward direction (see the dotted lines in Fig. \ref{fig:FigB2_Schw_Higher_Order_Images}). This brings us unavoidably to the class of $\mathrm{type\ C}$ orbits.

These are understood intuitively as follows. The total angular deflection experienced by photons emitted from their radial turning points (i.e., with exactly zero radial velocity at emission), $\slashed{\Delta}\vartheta_{\mathrm{tp}}(\bar{\eta}) = \slashed{\Delta}\vartheta(\bar{\eta}, r_{\mathrm{tp}}(\bar{\eta}))$, increases as photons appear increasingly closer to the shadow boundary $\bar{\eta} \rightarrow 0^+$ (see the right panel of Fig. \ref{fig:Fig13_Fig1_Schw_Strong_Lensing_Approximation_Appendix}). Thus, for this configuration of source and detector, we should be able to find a photon orbit with some angular momentum $\bar{\eta} = \bar{\eta}_{1;\mathrm{C^T}} > 0$ such that $\slashed{\Delta}\vartheta_{\mathrm{tp}}(\bar{\eta}_{1;\mathrm{C^T}}) = 3\pi/2$. This is the $\mathrm{type\ C^T}$ orbit and its radius $\bar{\eta} = \bar{\eta}_{1;\mathrm{C^T}}$ on the image plane, located at the intersection of the blue and dashed-purple (the ``turning point line'') curves in the right panel of Fig. \ref{fig:Fig6_Schw_Thin_Disk_Face_On}, demarcates the region that collects $n=1$ photons moving on $\mathrm{type\ E}$ orbits ($\bar{\eta} > \bar{\eta}_{1;\mathrm{C^T}}$) from that which collects all the other types of $n=1$ photon orbits. If a photon appears in the region $0 < \bar{\eta} < \bar{\eta}_{1;\mathrm{C^T}}$ and was emitted from its turning point, it would have been lensed through an angle larger than $3\pi/2$, and cannot participate in the $n=1$ image formation. For any $\bar{\eta} > 0$, only photons emitted with initially positive radial velocities undergo smaller angular deflections than those emitted with zero radial velocities. Thus, the photons forming the $n=1$ image in the region $0 < \bar{\eta} < \bar{\eta}_{1;\mathrm{C^T}}$ must have been emitted from outside the photon sphere in the radially outward direction; These are the $\mathrm{type\ C^+}$ orbits. 

Furthermore, particularly clearly visible in the right panel of Fig. \ref{fig:Fig6_Schw_Thin_Disk_Face_On} are the photons that appear on ($\bar{\eta} = 0$) and just inside ($\bar{\eta} \lesssim 0$) the shadow boundary, all of which originated from just outside the photon sphere (also all emitted with positive radial velocities): These are the $\mathrm{type\ C^0}$ and the $\mathrm{type\ C^-}$ photons respectively. Photons that were emitted from just inside the photon sphere and which appear inside the shadow boundary are also of $\mathrm{type\ C^-}$. The blue-shaded region of this panel houses the $\mathrm{type\ E}$ and all $\mathrm{type\ C}$ orbits whereas the red-shaded region is composed only of the $\mathrm{type\ C^-}$ and the $\mathrm{type\ A}$ photon orbits. 

In the previous paragraph, we have discussed the distinctive features of the different $\mathrm{type\ C}$ photon orbits that form the $n=1$ image, namely where they are sourced from in the bulk and with what velocities, and also where they appear on the image plane. We now provide a simplified summary of the organization of photon orbits of all types on the image plane. Photons that form the inner photon ring ($\bar{\eta} < 0$) have orbits of $\mathrm{type\ A}$ or $\mathrm{C^-}$ whereas those that form the outer photon ring can be of $\mathrm{type\ C^+, C^T},$ or $\mathrm{E}$, in the sequence of increasing distance from the center of the image plane. The $\mathrm{type\ C^0}$ photon appears exactly on the shadow boundary ($\bar{\eta} = 0$) and demarcates the outer and inner sections of the photon subring. Finally, we note that this organization is generically true for arbitrary higher-order images and also holds qualitatively for arbitrary relative inclinations of the source and the observer and for arbitrarily geometrically-thick sources. 


\section{Universal Scaling Relations of Intensity and Flux Density in the Photon Ring}
\label{app:AppD_Flux}

In this section, we will derive the scaling of the flux density through photon subrings with the order of the subring for a general morphology of the emitting region, extending previous results that restrict the emission to the equatorial plane.

Neglecting scattering effects, the specific intensity ($\mathrm{W}~\mathrm{m}^{-2}\mathrm{sr}^{-1}\mathrm{Hz}^{-1}$) at a point $(\eta, \psi)$ on the image plane due to emission from an optically-transparent (negligible absorption) region is determined by integrating the (appropriately simplified) radiative transfer equation as \citep{Rybicki+1986, Younsi+2012} $I_\nu(\eta, \psi) = \fint_{0}^{\lambda_f} \Gamma^2 j_\nu~\mathrm{d}\lambda$, where $\Gamma = \nu_{\mathrm{o}}/\nu_{\mathrm{e}} = k_\alpha u^\alpha_{\mathrm{o}}/k_\alpha u^\alpha_{\mathrm{e}}$ is the redshift factor, which accounts for both gravitational and Doppler redshifts, and $j_\nu$ is the monochromatic emission coefficient ($\mathrm{W}~\mathrm{m}^{-3}\mathrm{sr}^{-1}\mathrm{Hz}^{-1}$). The slash indicates, as usual, that the integral is to be evaluated along the photon orbit $x^\mu(\lambda)$ that terminates on the image plane at $(\eta, \psi)$, and the quantities in the integrand depend on the orbit implicitly as $\Gamma = \Gamma(\lambda) = \Gamma(\eta, x^\alpha(\lambda))$ and $j_\nu = j_\nu(\lambda) = j_\nu(\eta, x^\alpha(\lambda))$. 

Furthermore, when constructing the intensity profile on the image plane of an asymptotic static observer ($u_{\mathrm{o}}^\mu = \delta^\mu_{\ t}$), the redshift factor reduces to $\Gamma = -E/k_\alpha u^\alpha_{\mathrm{e}}$. Thus, adopting these reasonable simplifications, we can write the specific intensity at a point $(\eta, 0 \leq \psi < 2\pi)$ on the image plane of an asymptotic static observer as (see also \citealt{Jaroszynski+1997, Bambi2013, Shaikh+2019, Bauer+2022, Kocherlakota+2022}),
\begin{equation} \label{eq:I_nu_slashed}
I_\nu(\eta, \psi) = \fint_{0}^{\lambda_f} \Gamma^2 j_\nu~\mathrm{d}\lambda\,;\quad \Gamma = -E/(k_\alpha u^\alpha_{\mathrm{e}})\,.
\end{equation}
Since without loss of generality, we can restrict our considerations to meridional photon orbits ($\dot{\varphi} = 0$) in spherically-symmetric spacetimes (see Appendix \ref{app:AppA_Planarity}), the dependence of the integrand on $\varphi(\lambda)$ is trivial (in both $\Gamma$ and $j_\nu$), i.e., it is simply determined by the image plane polar angle $\psi$. Furthermore, in the absence of (radial) turning points, there is a bijective map between the affine parameter $\lambda$ and the radial coordinate $r$ along the orbit, which allows expressing $t(\lambda)$ and $\vartheta(\lambda)$ as $t(\lambda) = t(r)$ and $\vartheta(\lambda) = \vartheta(r)$ instead. Thus, on sections of the photon orbit with no turning points, we can write $\Gamma = \Gamma(\eta, \psi, r(\lambda))$ and $j_\nu = j_\nu(\eta, \psi, r(\lambda))$. With all this, similar to eq. \ref{eq:del_vartheta}, we can ``unslash'' the integral in eq. \ref{eq:I_nu_slashed} and rewrite it simply as (cf.  \citealt{Kocherlakota+2022}),
\begin{align} \label{eq:I_nu}
I_\nu(\eta, \psi) = 
\begin{cases}
\int_{r_{\mathrm{H}}^+}^\infty \Gamma^2 j_\nu/\sqrt{\mathscr{R}}~\mathrm{d}r\,,
& \eta < \eta_{\mathrm{PS}}\\ 
\int_{r_{\mathrm{PS}}^+}^\infty \Gamma^2 j_\nu/\sqrt{\mathscr{R}}~\mathrm{d}r\,,
& \eta = \eta_{\mathrm{PS}}\\
- \int_\infty^{r_{\mathrm{tp}}(\eta)} \Gamma^2/\sqrt{\mathscr{R}}~\mathrm{d}r +  \int_{r_{\mathrm{tp}}(\eta)}^\infty \Gamma^2/\sqrt{\mathscr{R}}~\mathrm{d}r\,,
& \eta > \eta_{\mathrm{PS}}
\end{cases}\,,
\end{align}
where the lower limits $r_{\mathrm{H}}^+$ and $r_{\mathrm{PS}}^+$ indicate that the integral is evaluated from just outside the event horizon and from just outside the photon sphere respectively. Written this way, eq. \ref{eq:I_nu} cannot be put in the form of the path-dependent general integral given in \ref{eq:general_Path_Dependent_Integral}. Nevertheless, we can still rewrite it in terms of the path-independent general integral defined in eq. \ref{eq:general_Path_Independent_Integral} with $\dot{Q} = \Gamma^2j_\nu$ (compare against eq. 12 of \citealt{Kocherlakota+2022}). 

We are now eminently poised to use the universal relations obtained in eq. \ref{eq:Universal_Relation} to find the leading-order behaviour in $\bar{\eta}$ of the specific intensity profile in the photon ring. For an emitting region extending over $r_{\mathrm{in}} \leq r \leq r_{\mathrm{out}}$, i.e., the emission coefficient $j_\nu$ takes nonzero values only in this range, for an outer boundary outside the photon sphere $r_{\mathrm{out}} > r_{\mathrm{PS}}$, depending on the location of the inner boundary $r_{\mathrm{in}}$, we can write (via eq. \ref{eq:Divergent_Slashed_v2}),
\begin{equation} \label{eq:Intensity_Scaling_PR}
I_\nu(\bar{\eta}, \psi) \approx 
\left\{
\begin{alignedat}{3}
& -\frac{\Gamma^2(\eta_{\mathrm{PS}}, \psi, r_{\mathrm{PS}})j_\nu(\eta_{\mathrm{PS}}, \psi, r_{\mathrm{PS}})}{\hat{\kappa}_{\mathrm{PS}}}\left[\ln{|\bar{\eta}|}+K_1(\psi)\right]\,,\quad
&&  r_{\mathrm{in}} \leq r_{\mathrm{PS}}\,; \quad 
&& \bar{\eta} > 0 \\
& -\frac{\Gamma^2(\eta_{\mathrm{PS}}, \psi, r_{\mathrm{PS}})j_\nu(\eta_{\mathrm{PS}}, \psi, r_{\mathrm{PS}})}{\hat{\kappa}_{\mathrm{PS}}}\left[\ln{|\bar{\eta}|}+K_2(\psi)\right]\,,\quad
&&  r_{\mathrm{in}} < r_{\mathrm{PS}}\,; \quad 
&& \bar{\eta} < 0 \\
& -\frac{\Gamma^2(\eta_{\mathrm{PS}}, \psi, r_{\mathrm{PS}})j_\nu(\eta_{\mathrm{PS}}, \psi, r_{\mathrm{PS}})}{\hat{\kappa}_{\mathrm{PS}}}\left[\frac{\ln{|\bar{\eta}|}}{2}+K_3(\psi)\right]\,,\quad
&& r_{\mathrm{in}} = r_{\mathrm{PS}}\,; \quad 
&& \bar{\eta} < 0 \\
& \tilde{K}(\psi)\,,\quad
&& r_{\mathrm{in}} > r_{\mathrm{PS}}
\end{alignedat}
\right.\,,
\end{equation}
where the $K_i(\psi)$ and $\tilde{K}(\psi)$ are independent of $\bar{\eta}$, and we will not devote any further attention to these. 

Therefore, the dependence of the specific intensity profile in the photon ring on the plasma physics is only through the values of the redshift and the emission coefficient evaluated in the plasma frame for photons on the circular orbits. 

We see from above that in the absence of substantial emission from inside or on the photon sphere, the characteristic logarithmic scaling with $\bar{\eta}$ is absent. This is interesting to compare against the figures in \cite{Narayan+2019} and \cite{Kocherlakota+2022}. Sec. 5.1 of \cite{Bauer+2022} presents a qualitatively similar result as in eq. \ref{eq:Intensity_Scaling_PR}.

When the emission is sourced by a hot and turbulent plasma that is being accreted onto the central compact object, $j_\nu(\lambda)$ is in general a function of time, and computing the intensity profile at a particular time in the frame of the observer via eq. \ref{eq:I_nu} necessitates accounting for the retarded-time state of the plasma or, equivalently, the retarded emission coefficient (``slow-light''; cf. \citealt{Bronzwaer+2018}). All of this is already encoded in $j_\nu(\lambda)$ and we are able to bypass a (relatively more) tedious numerical computation due to the analysis presented in Appendix \ref{app:AppC_Analytic_Approximations} in arriving at eq. \ref{eq:Intensity_Scaling_PR}. 

The specific flux or flux density ($\mathrm{W}~\mathrm{m}^{-2}\mathrm{Hz}^{-1}$) on the image plane through a ring bounded by two closed curves $\eta_{\mathrm{in}}(\psi)$ and $\eta_{\mathrm{out}}(\psi)$ is then (cf. also \citealt{Bisnovatyi-Kogan+2022}),
\begin{align} \label{eq:F_nu}
F_\nu  &=\ \frac{1}{D^2}\int_0^{2\pi}\int_{\eta_{\mathrm{in}}(\psi)}^{\eta_{\mathrm{out}}(\psi)}  I_\nu(\eta, \psi)~\eta~\mathrm{d}\eta~\mathrm{d}\psi
=\ \frac{\eta_{\mathrm{PS}}^2}{D^2}\int_0^{2\pi}\int_{\bar{\eta}_{\mathrm{in}}(\psi)}^{\bar{\eta}_{\mathrm{out}}(\psi)} I_\nu(\bar{\eta}, \psi)\left[1+\bar{\eta}\right]\mathrm{d}\bar{\eta}~\mathrm{d}\psi \,, 
\end{align}
where $D$ is the distance of the ultracompact object from the observer. In particular, when computing the flux density through a photon subring, where the bounding curves are close to each other, $|\bar{\eta}_{\mathrm{out}} - \bar{\eta}_{\mathrm{in}}| \ll 1$, and are also close to the shadow boundary, $|\bar{\eta}_{\mathrm{in}}, \bar{\eta}_{\mathrm{out}}| \ll 1$, we can use eq. \ref{eq:Intensity_Scaling_PR} to simplify eq. \ref{eq:F_nu} as,
\begin{equation} \label{eq:F_nu_subring}
F_\nu \approx 
\begin{cases}
-\frac{\eta_{\mathrm{PS}}^2}{D^2}\frac{\sqrt{g_{\mathrm{PS}}}}{\kappa_{\mathrm{PS}}}\int_0^{2\pi}\Gamma^2(\eta_{\mathrm{PS}}, \psi, r_{\mathrm{PS}})j_\nu(\eta_{\mathrm{PS}}, \psi, r_{\mathrm{PS}})\left[K_i(\psi) + \ln{(\bar{\eta}_{\mathrm{in}}(\psi)}\right]\left[\bar{\eta}_{\mathrm{out}}(\psi) - \bar{\eta}_{\mathrm{in}}(\psi)\right]~\mathrm{d}\psi\,,
& r_{\mathrm{in}} \leq r_{\mathrm{PS}}\\
\frac{1}{D^2}\int_0^{2\pi}\tilde{K}(\psi) \left[\bar{\eta}_{\mathrm{out}}(\psi) - \bar{\eta}_{\mathrm{in}}(\psi)\right]~\mathrm{d}\psi\,,
& r_{\mathrm{in}} > r_{\mathrm{PS}}
\end{cases}\,.
\end{equation}
If we define $\bar{w}(\psi) = \bar{\eta}_{\mathrm{out}}(\psi)  - \bar{\eta}_{\mathrm{in}}(\psi)$, then for approximately concentric curves, $|\mathscr{M}_\psi (= \partial_\psi\bar{w})| \ll 1$, the above reduces to,
\begin{align} \label{eq:F_nu_subring_approx_concentric}
F_\nu \approx [J]_\psi \left(\bar{\eta}_{\mathrm{out}} - \bar{\eta}_{\mathrm{in}}\right)\,,
\end{align}
where we have used $[\cdot]_\psi$ to denote integration over the image plane polar angle,
\begin{align}
[J]_\psi = 
\begin{cases}
-\frac{\eta_{\mathrm{PS}}^2}{D^2}\frac{\sqrt{g_{\mathrm{PS}}}}{\kappa_{\mathrm{PS}}}\int_0^{2\pi}\Gamma^2(\eta_{\mathrm{PS}}, \psi, r_{\mathrm{PS}})j_\nu(\eta_{\mathrm{PS}}, \psi, r_{\mathrm{PS}})\left[K_i(\psi) + \ln{(\bar{\eta}_{\mathrm{in}}(\psi)}\right]~\mathrm{d}\psi\,,
& r_{\mathrm{in}} \leq r_{\mathrm{PS}}\\
\frac{1}{D^2}\int_0^{2\pi}\tilde{K}(\psi)~\mathrm{d}\psi\,,
& r_{\mathrm{in}} > r_{\mathrm{PS}}\\
\end{cases}\,,
\end{align}
This last approximation may not be appropriate when considering the shapes of photon subrings cast on the image plane of a highly-inclined observer (cf. Fig. \ref{fig:Fig3_Schw_Hotspots} below), and the perpendicular magnification $\mathscr{M}_\psi$ (cf. \citealt{Ohanian1987}) can play a key role in determining the subring asymmetry. Now, from eq. \ref{eq:F_nu_subring_approx_concentric} we can obtain the following general (i.e., for all $r_{\mathrm{in}}$) subring flux density scaling relation,
\begin{align} \label{eq:Flux_Scaling_Photon_Subring}
\frac{F_{\nu; n+1}}{F_{\nu; n}} &\approx\ \frac{\bar{\eta}_{n+1; \mathrm{out}} - \bar{\eta}_{n+1; \mathrm{in}}}{\bar{\eta}_{n; \mathrm{out}} - \bar{\eta}_{n; \mathrm{in}}} = \frac{\bar{w}_{n+1}}{\bar{w}_n} = \frac{w_{n+1}}{w_n} \approx \mathrm{e}^{-\gamma_{\mathrm{PS}}}\cdot\mathrm{e}^{\pm\gamma_{\mathrm{PS}}(2\vartheta_{\mathrm{e}}/\pi-1)}\,. 
\end{align}
We can see from above that in general (independently of the emitting region morphology) the photon subring flux density ratio is simply the ratio of the widths $w_n = \eta_{\mathrm{PS}}\bar{w}_n$ of the subrings (cf. also \citealt{Johnson+2020}). In the above, we should choose the positive sign ($+$) for even $n$ and the negative sign ($-$) for odd $n$ (cf. eq. \ref{eq:Angle_Deflection_Higher_Order_Images_n}).


\section{Photon Ring Calibration Factors}
\label{app:AppE_PR_Calibration}

As discussed above, the properties of the photon subrings, such as their sizes and widths, depend on the material properties of the emission zone such as its morphology, associated plasma emissivity, optical depth, velocity, magnetic fields etc. Furthermore, since photon subrings are higher-order images of the emitting material on the image plane caused by strong gravitational lensing, the spacetime geometry has a role in shaping the photon ring as well. While accessing increasingly higher-order images allows disentangling gravitational effects from other effects with increasing ease, to quantify the impact of the diversity of non-gravitational effects on photon ring characteristics, and to cleanly delineate the influence of non-gravitational physics from spacetime geometry, we leverage the fruitful vocabulary of calibration factors developed in \cite{EHTC+2022f}. 

The $\alpha_1-$calibration factor introduced there related the diameter $d_{\mathrm{m}}$ of the emission ring in the image of Sgr A$^\star$ to the diameter of its shadow boundary $d_{\mathrm{sh}}$ as, $\alpha_1 = \Braket{d_{\mathrm{m}}}_\psi/\Braket{d_{\mathrm{sh}}}_\psi$, where we have used $\braket{d}_\psi$ to indicate the median value of a polar curve $d(\psi)$ over the image plane polar angle $0 \leq \psi < 2\pi$. This calibration factor provides insights into the physical state of the accreting system: Images of accreting BHs for which $\alpha_1 - 1$ is close to zero are (retarded time) snapshots of the dynamical flow when the largest amount of emission (the emissivity peak) is sourced extremely close to the photon shell in the bulk (see also \citealt{Ozel+2021, Younsi+2021, Kocherlakota+2022}). 

In a similar vein, we now introduce the subring diameter calibration factors $\alpha_{1; n}$ as 
\begin{equation} \label{eq:alpha_1n_PR}
\alpha_{1; n} := \Braket{d_{n}}_\psi/\Braket{d_{\mathrm{sh}}}_\psi = 1 + \Braket{\bar{\eta}_{n; \mathrm{out}}}_\psi\,,
\end{equation}
where $d_n(\psi) = \eta_{n; \mathrm{out}}(\psi) + \eta_{n; \mathrm{out}}(\psi+\pi)$ is the diameter of the $n^{\mathrm{th}}$ photon subring, with $(\eta, \psi) = (\eta_{n; \mathrm{out}}(\psi), \psi)$ describing the outer edge of the order$-n$ image of the emitting source, or equivalently, the order$-n$ image of its outer boundary. In writing the above, we have used the fact that in static and spherically-symmetric spacetimes the shadow boundary curve is perfectly circular, $\braket{d_{\mathrm{sh}}}_\psi = 2\eta_{\mathrm{PS}}$. We use the median here for consistency with \cite{EHTC+2022f} but this can be replaced with any characteristic measure of the diameter in principle. As above, if $\alpha_{1; n} - 1$ is small, then the emissivity peak is located close to the photon shell in the bulk. 

We have established above (cf. Sec. \ref{sec:Sec5_Photon_Rings}) that $\alpha_{1; n} - 1$ is typically substantially smaller than $\alpha_{1} - 1$, meaning that a measurement of the diameter of the first subring will yield a more precise inference of the shadow diameter than is currently available. 

Furthermore, the fractional variation in the subring diameter due to varying emitting region morphology is simply the difference between the maximum and minimum values that the relevant subring calibration factor takes over the range of morphological parameters,
\begin{align} \label{eq:Fractional_Variation_Subring_Diameter}
\left(\mathrm{max.}\left[\Braket{d_{n}}_\psi\right] - \mathrm{min.}\left[\Braket{d_{n}}_\psi\right]\right)/\Braket{d_{\mathrm{sh}}}_\psi  =\ \mathrm{max.}\left[\alpha_{1; n}\right] - \mathrm{min.}\left[\alpha_{1; n}\right] =: \Delta\alpha_{1; n}\,. 
\end{align}
From Sec. \ref{sec:Sec5_Photon_Rings} we see that the fractional subring diameter variation diminishes exponentially with increasing image order as $\Delta\alpha_{1; n+1} \approx \mathrm{e}^{-\gamma_{\mathrm{PS}}}\Delta\alpha_{1; n}$, in line with our expectation, meaning that the variations in the emitting region morphology become concomitantly suppressed. 

Combining the two statements above, with increasing order of image, on the one hand, we can obtain increasingly better estimates of the shadow boundary, whereas, on the other, the impact of the non-gravitational physics on determining the subring diameter becomes increasingly unimportant. Equivalently, by measuring the diameters of increasingly higher-order photon subrings (i.e., with increasing $n$), we obtain increasingly accurate ($\alpha_{1; n} - 1 \rightarrow 0$) as well as increasingly precise ($\Delta\alpha_{1; n} \rightarrow 0$) estimates of the shadow boundary diameter $d_{\mathrm{sh}}$, which depends purely on the spacetime geometry. 

Finally, due to the approximate \eqref{eq:n_n+1_Lensing_Lyapunov_Exp_Inclination} scaling relations between the fractional diameters of the subrings (see eq. \ref{eq:n_n+1_Lensing_Lyapunov_Exp_Face_On}),
\begin{equation} \label{eq:Diameter_Scaling_Reln}
(d_{\mathrm{n+1}}/d_{\mathrm{sh}} - 1)/(d_{\mathrm{n}}/d_{\mathrm{sh}} - 1) 
= \bar{\eta}_{n+1; \mathrm{out}}/\bar{\eta}_{n; \mathrm{out}} 
\approx \mathrm{e}^{-\gamma_{\mathrm{PS}}}\,,
\end{equation}
the calibration factors \eqref{eq:alpha_1n_PR} corresponding to consecutive subrings will also obey an approximate scaling relation,
\begin{equation} \label{eq:Calib_Scaling_Reln}
(\alpha_{1; n+1} - 1)/(\alpha_{1; n} - 1) \approx \mathrm{e}^{-\gamma_{\mathrm{PS}}}\,.
\end{equation}
Thus, a measurement of two subring diameters will yield an approximate measurement of the lensing Lyapunov exponent $\gamma_{\mathrm{PS}}$. 

\bigskip

\bsp	
\label{lastpage}
\end{document}